%% file: MSSnop.tex
\documentclass[oneside]{amsart}

\usepackage[process=all,crop=pdfcrop]{pstool} 
\usepackage[usenames,dvipsnames]{xcolor}
\usepackage{amssymb}
\usepackage{graphicx,psfrag}
\usepackage{xspace}
\usepackage{nicefrac}
\usepackage{mathtools}
\usepackage{tikz}
\usepackage{xcolor}
\usepackage{url}
\usepackage[all]{xy}
\usepackage{soul} 

\input{header}

\MakeLemmaReference{WVT}{lem:wvt}
\MakeLemmaReference{Worldline Relocation}{lem:worldline-relocation}
\MakeLemmaReference{Observer Rotation}{lem:relocation}
\MakeLemmaReference{Rotated Observer Lines}{lem:rotated-observer-lines}
\MakeLemmaReference{Observer Line Intersections}{lem:observer-line-intersections}
\MakeLemmaReference{Transformed Observer Lines}{lem:transformed-observer-lines}
\MakeLemmaReference{Horizontal Rotation}{lem:vanrotacio}
\MakeLemmaReference{Same-Slope Rotation}{lem:2vanrotacio}
\MakeLemmaReference{Observer Line Intersections}{lem:observer-line-intersections}
\MakeLemmaReference{Observer Lines Lemma}{lem:observer-lines}

\MakeLemmaReference{Speed}{lem:speed}
\MakeLemmaReference{Triangulation}{lem:triangulation}
\MakeLemmaReference{Triads}{lem:triads}
\MakeLemmaReference{Plane-to-Plane}{lem:plane-to-plane}
\MakeLemmaReference{Infinite Speeds $\Rightarrow$ Lines are Observer Lines}{lem:straight-lines-are-observer-lines}
\MakeLemmaReference{Line-to-Line Lemma}{lem:line-to-line}

\MakeLemmaReference{IOb$_o$}{lem:iobo2}
\MakeLemmaReference{Affine}{lem:affine}
\MakeLemmaReference{$\triv = \bigcap\kiso$}{lem:xy}
\MakeLemmaReference{Equal Worldlines}{lem:equal-worldlines}
\MakeLemmaReference{Colocate}{lem:colocate}
\MakeLemmaReference{$tx$-Plane Lemma}{lem:txplane}

\MakeLemmaReference{Translation to IOb$_o$}{lem:iobo}
\MakeLemmaReference{Vertical Plane Rotation}{lem:letezikrotacio}
\MakeLemmaReference{LinTriv $\Rightarrow$ Same Speed}{lem:lintriv-implies-same-speed}
\MakeLemmaReference{Same-Speed Lemma}{lem:same-speed-hard}

\MakeLemmaReference{Configuration}{lem:configuration}
\MakeLemmaReference{Same Speed Easy}{lem:same-speed-easy}
\MakeLemmaReference{Quadratic IVT}{lem:quadratic}
\MakeLemmaReference{Fundamental Lemma}{lem:fundamental}

\MakeLemmaReference{Rest}{lem:rest}
\MakeLemmaReference{Observer Origin}{lem:observer-origin}
\MakeLemmaReference{Median Observer}{lem:median-observer}
\MakeLemmaReference{$\kappa$ is unique}{lem:k-is-unique}
\MakeLemmaReference{Main Lemma}{lem:main}

\MakeLemmaReference{Characterisation}{thm:characterisation}
\MakeLemmaReference{Model Construction}{thm:group-instantiation}
\MakeLemmaReference{Determination}{thm:classification1}
\MakeLemmaReference{Classification}{thm:classification2}
\MakeLemmaReference{Consistency}{thm:consistency}

\MakeLemmaReference{Alternatives for \W}{cor:main}
\MakeLemmaReference{Satisfaction}{lem:group}

\begin{document}

\title{Groups of Worldview Transformations Implied by Isotropy of Space}

\author{Judit X. Madar\'asz}
\address{Judit X. Madar\'asz, R\'enyi Institute, Hungary}

\author{Mike Stannett}
\address{Mike Stannett, University of Sheffield, UK}

\author{Gergely Sz\'ekely} \address{Gergely Sz\'ekely, R\'enyi
  Institute, Hungary \and University of Public Service, Hungary}


\date{}

\begin{abstract}
Given any Euclidean ordered field, $\Q$, and any `reasonable' group,
$\G$, of (1+3)-dimensional spacetime symmetries, we show how to
construct a model $\M_{\G}$ of kinematics for which the set $\W$ of
worldview transformations between inertial observers satisfies $\W =
\G$. This holds in particular for all relevant subgroups of $\gal$
$\cpoi$, and $\ceucl$ (the groups of Galilean, Poincar\'e and
Euclidean transformations, respectively, where $c \in Q$ is a
model-specific parameter corresponding to the speed of light in the
case of Poincar\'e transformations).
  
In doing so, by an elementary geometrical proof, we demonstrate our
main contribution: spatial isotropy is enough to entail that the set
$\W$ of worldview transformations satisfies either $\W \subseteq \gal$, $\W
\subseteq \cpoi$, or $\W \subseteq \ceucl$ for some $c > 0$. So
assuming spatial isotropy is enough to prove that there are only 3
possible cases: either the world is classical (the worldview
transformations between inertial observers are Galilean
transformations); the world is relativistic (the worldview
transformations are Poincar\'e transformations); or the world is
Euclidean (which gives a nonstandard kinematical interpretation to
Euclidean geometry).  This result considerably extends previous
results in this field, which assume a priori the (strictly stronger)
special principle of relativity, while also restricting the choice of
$\Q$ to the field $\mathbb{R}$ of reals.

As part of this work, we also prove the rather surprising result that, for any $\G$ containing translations and rotations fixing the time-axis $\taxis$, the requirement that $\G$ be a subgroup of one of the
groups $\gal$, $\cpoi$ or $\ceucl$ is logically equivalent to the
somewhat simpler requirement that, for all $g \in \G$: $g[\taxis]$ is
a line, and if $g[\taxis] = \taxis$ then $g$ is a trivial
transformation (\ie $g$ is a linear transformation that preserves
Euclidean length and fixes the time-axis setwise).
\end{abstract}

\maketitle

\section{Introduction}

Physical theories conventionally define coordinate systems and
transformations using values and functions defined over the field of
reals, $\mathbb{R}$. However, this assumption is not well-founded in
physical observation because all physical measurements yield only
finite-accuracy values --- even quantum electrodynamics (QED), one of
the most precisely tested physical theories, is only accurate to
around 12 decimal digits~\cite{QEDAccuracy}. Since we have no
empirical reason to make this assumption, it is worth investigating
what happens to our expectations of physical theories if we generalize
by assuming less about the physical quantities used in
measurements. In this paper, we assume only that every positive
element in the ordered field of quantities has a square root, but it
is worth noting that special relativity can also be modelled over the
field of rational numbers~\cite{MSzRac}, in which even this assumption
fails. It remains an open question whether the new results presented
here generalize over arbitrary ordered fields.

Starting in 1910, Ignatovsky's \cite{Ign1910c,Ignatowsky,Ign1911b}
attempt to derive special relativity assuming only Einstein's
principle of relativity initiated a new research direction
investigating the consequences of assuming the principle of relativity
without Einstein's light postulate. However, Frank and Rothe
\cite{Frank+Rothe} quickly identified (1911) that hidden assumptions
were implicitly used by both Einstein and Ignatovsky, and it is still
not uncommon over a century later to find hidden assumptions in
related works.

One notable investigation was that of Borisov
\cite{Borisov1978} (see also \cite[\S 10, pp.~60-61]{Guts}). Borisov
explicitly introduced all the assumptions used in his framework
investigating the consequences of the principle of relativity. Then he
showed that there are basically two possible cases: either the world is
classical and the worldview transformations between inertial observers
are Galilean; or the world is relativistic and the worldview
transformations are Poincar\'e transformations.\footnote{\,Metric geometries corresponding to these two structures also appear
  among Cayley-Klein geometries; see, \eg 
  \cite{Struve16} and \cite[\S 6]{Pambuccian17}.}

In \cite{OurBorisov}, we made Borisov's framework even more
explicit using first-order logic, and investigated the role of his
assumption that the structure of physical quantities is the field of
real numbers. We showed that over non-Archimedean fields there is
a third possibility: the worldview transformations can also be
Euclidean isometries.\footnote{\,That the principle of relativity is
  consistent with world view transformations being Euclidean
  isometries has previously been shown by Gyula D\'avid~\cite{dgyNoPh}.}

In this paper, we present a general axiom system for kinematics using
a simple language talking only about quantities, inertial observers
(coordinate systems), and the worldview transformations between
them. Our axiom system is based on just a few natural assumptions, \eg
instead of assuming that the structure of physical quantities is the
field of real numbers we assume only that it is an ordered field $\Q$
in which all non-negative values have square roots. Using this
framework, we investigate what happens if instead of the principle of
relativity we make the weaker assumption that space is isotropic. We
show that isotropy is already enough to ensure that the worldview
transformations are either Euclidean isometries, or Galilean or
Poincar\'e transformations; see \CiteTheorem{thm:classification2}.

The investigation presented in this paper is part of the Andr\'eka--N\'emeti
school's general project of logic-based axiomatic foundations of
relativity theories, see \eg
\cite{AMNsamples,logst,Synthese,Comparing}. Friend and Molinini \cite{Friend15, FriMol15}
discuss the significance of this project and the underlying methodology
from the viewpoints of epistemology and explanation in science. One important feature of using a first-order logic-based axiomatic framework is that it helps avoid hidden assumptions, which is fundamental in foundational analyses of this nature. Another feature is that it opens up the possibility of machine verification of the results, 
see~\eg~\cite{StannettNemeti,GBT15}.

\section{Framework}\label{sec:framework}
We are concerned in this paper with two sorts of objects,
\lemph{(inertial)} \semph{observers} and \semph{quantities}, which we
represent as elements of non-empty sets $\IOb$ and $\Q$,
respectively. 

Observers are interpreted to be labels for inertial coordinate systems. Quantities
are used to specify coordinates, lengths and related quantities, and
we assume that $\Q$ is equipped with the usual binary operations,
$\cdot$ (multiplication) and $+$ (addition); constants, $0$ and $1$
(additive and multiplicative identities); and a binary relation,
$\leq$ (ordering).

Although the results presented here can also be generalized to
higher-dimensional spaces (though not necessarily lower-dimensional
ones --- see Sect. \ref{sec:discussion}), we assume for definiteness
that observers inhabit 4-dimensional spacetime, $\Q^4$, and locations
in spacetime are accordingly represented as $4$-tuples over $\Q$.  We
often write $\vvp$, $\vvq$ and $\vvr$ to denote generic spacetime
locations.

For each pair of observers $k, h \in \IOb$, we assume the existence of
a function $\w{k}{h} \colon \Q^4 \to \Q^4$, called the
\semph{worldview transformation}\lemph{ from the worldview of $h$ to
  the worldview of $k$}, which we interpret as representing the idea
that observers may see (\ie coordinatize) the same events, but at
different spacetime locations: whatever is seen by $h$ at $\vvp$ is
seen by $k$ at $\w{k}{h}\take\vvp$.\footnote{\,In more general
  theories, for example in general relativity, this relation
  need not be a function or even defined on the whole $\Q^4$, because
  an event seen by $k$ may be invisible to $h$ or may appear at one or
  more different spacetime locations from $h$'s point of view, but in
  this paper we assume that all observers completely and unambiguously
  coordinatize the same universe --- they all see the same events,
  albeit in different locations relative to one another.}

Formally, this framework corresponds to using a two
  sorted first-order language where the models are of the following form
  \begin{equation*}
    \M=( \IOb,\Q, +,\cdot,0,1,\le,\w{}{}), 
  \end{equation*}
where: $\IOb$ and $\Q$ are two sorts; $+$ and $\cdot$ are binary
operations on $\Q$; $0$ and $1$ are constants on $\Q$; $\le$ is a
binary relation on $\Q$; and $\w{}{}$ is a function from
$\IOb\times\IOb\times\Q^4$ to $\Q^4$. In this language, the worldview
transformation between fixed observers $k$ and $h$ can be introduced
as:
\begin{equation*}
  \w{k}{h}\take{t,x,y,z}\de\w{}{}(k,h,t,x,y,z).
\end{equation*}

\section{Axioms}
\label{sec:kin-axioms}

In this section, we describe the general axiom system, $\KIN$, used to represent kinematics in this paper. Additional axioms representing spatial isotropy and the special principle of relativity will be introduced in Section~\ref{sec:spr-iso-axioms}.

\subsection{Quantities}
We assume that $(\Q, +, \cdot,0,1,\leq)$ exhibits the most fundamental
 algebraic properties expected of the real numbers
($\mathbb{R}$), so that calculations can be performed and results
compared with one another. We also assume that square-roots are
defined for non-negative values (\ie that $\Q$ is a \semph{Euclidean
  field} \cite{DefnOfEuclideanField}).

\begin{newaxiom} \label{ax:efield} 
\item[\ul{\AxEField}] $(\Q, +, \cdot,0,1, \leq)$ is a Euclidean
  field, \ie a linearly ordered field in which every non-negative
  element has a square root.
\end{newaxiom}
Assuming \AxEField also means that the derived operations of
subtraction ($-$), division ($/$), square root ($\sqrt{\phantom{o}}$),
dot product of vectors ($\cdot$), Euclidean length of vectors,
etc.,\ are well-defined on their domains, and allows us to assume the usual vector space
structure of $\Q^4$ over $\Q$. We will generally omit the
multiplication symbol.

\subsection{Worldview transformations}
The following axiom states informally that: (i) the worldview transformation from an
observer's worldview to itself is just the identity
transformation, $\id \colon \Q^4 \to \Q^4$; and (ii) switching from $k$'s
worldview to $h$'s and then to $m$'s has the same effect as switching
directly from $k$'s worldview to $m$'s.
\begin{newaxiom}
\item[\ul{\AxWvt}]
For all $k,h,m \in \IOb$:
\begin{itemize}
\item[(i)]  $\w{k}{k} = \id$;
\item[(ii)] $\w{m}{h} \circ \w{h}{k} = \w{m}{k}$.
\end{itemize}

\end{newaxiom}

\subsection{Lines, worldlines and motion}
By assumption, all of the locations under discussion in this paper are
points in $\Q^4$. We often write $(t,x,y,z)$ to indicate the
coordinates of a generic point in $\Q^4$. Given any $n > 0$ and  $\vvp=(p_1,p_2,\ldots,p_n)\in\Q^n$, its 
\semph{squared length}, $|\vvp|^2$, is defined by
  \begin{equation*}
    |\vvp|^2\de p_1^2 + \ldots + p_n^2.
  \end{equation*}
(This is just the standard Euclidean squared length of \vvp.) 

To simplify our notation, we write $\origin \de (0,0,0,0)$ for the
zero-vector (origin) in $\Q^4$. More generally, we sometimes write $\vv 0$ for any tuple of zeroes (the length will always be clear from context). We define
the \semph{time-axis}, \taxis, and the \semph{present simultaneity}, \saxis, 
to be the set
\begin{equation*}
  \taxis \de \{(t,0,0,0) : t \in \Q\}.
\end{equation*}
and the spatial hyperplane
\begin{equation*}
  \saxis \de \{(0,x,y,z) : x,y,z \in \Q\},
\end{equation*}
respectively. We write \tunit for the unit time vector $(1,0,0,0)$, and likewise $\xunit \de (0,1,0,0)$, $\yunit \de (0,0,1,0)$ and $\zunit \de (0,0,0,1).$ If $\vvp=(t,x,y,z)\in\Q^4$, we call $\vvp_t \de t$ the \semph{time component}, and $\vvp_s \de (x,y,z)$ the \semph{space component}, of $\vvp$. 
Finally, if $t \in \Q$ and $\vv s \in \Q^3$, we write $(t,\vv s)$ for the point with time component $t$ and space component $\vv s$.

The \semph{worldline} \lemph{of observer $h$ according to observer $k$} is defined as
\[
\wl_k (h) \de \w{k}{h}\takeset\taxis .
\]
In particular, if we assume \AxWvt and take $k = h$, we have $\wl_h(h)
= {\w{h}{h}\takeset\taxis} = \taxis$. This corresponds to the
convention that observers consider themselves to be at the
spatial origin relative to which measurements are made: from their own
viewpoint their worldline is the time-axis; and $\wl_k(h) =
\w{k}{h}[\taxis] = \w{k}{h}[\wl_h(h)]$ describes the same worldline
but from $k$'s point of view.

When we say that one observer moves \lemph{inertially} with respect to
another, we mean that neither of them accelerates relative to the
other, so that linear motions seen by one remain linear when seen by
the other. Since each observer considers its own worldline to be the
line $\taxis$, we would expect all inertial observers
to agree that each others' world lines are lines.

Formally, a subset $\ell\subseteq\Q^4$ is a \semph{line}
if{}f there are $\vvp,\vvv\in\Q^4$, where $\vvv \neq \origin$ and
$\ell=\{\vvp+\lambda\vvv\: :\: \lambda\in\Q\}.$ The next axiom states
that worldlines of observers according to observers are
lines.
\begin{newaxiom}
\item[\ul{\AxLine}]  For every $k, h \in\IOb, \wl_k(h)$ is a line.
\end{newaxiom}

According to \AxLine, the worldlines of observers are 
lines, and by \AxWvt each observer considers its own worldline to be
the time-axis; we can therefore express the idea that \lemph{observer
  $k$ is }\semph{moving according to} \lemph{observer $m$} by saying
that $\wl_{m}(k)$ is not parallel to \taxis,\footnote{As one would
  expect, being in motion relative to another observer --- and likewise
  being at rest --- are symmetric relations; see \CiteLemma{lem:rest}.}
or more simply, that $\w{m}{k}$ takes the time-unit vector $\tunit$
and the zero-vector $\origin$ to coordinate points having different
spatial components, \ie $\w{m}{k}\take{\tunit}_s \neq
\w{m}{k}\take{\origin}_s$. In the same spirit, we say that $k$
\semph{is at rest according to} $m$ if{}f $\w{m}{k}(\tunit)_s =
\w{m}{k}(\origin)_s$.

We will sometimes need to assume explicitly the existence of observers moving relative to one another, which we express using the following formula: 
\begin{newaxiom}
\item[\ul{\EMovingIOb}] There are observers $m,k\in\IOb$ such
  that $\w{m}{k}\take{\tunit}_s \neq \w{m}{k}\take{\origin}_s$.
\end{newaxiom}

\subsection{Trivial transformations}

We say that a linear transformation $T:\Q^4\to\Q^4$ is a \semph{linear
  trivial transformation} provided it fixes (setwise) both the
time-axis and the present simultaneity, and preserves squared lengths
in both, \ie
\begin{itemize}
\item 
  if $\vvp\in\taxis$, then $T(\vvp)\in\taxis$ and $T(\vvp)_t^2=\vvp_t^2$;
   and
\item
  if $\vvp\in\saxis$, then $T(\vvp)\in\saxis$ and
$|T(\vvp)_s|^2=|\vvp_s|^2$. 
\end{itemize}

\begin{rem}\label{rem:triv}
Assuming \AxEField, the statement that $T$ is a linear trivial
transformation is equivalent to the statement that $T$ is a linear
transformation that preserves Euclidean length and fixes the
time-axis setwise.\footnote{\,This claim follows by \CiteLemma{lem:xy}, but can also be proven directly. Suppose $T$ is linear, preserves Euclidean length and fixes \taxis setwise. It follows immediately that $T(\tunit) = \pm\tunit$. Now choose any $(0,\vv s) \in \saxis$, and suppose $T(0,\vv s) = (t',\vv s')$. Then $|T(\pm 1,\vv s)|^2 = |T(0,\vv s) \pm T(\tunit)|^2 = (t' \pm 1)^2 + |\vv s'|^2$. Since $|(1,\vv s)|^2 = |(-1,\vv s)|^2$ and $T$ preserves Euclidean length, we therefore require $(t' + 1)^2 + |\vv s'|^2 = (t' - 1)^2 + |\vv s'|^2$, whence $t' = 0$. Thus, $T$ also fixes \saxis, so it is a linear trivial transformation. The converse is trivial.}\QED
\end{rem}

A map $f:\Q^4\to\Q^4$ is a \semph{translation} if{}f there is
$\vvq\in\Q^4$ such that $f(\vvp)=\vvp+\vvq$ for every $\vvp\in\Q^4$. We
write $\tran$ for the \emph{set of all translations}.

A transformation is called a
\semph{trivial transformation} if it is a linear
trivial transformation composed with a translation. We
write $\triv$ for the \emph{set of all trivial transformations}.

We say that two observers $k$ and $k'$ are \semph{co-located} if they
consider themselves to share the same worldline: $\wl_k(k)=\wl_k(k')$
(assuming \AxWvt, this relationship is
symmetric; see \CiteLemma{lem:equal-worldlines}).  The following axiom
says that, if observers $k$ and $k'$ are co-located, then their
worldviews are related to one another by a trivial transformation. In
other words, even though inertial observers following the same
worldline may use different coordinate systems, these coordinate
systems can only differ by using a different orthonormal basis for
coordinatizing space and/or a different direction and origin of
time.\footnote{\,By $\AxWvt$, if $k$ and $k'$ are co-located, \ie
  $\wl_k(k)=\wl_k(k')$, then $\w{k}{k'}\takeset{\taxis}=\taxis$. This
  is why we do not need to assume explicitly in the statement of
  $\AxColocate$ that co-located observers share the same time-axis.}
\begin{newaxiom} 
\item[\ul{\AxColocate}] For all $k, k' \in \IOb$, if
  $\wl_k(k)=\wl_k(k')$, then $\w{k}{k'} \in \triv$.
\end{newaxiom}

\subsection{Spatial rotations.}
A linear trivial transformation
$R:\Q^4\to\Q^4$ is called a \semph{spatial rotation} if{}f it
preserves the direction of time and the orientation of space, \ie
$R(\tunit)=\tunit$ and the determinant of $3\times 3$ matrix
$[R(\xunit)_s,R(\yunit)_s,R(\zunit)_s]$ is positive.\footnote{\,This can
  be expressed in our formal language without any assumption about the
  structure of quantities as:
  $R(\xunit)_2R(\yunit)_3R(\zunit)_4+R(\xunit)_4R(\yunit)_2R(\zunit)_3+R(\xunit)_3R(\yunit)_4R(\zunit)_2>R(\xunit)_4R(\yunit)_3R(\zunit)_2+R(\xunit)_2R(\yunit)_4R(\zunit)_3+R(\xunit)_3R(\yunit)_2R(\zunit)_4$,
  here $R(\vvp)_2$, $R(\vvp)_3$, and $R(\vvp)_4$ denotes the second,
  third and fourth component of $R(\vvp)\in\Q^4$, \ie if
  $R(\vvp)=(t,x,y,z)$, then $R(\vvp)_2=x$, $R(\vvp)_3=y$, and
  $R(\vvp)_4=z$.} We denote the \emph{set of all spatial rotations}
by $\srot$.

The following axiom says that translated and spatially rotated
versions of any inertial coordinate system are also inertial
coordinate systems.\footnote{\,The quantification over $T$ in \AxRelocate appears at first sight to be second-order. However, because translations are determined by the image of
    the origin, while spatial rotations are determined by the images of
    the three spatial unit vectors, this axiom can be formalized in our
    first-order logic language by quantifying over the 4 parameters
    representing the image of the origin and the 12 parameters
    representing the images of the three spatial unit vectors.}

\begin{newaxiom}
\item[\ul{\AxRelocate}] For all $k\in\IOb$ and for
  all
  $T\in\tran\cup\srot$, there is $h\in\IOb$ such that $\w{k}{h}=T$.
\end{newaxiom}

The underlying axiom system with which we are concerned in this paper is
\[
   \KIN \de \{\AxEField, \AxWvt, \AxLine, \AxRelocate, \AxColocate \},
\]
 which defines our basic theory of the \semph{kinematics} of inertial observers.

\section{The special principle of relativity, isotropy and set of worldview transformations}
\label{sec:spr-iso-axioms}
There are many different formal interpretations of the
principle of relativity \cite{Gom15,GSz15,MSS3SPR}. In this paper, we
interpret the \semph{special principle of relativity} (\emph{SPR}) to
mean that all inertial observers agree as to how they are related to
other observers, so that no observer can be distinguished from any
other in terms of the things they can and cannot (potentially)
observe. We express this via the following axiom:
\begin{newaxiom}
\item[\ul{\AxSpr}]
For every $k, k^*, h \in\IOb$, there exists $h^*\in\IOb$ such that $\w{k}{h}=\w{k^*}{h^*}$,
\end{newaxiom}
that is, given observers $k, k^*, h$, there must (potentially) be some $h^*$ which is related to $k^*$ in exactly the same way that $h$ is related to $k$, \ie the geometrical structure of spacetime cannot forbid such an observer.

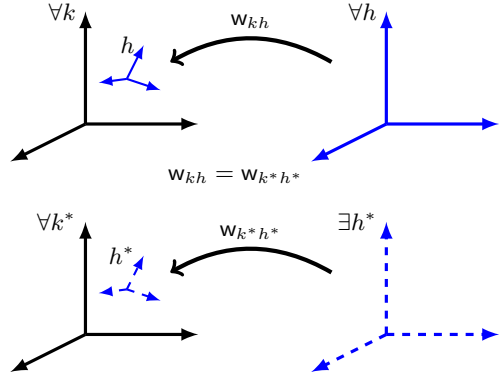
\begin{figure}[!htb]
  \begin{center}
    \input{axspr.tikz}
    \caption{Isotropy and the special principle of relativity. The special principle, \AxSpr, says that given any $k$, $h$ and $k^*$, there exists an $h^*$ that is related to $k^*$ the same way that $h$ is related to $k$ (\ie there are no distinguished inertial coordinate systems). Spatial isotropy, \AxIso, is similar, except that we only require $h^*$ to exist when $\w{k}{k^*}$ is a spatial rotation (\ie rotating ones spatial coordinate system has no effect on what can and cannot potentially be seen).}
    \label{fig:axspr}
  \end{center}
\end{figure}

In contrast, \semph{isotropy} refers to the weaker constraint that there is no distinguished direction in space, \ie no matter which direction we face, we should be able to perform the same experiments and observe the same outcomes. Isotropy can be expressed in much the same way as SPR, except that we only require equivalence as to what can be observed ($h$) when the relevant observers ($k$ and $k^*$) are related via a spatial rotation (see Figure~\ref{fig:axspr}):

\begin{newaxiom}
\item[\ul{\AxIso}]
For every $k, k^*, h \in\IOb$, if $\w{k}{k^*} \in \srot$, there exists $h^*\in\IOb$ such that $\w{k}{h}=\w{k^*}{h^*}$.
\end{newaxiom}

In order to investigate these ideas, we will need to consider various
sets of worldview transformations, and attempt to establish both their
algebraic properties and the relationships between them. The set
$\W_k$ of worldview transformations associated with a specific
observer $k \in \IOb$ will be defined by
\begin{equation*}
  \W_k\de \{ \w{k}{h}: h\in \IOb \}
\end{equation*}
and the set of \emph{all} worldview transformations is then given by
\begin{equation*}
  \W \de \{ \w{k}{h}: k,h\in \IOb\} = \bigcup \{ \W_k : k \in \IOb \} .
\end{equation*}

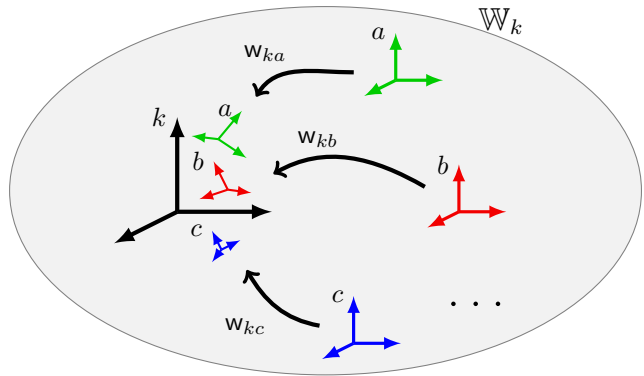
\begin{figure}[!hbt]
  \begin{center}
    \input{Wk2.tikz}
    \caption{The set $\W_k$ of all worldview transformations into $k$'s coordinate system. For each observer $a, b, c, \dots$, the set $\W_k$ contains the associated transformation $\w{k}{a}, \w{k}{b}, \w{k}{c}, \dots$.}
  \end{center}
\end{figure}

Using these notations \AxSpr can be reformulated as saying that all
inertial observers have essentially the same worldview, \ie $\W_k =
\W_{k^*}$ for all $k, k^* \in \IOb$. Although it is not immediately
obvious that any $\W_k$ can form a group, if we assume \AxWvt it can
be proven that \AxSpr is equivalent to saying that there is at least
one $k$ for which $\W_k$ forms a group under composition, which is
itself equivalent to saying that $\W_k=\W$.  For the proof of this and
other equivalent formulations of \AxSpr, see
\cite[Prop.~2.1]{OurBorisov}. Similarly, \AxIso is equivalent to
saying, for all $k,k^* \in \IOb$, if $\w{k}{k^*} \in \srot$, then
$\W_k = \W_{k^*}$.

\begin{rem}
We have already noted that \AxSpr entails \AxIso, so that the special
principle of relativity is at least as strong assumption as spatial
isotropy. In fact, it is strictly stronger, because $\W$ is a group in
all models of $\KIN + \AxIso$, but $\W_k$ need not be. In particular,
therefore, $\KIN + \AxIso$ does not imply $\AxSpr$. This remains true
even if we add the restriction that $(\Q,+,\cdot,0,1,\le)$ is the
ordered field of real numbers. However, if we add the assumption that
co-located observers agree on the direction of time, then it can be
shown that $\KIN + \AxIso$ implies $\AxSpr$.
\end{rem}

For easy reference, Table \ref{tab:axioms} summarizes the axioms used in this paper and discussed above.

\begin{table}[!htb]
\caption{Our axioms and their intuitive meanings.}
\begin{tabular}{|c|l|l|l|} \hline
\KIN   &  \textsf{Axiom} & \textsf{Description} \\ \hline\hline
\tick  &  \AxEField & \entrypar{the set $\Q$ of quantities is an ordered field in which all non-negative values have square roots} \\[6pt]\hline
\tick  &  \AxWvt & \entrypar{$\w{k}{k}$ transforms $k$'s worldview to itself identically; and  going from $k$'s
worldview to $h$'s and then to $m$'s is same as going directly
from $k$'s worldview to $m$'s} \\[6pt]\hline
\tick    & \AxLine & \entrypar{inertial observers see each other's
  worldlines as lines} \\[6pt]\hline
\tick  &    \AxColocate & \entrypar{if two observers are co-located, their 
worldviews are trivially related to one another} \\[6pt]\hline
\tick  &  \AxRelocate & \entrypar{translated and spatially rotated versions of inertial coordinate systems are also inertial} \\[6pt]\hline
\ptick &  \AxSpr  & \entrypar{the special principle of relativity} \\[6pt]\hline
\ptick & \AxIso  & \entrypar{isotropy of space} \\[6pt]\hline
\end{tabular}
\label{tab:axioms}
\end{table}

\section{Main theorems}
\label{sec:main-results}

First let us introduce the transformations that will be used in this
paper to characterize the worldviews of observers.  In this section,
we assume that $(\Q,+,\cdot,0,1)$ is a field. Table \ref{tab:groups}
summarizes the various transformation groups referred to in the
theorems.
\begin{table}[!htb]
\caption{Transformation groups considered in this paper.}
  \begin{tabular}{|c|lcr|} \hline 
\tran     & translations                   &   & \\ \hline
\srot     & spatial rotations              &   & \\ \hline
\triv     & trivial transformations        &   & \\ \hline
\kiso     & $\kappa$-isometries            &   & \\ \hline
\cpoi     & $c$-Poincar\'e transformations & = & $\kiso[\nicefrac{1}{c^2}]$ \\ \hline
\ceucl    & $c$-Euclidean transformations  & = & $\kiso[\nicefrac{-1}{c^2}]$ \\ \hline
\gal      & Galilean transformations       & = & \kiso[0] \\ \hline
\end{tabular}
\label{tab:groups}
\end{table}

\subsection{$\kappa$-isometries}

Given $\vvp = (t,x,y,z)$, the \semph{(squared) $\kappa$-length} of
$\vvp$ is defined by
\[
 \knorm{(t,x,y,z)}^2 \de t^2 - \kappa (x^2 + y^2 + z^2) ,
\]
or in other words,
\begin{equation*}
\knorm{\vvp}^2\de \vvp_t^2-\kappa|\vvp_s|^2.
\end{equation*}
Taking $\kappa = 1$ gives the squared \emph{Minkowski length}
$\knorm[1]{\vvp}^2 = t^2 - (x^2 + y^2 + z^2)$ of $\vvp$, while $\kappa
= -1$ gives its squared \emph{Euclidean length}, $\knorm[-1]{\vvp}^2
=\norm{\vvp}^2 = {t^2 + (x^2 + y^2 + z^2)}$.


\begin{defn}[$\kappa$-isometry, $\kappa \neq 0$]
If $\kappa \neq 0$, we call a linear transformation $f:\Q^4\rightarrow\Q^4$ a \semph{linear $\kappa$-isometry} provided it preserves $\kappa$-length, \ie  for every $\vvp\in\Q^4$, 
\[
    \knorm{f(\vvp)}^2 = \knorm{\vvp}^2.
\]
\end{defn}

In the case of $\kappa=0$, we require more than simply preserving $0$-length, for while $0$-length takes account of temporal extent, it ignores spatial structure. We therefore need to add an extra condition to the definition of $0$-isometry to ensure that spatial structure is also respected when considering points with equal time coordinates.\footnote{Although every $0$-isometry preserves $0$-length, the converse is not true.}

\begin{defn}[$\kappa$-isometry, $\kappa = 0$]
Let $f:\Q^4\rightarrow\Q^4$ be a linear transformation. We call $f$ a \semph{linear $0$-isometry} provided, for every $\vvp\in \Q^4$,
\begin{equation} \label{eqn:kiso-defn}
  f(\vvp)_t^2 =  \vvp_t^2 
         \text{ and } 
         \left( \vvp_t=0 \ \Rightarrow\ \norm{ f(\vvp)_s }^2 = \norm{ \vvp_s }^2 \right).
\end{equation}
\end{defn}

We call the composition of a linear $\kappa$-isometry and a
translation a \semph{$\kappa$-isometry}, and write $\kiso$ for the
\emph{set of all $\kappa$-isometries}. 

\begin{defn}[\cpoi, \ceucl and \gal]
For $c>0$, $\nicefrac{1}{c^2}$-isometries will be called
\semph{$c$-Poincar\'e transformations} and
$\nicefrac{-1}{c^2}$-isometries will be called \semph{$c$-Euclidean
  isometries}. Parameter $c$ in $c$-Poincar\'e transformations
corresponds to the ``speed of light''. A $0$-isometry is also called a
\semph{Galilean symmetry}. We denote these sets of transformations by
\cpoi, \ceucl and \gal, respectively.
\end{defn}
It is easily verified that each of these sets forms a group
  under function composition. In general, when we
speak about a set $\G$ of transformations as a group, we mean
  $\G$ under function composition, \ie $(\G,\circ)$. As usual, we
write $\GH\le\G$ to mean that \GH is a subgroup of \G, and $\GH<\G$ to
mean that the inclusion is proper.

We note that 1-Poincar\'e transformations form the usual group \poi of
Poincar\'e transformations and 1-Euclidean isometries form the usual
group \eucl of Euclidean isometries.  Notice also that trivial
transformations, translations and spatial rotations are
$\kappa$-isometries for all values of $\kappa$.  Moreover, by
\CiteLemma{lem:xy},
\begin{equation}
\label{eq:tartalmazasok}
  \tran\cup\srot\subset\triv=\bigcap_{\kappa\in\Q}\kiso=\kiso[x]\cap\kiso[y]
 \end{equation}
for any two distinct $x,y\in\Q$. It follows immediately that
$\tran\cup\srot \subset \cpoi\cap\ceucl\cap\gal$.

\subsection{The theorems}
Our first result, \CiteTheorem{thm:characterisation}, tells us that if
space is isotropic then all worldview transformations are
$\kappa$-isometries for some $\kappa$, and shows how to calculate the
value of $\kappa$ in the case that two observers can be found which
move relative to one another.

\begin{Theorem}{thm:characterisation}
Assume $\KIN+ \AxIso$. Then there is a $\kappa \in \Q$ such that the
set of worldview transformations is a set of $\kappa$-isometries, \ie
\[
   \mathbb{W} \subseteq \kiso.
\]
In other terms,
\[
\mbox{either $\W \subseteq \cpoi$,
$\W \subseteq \gal$, or $\W \subseteq \ceucl$ for some $c >
0$.}
\]

Moreover,
\begin{itemize}
\item if $\lnot\EMovingIOb$ is assumed, then $\W \subseteq \triv$;
\item if \EMovingIOb is assumed, this $\kappa$ is uniquely determined
  by the $\w{m}{k}$-images of $\origin$ and $\tunit$ where $m$ and $k$
  are observers moving relative to one another, and can be calculated
  as
  \begin{equation*}
    \kappa=\frac{\left|\w{m}{k}\take{\tunit}_t-
      \w{m}{k}\take{\origin}_t\right|^2-1}{\left|\w{m}{k}\take{\tunit}_s-
      \w{m}{k}\take{\origin}_s\right|^2} .
  \end{equation*}
  \qed
\end{itemize}
\end{Theorem}

For all positive $c\in\Q$, the group $\cpoi$ is isomorphic to group
$\poi$ (via natural inner automorphisms of the affine group,
representing the effects of changing the spatial or temporal units of
measurements) and similarly group $\ceucl$ is isomorphic to the
Euclidean transformation group $\eucl$ (via the same inner
automorphisms); see \cite[Prop.~6.9]{OurBorisov}. So essentially there
are only three nontrivial cases: either all the worldview
transformations are relativistic; all of them are classical; or all of
them are Euclidean isometries.  Subject to this constraint, however,
\CiteTheorem{thm:group-instantiation} says that all `reasonable'
transformation groups (groups containing the translations and spatial
rotations, which we know must be present) can occur as the group of
worldview transformations in a model of $\KIN+\AxSpr$.

To present a general model construction, let us write $\sym(\Q^4)$ for
the set of all permutations of $\Q^4$. Given any transformation group
$\G \le \sym(\Q^4)$, we define a model $\M_\G$ of our language by
taking $\IOb := \G$ and $\w{m}{k} := m\circ k^{-1}$ for $k,m \in
\G$. \CiteTheorem{lem:group} connects the axioms of $\KIN$ to
properties of $\G$.

\begin{Theorem}{lem:group}
Let $\G\le\sym(\Q^4)$. Then
  \begin{itemize}
  \item[(a)] $\M_\G$ satisfies \AxWvt, \AxSpr and $\W=\G$.
  \item[(b)] $\M_{\G}$ satisfies $\AxRelocate$ if{}f
    $\srot\cup\tran\subseteq \G$.
  \item[(c)] $\M_{\G}$ satisfies $\AxLine$ if{}f $g[\taxis]$ is
    a line for all $g\in\G$.
  \item[(d)] $\M_{\G}$ satisfies $\AxColocate$ if{}f $g\in\triv$
whenever $g\in\G$ and $g[\taxis]=\taxis$.
  \end{itemize}
\end{Theorem}

\begin{Theorem}{thm:group-instantiation}
  Assume \AxEField.  Let $\G$ be a group such that
  \begin{itemize}
  \item $\srot\cup\tran \subseteq \G\le\cpoi$ for some $c\in\Q$; or
  \item $\srot\cup\tran \subseteq \G\le\ceucl$ for some $c\in\Q$; or
  \item $\srot\cup\tran \subseteq \G\le\gal$.
  \end{itemize}
  Then $\M_{\G}$ is a model of $\KIN+\AxSpr$ for which $\W=\G$.
\end{Theorem}
By \CiteTheorem{thm:characterisation},
\CiteTheorem{thm:group-instantiation} and \CiteTheorem{lem:group},
in order to determine whether a group of symmetries has to be a
subgroup of one of the groups \cpoi, \ceucl and \gal, it is sufficient
to consider its members' actions on \taxis:

\begin{Theorem}{thm:classification1}
Let $(\Q,+,\cdot,0,1,\le)$ be a Euclidean field, and let $\G$ be a
group satisfying $\srot\cup\tran \subseteq \G \le \sym(\Q^4)$. Then\\

\noindent
  \begin{tabular}{ccc}
\parbox{.42\linewidth}{\centering (i) For all $ g \in \G$, $g[\taxis]$
  is a line, and\newline if $g[\taxis]=\taxis$, then $g \in \triv$.}
& $\ \; \Longleftrightarrow \ $ &
\parbox{.42\linewidth}{\centering (ii)
$\G\le \cpoi$, $\G\le \ceucl$ or $\G\le \gal$\newline
for some positive $c \in \Q$.
}
  \end{tabular}
\end{Theorem}

\begin{figure}[h!btp] 
\small
\psfragfig[keepaspectratio,width=\linewidth]{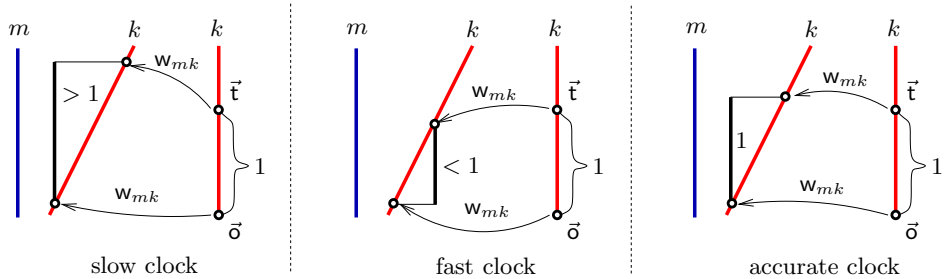}
{
\psfrag{oo}[rt][rt]{$\origin$}
\psfrag{tt}[lb][lb]{$\tunit$}
\psfrag{m}[b][b]{$m$}
\psfrag{k}[b][b]{$k$}
\psfrag{w}[b][b]{$\w{m}{k}$}
\psfrag{1}[l][l]{$1$}
\psfrag{G}[l][l]{$>1$}
\psfrag{L}[l][l]{$<1$}
\psfrag{E}[l][l]{$1$}
\psfrag{slow}[b][b]{slow clock}
\psfrag{fast}[b][b]{fast clock}
\psfrag{accurate}[b][b]{accurate clock}
}
\caption{$k$'s clock can be fast, slow or accurate
  according to $m$ \label{fig:time}}
\end{figure}

Our next result, \CiteTheorem{thm:classification2}, tells us that we can classify all possible models by looking at how observers' clocks run relative to one another. Based on the difference between the time components of the $\w{m}{k}$-image of
$\tunit$ and $\origin$, we can decide whether observer $k$'s clock is fast,
slow or accurate relative to observer $m$'s clock; see
Figure~\ref{fig:time}. Using these notions, we can capture the following
situations:

\begin{newaxiom}
\item[\ul{\ESlowClock}]
There are observers $m,k\in\IOb$ such
that $$\norm{ \w{m}{k}\take{\tunit}_t -\w{m}{k}\take{\origin}_t }>1.$$
\end{newaxiom}

\begin{newaxiom}
\item[\ul{\EFastClock}]
 There are observers $m,k\in\IOb$ such that
 $$\norm{ \w{m}{k}\take{\tunit}_t -\w{m}{k}\take{\origin}_t } < 1.$$
\end{newaxiom}

\begin{newaxiom}
\item[\ul{\EMovingAccurateClock}]
There are observers $m,k\in\IOb$ such that
$$\w{m}{k}\take{\tunit}_s\neq\w{m}{k}\take{\origin}_s  \text{ and } \norm{ \w{m}{k}\take{\tunit}_t - \w{m}{k}\take{\origin}_t } =1.$$
\end{newaxiom}

\begin{newaxiom}
\item[\ul{\AMovingClockSlow}] For all observers $m,k\in\IOb$,
  \begin{equation*}
    \text{if } \w{m}{k}\take{\tunit}_s \neq
    \w{m}{k}\take{\origin}_s,\text{ then }\norm{\w{m}{k}\take{\tunit}_t -
    \w{m}{k}\take{\origin}_t} > 1.
  \end{equation*}
\end{newaxiom}

\begin{newaxiom}
\item[\ul{\AMovingClockFast}] For all observers $m,k\in\IOb$,
$$\text{if } \w{m}{k}\take{\tunit}_s \neq \w{m}{k}\take{\origin}_s, \text{ then } \norm{\w{m}{k}\take{\tunit}_t - \w{m}{k}\take{\origin}_t} < 1.$$
\end{newaxiom}

\begin{newaxiom}
\item[\ul{\AClockAccurate}] For all observers $m,k\in\IOb$,
$\norm{ \w{m}{k}\take{\tunit}_t - \w{m}{k}\take{\origin}_t } = 1$. 
\end{newaxiom}

\begin{Theorem}{thm:classification2}
  Assume $\KIN + \AxIso$. Then precisely one of the following four
  cases holds:
    \begin{enumerate}
    \item There exists a slow clock (\ESlowClock). In this case, there
      exists a moving observer (\EMovingIOb), all moving clocks are
      slow (\AMovingClockSlow), and
      $$\W \subseteq \cpoi \text{ for some positive }c \in \Q.$$
    \item There exists a fast clock (\EFastClock). In this case, there
      exists a moving observer (\EMovingIOb), all moving clocks are
      fast (\AMovingClockFast), and
      $$\W \subseteq \ceucl\text{ for some positive }c \in \Q.$$
    \item There exists a moving accurate clock
      (\EMovingAccurateClock). In this case, all clocks are accurate
      (\AClockAccurate) and
      $$\W \subseteq \gal.$$
   \item There are no moving observers ($\lnot\EMovingIOb$). In this
     case,
     $$\W \subseteq \triv.$$
  \end{enumerate}   
\end{Theorem}

\noindent By \CiteTheorem{thm:consistency}, all of these
   situations can indeed arise.
 
\begin{Theorem}{thm:consistency}
  The following axiom systems are all consistent (they all have models):
  \begin{enumerate}
  \item $\KIN+\AxSpr+\ESlowClock$,
  \item $\KIN+\AxSpr+\EFastClock$,
  \item $\KIN+\AxSpr+\EMovingAccurateClock$,
  \item $\KIN+\AxSpr+\lnot\EMovingIOb$.
  \end{enumerate}
\end{Theorem}


\section{Subsidiary theorems and lemmas} \label{sec:lemmas}

Because we use only a small number of basic axioms, we have a large
number of intermediate lemmas to prove before we can prove
our main theorems. This section is accordingly split into six subsections, each focussing on a key stage in the overall proof of our main findings. Each stage builds on its predecessor(s) and together they establish the following subsidiary theorems. Informally stated, they assert (subject to various conditions) that:

\begin{description}
\item[\CiteTheorem{lem:observer-lines}] ~
  \\ If $\ell$ is a possible worldline, then all lines of the same slope as $\ell$ are also possible worldlines.
\item[\CiteTheorem{lem:line-to-line}] ~
  \\ Each worldview transformation is a bijection taking lines to lines, planes to planes and hyperplanes to hyperplanes.
\item[\CiteTheorem{lem:txplane}] ~
  \\ If $\w{k}{m}$ maps the $tx$-plane to itself, then it also maps the $yz$-plane to itself; moreover, if $\w{k}{m}$ is linear, there is some positive $\lambda$ such that $|\w{k}{m}(\vvp)| = \lambda|\vvp|$ for all $\vvp$ in the $yz$-plane.
\item[\CiteTheorem{lem:same-speed-hard}] ~
  \\ Suppose at least one observer considers $h$ and $k$ to be travelling with the same speed. Then $\w{h}{k}$ is a $\kappa$-isometry for some $\kappa$.
\item[\CiteTheorem{lem:fundamental}] ~
  \\ Suppose no observers move with infinite speed, and that $\speed_k(m) = u > 0$. Then there exists $\varepsilon >0$ for which, given any positive $v \le u+\varepsilon$, there is some $h$ with $\speed_k(h) = v$ and $\speed_m(h) = \speed_m(k)$.
\item[\CiteTheorem{lem:main}] ~
  \\ There exists at least one observer $k$ and one $\kappa$ for which all worldview transformations $\w{m}{k}$ involving observers $m$ who agree with $k$ about the origin are $\kappa$-isometries.
\end{description}

The order of implications in the proofs that follow is:

\begin{center}
\xymatrix{
 &&& \text{Same-Speed}         \ar[dr]        &   \\
   \text{ \setlength\tabcolsep{0 pt}\begin{tabular}{c}Observer\\Lines\end{tabular}\setlength\tabcolsep{6 pt} }       \ar[r]         & 
   \text{Line-to-Line}         \ar[r]         & 
   \text{$tx$-Plane} \ar[ur]\ar[dr] &&
   \text{Main}                                    \\
 &&& \text{Fundamental}        \ar[ur]        & 
}
\end{center}

\subsection{Observer Lines Lemma}

\InformalStatement{If $\ell$ is a possible worldline, then all lines with the same slope as $\ell$ are also
possible worldlines.}

We say that a subset $\ell \subseteq Q^4$ is an \semph{observer line}
\lemph{for $k$} if there is some observer $h$ for which $\ell =
\wl_k(h)$, and write $\oblines(k)$ for the set of $k$-observer
lines. We say that $\ell$ is an \emph{observer line} if there is some
$k$ for which it is an observer line. By \AxLine, all observer lines
are lines (because they are worldlines). In this section, we prove
that if $k$ can see an observer travelling along a worldline, then
every other line with the same slope is also a worldline as far as $k$
is concerned; there are none of these lines from which observers are
banned.

Now suppose \AxEField holds. If $\ell$ is a line and \vvp, \vvq are
distinct points in $\ell$, we define its \semph{slope} by
\[
  \slope(\ell) \de \begin{cases}
	   \nicefrac{\norm{\vvp_s - \vvq_s} }{ \norm{\vvp_t - \vvq_t} } 
& \text{ if $\vvp_t \neq \vvq_t$, } \\
		 \infty & \text{ otherwise .}
	\end{cases}
\]

\begin{Theorem}{lem:observer-lines}
Assume \AxEField, \AxWvt, \AxRelocate, \AxLine and \AxIso. Suppose either
\begin{itemize}
\item[(a)] $\slope(\ell) = \slope(\ell') \neq \infty$; or else
\item[(b)] $\slope(\ell) = \slope(\ell') = \infty$ and there exist $\vvp \in
\ell$ and $\vvq \in \ell'$ whose time coordinates are equal.
\end{itemize}
Then for any observer $k$, we have $\ell \in \oblines(k)$ if{}f $\ell' \in \oblines(k)$. \qed
\end{Theorem}

\begin{figure}[h!btp] 
\small
\psfragfig[keepaspectratio,width=\linewidth]{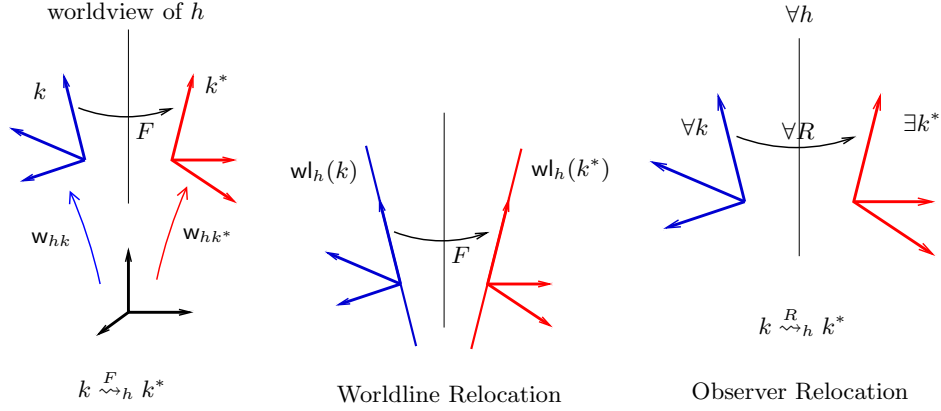}
{
\psfrag{h-vw}[b][b]{worldview of $h$}
\psfrag{F}[t][t]{$F$}
\psfrag{whk}[l][l]{$\w{h}{k^*}$}
\psfrag{whk*}[r][r]{$\w{h}{k}$}
\psfrag{k*}[b][b]{$k$}
\psfrag{k}[b][b]{$k^*$}
\psfrag{whk}[l][l]{$\w{h}{k^*}$}
\psfrag{whk*}[r][r]{$\w{h}{k}$}
\psfrag{wlreloc}[b][b]{Worldline Relocation}
\psfrag{reloc}[b][b]{$k\mappedto{R}{h}k^*$}
\psfrag{Ftr}[b][b]{$k\mappedto{F}{h}k^*$}
\psfrag{wk*}[r][r]{$\wl_h(k)$}
\psfrag{wk}[l][l]{$\wl_h(k^*)$}
\psfrag{relocation}[b][b]{Observer Relocation}
\psfrag{fk}[l][l]{$\exists k^*$}    
\psfrag{fR}[t][t]{$\forall R$}
\psfrag{ek*}[r][r]{$\forall k$}
\psfrag{fh}[b][b]{$\forall h$}
}
\caption{\label{fig:F-tr-Relocation} 
$F$-transforms (left) describe how $h$ can transform what it considers to be $k$'s worldview --- and worldline (middle) --- into $k^*$'s (Definition~\ref{def:F-transformations}, \CiteLemma{lem:worldline-relocation}). \CiteLemma{lem:relocation} tells us that all spatial rotations can be interpreted as $F$-transforms (right).}
\end{figure}

In order to prove this result, we require various supporting lemmas (the more elementary ones are re-used in subsequent proofs). These lemmas refer to a concept we call \emph{$F$-transformation} that relates the worldviews of any two observers via that of a third (see Figure~\ref{fig:F-tr-Relocation}). To illustrate the concept, suppose that I am observing two planets, $k$ and $k^*$, in the night sky. From my point of view, people living on those planets would see the world quite differently, but they nonetheless see the same world I do, so I ought to be able to find some function ($F$) that transforms ``what I think $k$ sees'' into ``what I think $k^*$ sees''. From my point of view, I can say that ``$k^*$ is an `$F$-transformed' version of $k$.''

\begin{defn}[$F$-transforms]
\label{def:F-transformations}
Given any bijection $F : \Q^4 \to\ Q^4$, we say that 
\lemph{$k^*$ is an \semph{$F$-transformed} version of $k$
  according to $h$}, and write $k \mappedto{F}{h} k^*$ if
\begin{equation}
  \w{h}{k^*} = F \circ \w{h}{k} .
\end{equation}
\QED\end{defn}

\begin{rem}
Assuming \AxWvt, $k \mappedto{\id}{h} k^*$ is equivalent to
$\w{k^*}{k}=\id$, in particular $k \mappedto{\id}{h} k$; relations
$k\mappedto{F}{h} k^*$ and $k^*\mappedto{G}{h} k'$ imply
$k\mappedto{G\circ F}{h} k'$; and $k\mappedto{F}{h} k^*$ implies
$k^*\mappedto{F^{-1}}{h} k$.
\end{rem}

\subsubsection{Supporting lemmas}

Some of these initial lemmas are quite elementary, but they form the
bedrock of what follows, and we need to prove them formally to ensure
they definitely follow from our somewhat restricted first-order axiom
set. The supporting lemmas can be informally described as follows:

\begin{description}
\item[\CiteLemma{lem:wvt}]
~  \\ This describes various elementary properties concerning worldview transformations. We often use these results without further mention.
\item[\CiteLemma{lem:worldline-relocation}]
~  \\ If $h$ can $F$-transform $k$ into $k^*$, then that transformation maps $k$'s worldline into $k^*$'s.
\item[\CiteLemma{lem:relocation}]
~  \\ Every spatial rotation can be interpreted as an $F$-transform.
\item[\CiteLemma{lem:transformed-observer-lines}]
~  \\ If $\ell$ is an observer line for $k$, then $\w{h}{k}[\ell]$ is an observer line for $h$.
\item[\CiteLemma{lem:rotated-observer-lines}]
~  \\ If $\ell$ is an observer line for $k$, so is any spatially rotated copy of $\ell$.
\item[\CiteLemma{lem:vanrotacio}]
~  \\ This is a technical lemma telling us when one pair of mutually orthogonal horizontal vectors can be spatially rotated into another (where ``horizontal'' means ``orthogonal to the time-axis'').
\item[\CiteLemma{lem:2vanrotacio}]
~  \\ If two lines have the same slope and both pass through the origin, it is possible to spatially rotate one into the other.
\item[\CiteLemma{lem:observer-line-intersections}]
~  \\ Suppose two intersecting lines have the same slope. If one of them is an observer line for $k$, then so is the other.
\item[\CiteLemma{lem:triangulation}]
~ \\ Suppose $\taxis'$ is a line parallel to the time-axis, \taxis, and that $\vvp$ is not on $\taxis'$. Given any positive $\lambda$ we can find lines $\ell_1$ and $\ell_2$ which intersect at \vvp, meet $\taxis'$ at different points, and have the same slope, $\lambda$. In other words, we can find an isosceles triangle whose base is along $\taxis'$ and vertex at \vvp, and whose equal non-base sides both have slope $\lambda$.
\end{description}

\subsubsection{Proofs of the supporting lemmas}

\begin{Lemma}{lem:wvt}
Assume \AxWvt. Then, for every $k,h,m\in\IOb$,
\begin{itemize}
\item[(i)] $\wl_k(k)=\taxis$;
\item[(ii)] $\w{h}{k} \takeset{\wl_k(m)}  = \wl_h(m)$;
\item[(iii)] $\w{h}{k} : Q^4 \to Q^4$ is a bijection from $\Q^4$ onto itself;
\item[(iv)] $\w{h}{k}^{-1} = \w{k}{h}$.
\end{itemize}
\end{Lemma}
\begin{proof}

(i) $\wl_k(k)=\w{k}{k}[\taxis]=\id[\taxis]=\taxis$.

(ii) Since $\wl_k(m) = \w{k}{m}\takeset{\taxis} $, we have 
$    \w{h}{k} \takeset{\wl_k(m)} 
   =  \w{h}{k}\takeset{\w{k}{m}\takeset{\taxis} } 
   = \w{h}{m}\takeset{\taxis} 
   = \wl_h(m)$, 
as required.

(iii), (iv): It follows from $\w{k}{h} \circ \w{h}{k} = \w{k}{k} = \id$ 
and $\w{h}{k} \circ \w{k}{h} = \w{h}{h} = \id$
that 
$\w{k}{h}$ and $\w{h}{k}$ are mutual inverses, 
and hence that they are both bijections.
\end{proof}

\begin{Lemma}{lem:worldline-relocation}
Assume \AxWvt, and suppose $k \mappedto{F}{h} k^*$ for some bijection 
$F : Q^4 \to Q^4$. Then $F$ maps $\wl_h(k)$ onto $\wl_h(k^*)$;
see Fig.~\ref{fig:F-tr-Relocation} (middle).
\end{Lemma}
\begin{proof}
Recall that $k \mappedto{F}{h} k^*$ means $\w{h}{k^*} = F \circ \w{h}{k}$. So
\[
  \wl_h(k^*) = \w{h}{k^*}\takeset{\taxis}  = 
(F \circ \w{h}{k} ) \takeset{\taxis}   = F \takeset{\wl_h(k)} .
\]
\end{proof}

\begin{Lemma}{lem:relocation}
Assume \AxEField, \AxWvt, \AxRelocate and \AxIso.  Then given any
spatial rotation $R \in \srot$ and $k,h\in\IOb$, there exists an
observer $k^*$ such that $k \mappedto{R}{h} k^*$; see Fig.~\ref{fig:F-tr-Relocation} (right).
\end{Lemma}
\begin{proof}
By \AxRelocate, there exists an observer $h^*$ for which $\w{h}{h^*} = R$. 
Because $h$ and $h^*$ are related via a spatial rotation, \AxIso 
tells us there exists $k^* \in IOb$ which is related to $h^*$ the same way $k$ 
is related to $h$, \ie $\w{h^*}{k^*} = \w{h}{k}$. It follows immediately that
$
   \w{h}{k^*}  = \w{h}{h^*} \circ \w{h^*}{k^*}   = 
R \circ \w{h}{k}
$,
\ie  $k \mappedto{R}{h} k^*$, as claimed.
\end{proof}

\begin{Lemma}{lem:transformed-observer-lines}
Assume \AxWvt. Then $\ell \in \oblines(k)$ if{}f $\w{h}{k}[\ell] \in \oblines(h)$.
\end{Lemma}
\begin{proof}
This follows immediately from \CiteLemma{lem:wvt}, since all $k$-observer lines are
worldlines.
\end{proof}

\begin{Lemma}{lem:rotated-observer-lines} Assume \AxEField, \AxWvt, \AxRelocate and \AxIso. 
If $\ell \in \oblines(k)$ and $R \in \srot$ is any spatial rotation, then $R[\ell] \in \oblines(k)$.
\end{Lemma}
\begin{proof}
Choose $h \in \IOb$ such that $\ell = \wl_k(h)$. 
By \CiteLemma{lem:relocation}, there is some $h^* \in \IOb$ for which 
$h \mappedto{R}{k} h^*$, \ie $\w{k}{h^*} = R \after \w{k}{h}$. 
By \CiteLemma{lem:worldline-relocation}, we have that $\wl_k(h^*) = R[\wl_k(h)] = R[\ell]$, and this
worldline is in $\oblines(k)$, as required.
\end{proof}

\begin{figure}[h!btp] 
\small
\psfragfig[keepaspectratio,width=\linewidth]{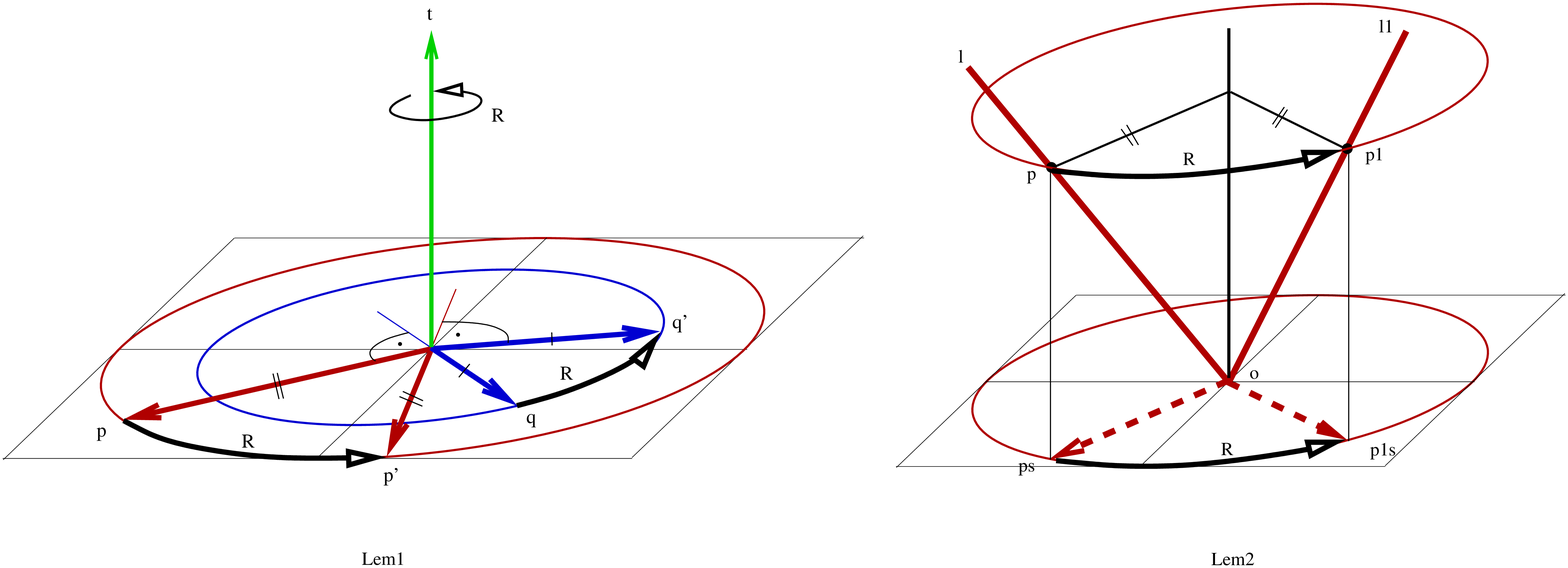}
{
\psfrag{t}[b][b]{$\tunit$}
\psfrag{p}[rt][rt]{$\vvp_1$}
\psfrag{p'}[t][t]{$\vvp_2$}
\psfrag{R}[lb][lb]{$R$}
\psfrag{q}[lt][lt]{$\vvq_1$}
\psfrag{q'}[l][l]{$\vvq_2$}
\psfrag{Lem1}[cb][cb]{\CiteLemma{lem:vanrotacio}}
\psfrag{l}[rb][rb]{$\ell_1$}
\psfrag{l1}[rb][rb]{$\ell_2$}
\psfrag{ps}[rt][rt]{$(0,(\vvp_1)_s)$}
\psfrag{p1s}[lt][lt]{$(0,(\vvp_2)_s)$}
\psfrag{o}[l][l]{$\origin$}
\psfrag{Lem2}[cb][cb]{\CiteLemma{lem:2vanrotacio}}
\psfrag{p1}[lt][lt]{$\vvp_2$}
}
\caption{\label{fig:rotation} Illustrations for \CiteLemma{lem:vanrotacio}
and
\CiteLemma{lem:2vanrotacio}.}
\end{figure}

\begin{Lemma}{lem:vanrotacio}
Let $(\Q,+,\cdot,0,1,\leq)$ be an ordered field and suppose
$\vvp_1,\vvq_1,\vvp_2,\vvq_2\in\Q^4$ satisfy:
\begin{itemize}
  \item[(a)] $\vvp_1$ and $\vvp_2$ have the same length, as do $\vvq_1$ and $\vvq_2$:\\
    $|\vvp_1|^2=|\vvp_2|^2$ and
    $|\vvq_1|^2=|\vvq_2|^2$;
  \item[(b)] $\vvp_1$ and $\vvq_1$ are horizontal and mutually orthogonal:\\
    $\vvp_1\cdot\tunit=\vvq_1\cdot\tunit=\vvp_1\cdot\vvq_1=0$; and
  \item[(c)] $\vvp_2$ and $\vvq_2$ are horizontal and mutually orthogonal:\\
    $\vvp_2\cdot\tunit=\vvq_2\cdot\tunit=\vvp_2\cdot\vvq_2 =0$.
\end{itemize}

Then there exists a spatial rotation
$R\in\srot$ such that $R(\vvp_1)=\vvp_2$ and $R(\vvq_1)=\vvq_2$; see
the left-hand side of Figure~\ref{fig:rotation}.

\end{Lemma}

\begin{proof}
Consider the linear map that takes $\alpha\tunit+\beta\vvp_1+\gamma\vvq_1$
to $\alpha\tunit+\beta\vvp_2+\gamma\vvq_2$. It is easy to see that this
map is a linear Euclidean isometry between two subspaces of $\Q^4$ which are each at most three-dimensional. Hence, by the refinement of Witt's
theorem~\cite[Thm 234.1, p.234]{mag} there is an extension
$R:\Q^4\to\Q^4$ which is a linear Euclidean isometry with determinant
1. This $R$ must be a spatial rotation, because $R(\tunit)=\tunit$.
\end{proof}

\begin{Lemma}{lem:2vanrotacio}
Let $(\Q,+,\cdot,0,1,\leq)$ be a Euclidean field. Assume
$\ell_1$ and $\ell_2$ are lines such that
$\slope\take{\ell_1}=\slope\take{\ell_2}$ and $\origin\in\ell_1\cap\ell_2$.
Then there exists $R\in\srot$ such that $R\takeset{\ell_1}=\ell_2$.
\end{Lemma}

\begin{proof} 
Let $\vvp_1\in\ell_1$ and $\vvp_2\in\ell_2$ be such that
$\vvp_1\neq\origin\neq\vvp_2$ and $(\vvp_1)_t=(\vvp_2)_t$, see the right-hand side of
Figure~\ref{fig:rotation}. Then
  $\norm{(0,(\vvp_1)_s)}^2=\norm{(0,(\vvp_2)_s)}^2$. Taking $\vvq_1 = \vvq_2 = \origin$, 
  \CiteLemma{lem:vanrotacio} now tells us there exists a spatial rotation $R$ that
  takes $(0,(\vvp_1)_s)$ to $(0,(\vvp_2)_s)$ and leaves $\origin$ fixed.  Since spatial rotations leave time coordinates unchanged, this $R$ takes
  $\vvp_1$ to $\vvp_2$, and since it also fixes the origin it must take $\ell_1$ to $\ell_2$.
\end{proof}

\begin{Lemma}{lem:observer-line-intersections}
Assume \AxEField, \AxWvt, \AxLine, \AxRelocate, and \AxIso.
If two lines $\ell_1, \ell_2$ intersect one another and have equal 
slope, then for any $k \in \IOb$ we have $\ell_1 \in \oblines(k)$ if{}f 
$\ell_2 \in \oblines(k)$.
\end{Lemma}
\begin{figure}[h!btp] 
\small
\psfragfig[keepaspectratio,width=0.8\linewidth]{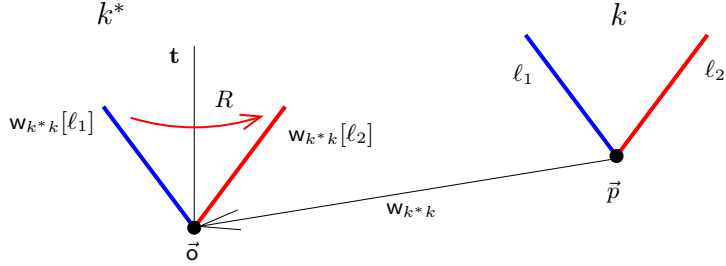}
{
\psfrag{t}[r][r]{$\taxis$}
\psfrag{o}[t][t]{$\origin$}
\psfrag{p}[t][t]{$\vvp$}
\psfrag{w}[t][t]{$w_{k^*k}$}
\psfrag{l1}[r][r]{$\ell_1$}
\psfrag{l2}[l][l]{$\ell_2$}
\psfrag{w2}[l][l]{$\w{k^*}{k}[\ell_2]$}
\psfrag{w1}[r][r]{$\w{k^*}{k}[\ell_1]$}
\psfrag{k}[t][t]{\Large $k$}
\psfrag{k*}[t][t]{\Large $k^*$}
\psfrag{R}[b][b]{$R$}
\psfrag{w}[t][t]{$\w{k^*}{k}$}
}
\caption{\label{fig:observer-line-intersections}
Illustration for the proof of 
\CiteLemma{lem:observer-line-intersections}.}
\end{figure}

\begin{proof}
Let \vvp be the point of intersection of $\ell_1$ and $\ell_2$, and
let $T$ be the translation taking \vvp to the origin, \origin. By
\AxRelocate, there exists some $k^* \in \IOb$ such that $\w{k^*}{k} =
T$; see Figure~\ref{fig:observer-line-intersections}.

Note first that the images of $\ell_1$ and $\ell_2$ under $\w{k^*}{k}$ are lines of equal slope because $\w{k^*}{k} = T$ is a translation, and translations map lines to lines and leave slopes unchanged. Moreover, both of these lines pass through $T(\vvp) = \origin$, so \CiteLemma{lem:2vanrotacio} tells us there exists a spatial rotation $R$ taking
$\w{k^*}{k}[\ell_1]$ to $\w{k^*}{k}[\ell_2]$.

The claim now follows. For suppose $\ell_1$ is a $k$-observer line; we have to show that $\ell_2$ is also a $k$-observer line. Since $\w{k^*}{k}[\ell_1] \in \oblines(k^*)$ 
by \CiteLemma{lem:transformed-observer-lines}, it follows  
that $\w{k^*}{k}[\ell_2] \in \oblines(k^*)$ as well, by 
\CiteLemma{lem:rotated-observer-lines}. Applying \CiteLemma{lem:transformed-observer-lines} in the opposite direction now tells us that $\ell_2 \in \oblines(k)$, as required. 

The converse follows by symmetry.
\end{proof}

\begin{Lemma}{lem:triangulation}
Assume \AxEField. Let  $\taxis'$ be a line parallel to the time-axis and 
let $\vvp$ be any point not on $\taxis'$. Given any positive 
$\lambda \in \Q$, there exist lines $\ell_1, \ell_2$ with

\begin{minipage}{0.45\textwidth}
\begin{itemize}
\item[(i)]   $\slope(\ell_1) = \slope(\ell_2) = \lambda$,
\item[(ii)]  $\vvp \in \ell_1 \cap \ell_2$,
\item[(iii)] $\ell_1 \cap \taxis' \neq \varnothing$,
\item[(iv)]  $\ell_2 \cap \taxis' \neq \varnothing$,
\item[(v)]   $\ell_1 \cap \ell_2 \cap \taxis' = \varnothing$.
\end{itemize}
\end{minipage}
\hfill
\begin{minipage}{0.45\textwidth}
\psfragfig[keepaspectratio,width=0.3\textwidth]{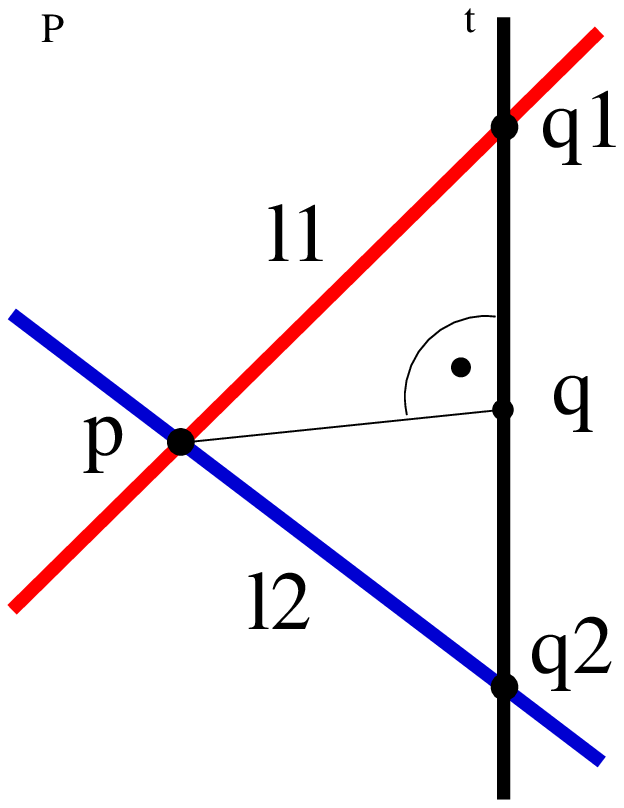}
{
\psfrag{P}[rb][rb]{$~$}
\psfrag{t}[rb][rb]{\small $\taxis'$}
\psfrag{l1}[rb][rb]{\small $\ell_1$}
\psfrag{l2}[rt][rt]{\small $\ell_2$}
\psfrag{p}[r][r]{\small $\vvp$}
\psfrag{q}[l][l]{\small $\vvq$}
\psfrag{q1}[lt][lt]{\small $\vvq_1$}
\psfrag{q2}[lb][lb]{\small $\vvq_2$}
}
\end{minipage}
\end{Lemma}

\begin{proof}
Let $\vvq \in \taxis'$ be the point on $\taxis'$ with $\vvq_t = \vvp_t$. We know that $\vvp_s \neq \vvq_s$ because $\vvp \not\in \taxis'$. 
Consider the points
\[
  \vvq_1 := \vvq + (\norm{\vvp_s-\vvq_s}/\lambda, 0,0,0) \quad \text{ and } \quad
  \vvq_2 := \vvq - (\norm{\vvp_s-\vvq_s}/\lambda, 0,0,0)
\]
and let $\ell_1$ be the line passing through $\vvp$ and $\vvq_1$, and $\ell_2$ the line passing through $\vvp$ and $\vvq_2$. Then direct calculation shows that $\ell_1$ and $\ell_2$ have the required properties.
\end{proof}

\subsubsection{Main proof} We now complete the proof of \CiteTheorem{lem:observer-lines}.



We use the word \emph{plane} in the usual Euclidean sense to mean a
$2$-dimensional slice of $\Q^4$, and refer to 3-dimensional `slices'
as \emph{hyperplanes}.  Formally, a subset $P\subseteq\Q^4$ is a
\semph{plane} if{}f there are linearly independent vectors $\vvv,\vvw
\neq \origin \in\Q^4$ and a point $\vvp\in\Q^4$, such that
$P=\{\vvp+\lambda\vvv+\mu\vvw\: :\: \lambda,\mu\in\Q\}$ (hyperplanes
are defined analogously).  By \AxEField, the usual properties of
Euclidean planes hold.  In particular, a plane $P$ can be specified by
giving a line $\ell \subseteq P$ and a point $\vvp \in P \setminus
\ell$, or three distinct non-collinear points $\vvp, \vvq, \vvr \in
P$, or two distinct but intersecting lines in $P$.  Moreover, given a
line $\ell \subseteq P$ and a point $\vvp \in P \setminus \ell$, there
is exactly one line $\ell_p$ through \vvp that is parallel to $\ell$
(indeed, if we assume \AxEField, the way in which we have defined
\emph{line} and \emph{plane} allows us to uniquely determine $\ell_p$
in the usual way once \vvp and $\ell$ are specified).

\begin{proof}[Proof of \CiteTheorem{lem:observer-lines}]
Let $\ell, \ell'$ be lines of equal slope: $\slope(\ell) = \slope(\ell')$. If $\ell = \ell'$, there is nothing to prove, so assume that $\ell \neq \ell'$. Also, if $\slope(\ell) = \slope(\ell') = 0$, then $\ell$ and $\ell'$ are both parallel to the time-axis, and it follows easily from \AxRelocate that $\ell, \ell' \in \oblines(k)$.

Suppose, therefore, that $\slope(\ell) = \slope(\ell') \neq 0$. 

Note first that there exist $\vvp, \vvq \in \Q^4$ such that 
\begin{equation*} \label{observer-lines.1}
  \vvp \in \ell, \quad 
  \vvq \in \ell', \quad
  \vvp \neq \vvq, \quad \text{and} \quad 
  \vvp_t = \vvq_t.
\end{equation*}
This is true by assumption for case (b), where $\slope(\ell) = \slope(\ell') = \infty$, 
and it is easy to see that such $\vvp, \vvq$ also exist in case (a) where $\slope(\ell) = \slope(\ell')$ is finite.\footnote{\,Pick any point $\vvp$ on $\ell$ that isn't on $\ell'$ and consider the `horizontal time slice' containing it; because $\ell'$ has finite slope, it must also pass through this time slice. Take $\vvq$ to be the corresponding point of intersection on $\ell'$.} 

Let $\hat\ell$ be the line containing $\vvp$ and $\vvq$. Because $\vvp, \vvq$ have the same time coordinate, $\slope(\hat\ell) = \infty$; see Figure~\ref{fig:observerlines}.

\begin{figure}[h!btp] 
\small
\psfragfig[keepaspectratio,width=\linewidth]{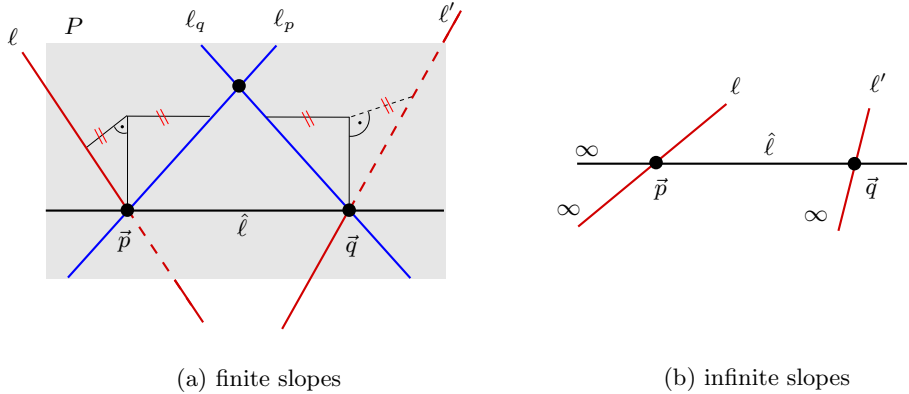}
{
\psfrag{P}[b][b]{$P$}
\psfrag{l1}[b][b]{$\ell_p$}
\psfrag{l2}[b][b]{$\ell_q$}
\psfrag{l}[b][b]{$\ell$}
\psfrag{l'}[b][b]{$\ell'$}
\psfrag{p}[t][t]{$\vvp$}
\psfrag{q}[t][t]{$\vvq$}
\psfrag{i}[rb][rb]{$\infty$}
\psfrag{qq}[lt][lt]{$\vvq$}
\psfrag{finite}[b][b]{(a) finite slopes}
\psfrag{infinite}[b][b]{(b) infinite slopes}
\psfrag{lh}[b][b]{$\hat{\ell}$}
\psfrag{if}[b][b]{$\infty$}
}
\caption{\label{fig:observerlines}
Illustration for the proof of \CiteTheorem{lem:observer-lines}} 
 \end{figure}

We now consider cases (a) and (b) in turn.

\textit{Case (a): finite slopes.} By assumption, $0 < \slope(\ell) =
\slope(\ell') \neq \infty$ and $\slope(\hat\ell) = \infty$.  Let $P$ be the plane containing $\hat\ell$ and parallel to
$\taxis$.%
\footnote{$P$ is parallel to $\taxis$ if{}f $P$ contains a line parallel to
$\taxis$.}  

Let $\taxis_p$ be the line parallel to \taxis which passes through \vvp, and notice that this line lies in $P$. Choose any point $\vvp' \in P \setminus \taxis_p$ and let $\lambda = \slope(\ell) = \slope(\ell')$. Then \CiteLemma{lem:triangulation} tells us that we can find two distinct lines which pass through $\vvp'$, lie in $P$ (because they meet both $\vvp'$ and $\taxis_p$), and have slope $\lambda$. Applying the translation taking $\vvp'$ to $\vvp$, the images of those two lines will still lie in $P$ and still have slope $\lambda$, but will intersect at $\vvp$. Similarly, we can find two distinct lines of slope $\lambda$ which lie in $P$ and pass through $\vvq$. Pick one of the lines passing through $\vvq$, and call it $\ell_q$. Since the two lines through \vvp are distinct, they cannot both be parallel to 
$\ell_q$ --- let $\ell_p$ be one that isn't. Since $\ell_p$ and $\ell_q$ are non-parallel lines lying in the same plane, they must intersect.

The claim now follows. For suppose $\ell \in \oblines(k)$. Then $\ell$ and $\ell_p$ are lines of equal slope which intersect at \vvp, so \CiteLemma{lem:observer-line-intersections} tells us that $\ell_p$ is also in $\oblines(k)$, whence (applying the same argument twice more) so are $\ell_q$ (because it meets $\ell_p$) and $\ell'$ (since it meets $\ell_q$).

\textit{Case (b): infinite slopes.} If $\slope(\ell) = \infty$, then $\ell$ and $\hat\ell$ are two lines of infinite slope which intersect at $\vvp$. Likewise, $\ell'$ and $\hat\ell$ are lines of infinite slope that intersect at $\vvq$. As before it now follows by 
\CiteLemma{lem:observer-line-intersections} that
\[
  \ell \in \oblines(k) \Longleftrightarrow
  \hat\ell \in \oblines(k) \Longleftrightarrow
  \ell' \in \oblines(k) .
\]

\medskip
In both cases, therefore, we have $\ell \in \oblines(k) \Longleftrightarrow \ell' \in \oblines(k)$, as required.
\end{proof}

\subsection{Line-to-Line Lemma}

\InformalStatement{Each worldview transformation is a bijection taking lines to lines, planes
to planes and hyperplanes to hyperplanes.}

\begin{Theorem}{lem:line-to-line}
Assume \AxEField, \AxWvt, \AxLine, \AxRelocate, \AxIso and \EMovingIOb. Then given any $k, h \in\IOb$, the worldview transformation $\w{h}{k}$ is a bijection that takes lines to lines, planes to planes, and hyperplanes to hyperplanes.
\end{Theorem}

\subsubsection{Supporting lemmas}

A number of the supporting lemmas refer to the concept of an \emph{observer line triad}:

\begin{defn}[Observer Line Triads]
If $\ell_1, \ell_2, \ell_3 \in \oblines(k)$ are three (necessarily coplanar) lines, each pair of which intersect in a point, and whose pairwise intersections are not collinear, we shall call the set $\{\ell_1, \ell_2, \ell_3\}$ an \emph{observer line triad for $k$}, or simply a \emph{$k$-triad}.
\QED\end{defn}

The lemmas can be described informally as follows:

\begin{description}
\item[\CiteLemma{lem:speed}]
~ \\ Speeds are well-defined, and the terms \emph{at rest} and \emph{in motion} have their expected meanings.
\item[\CiteLemma{lem:triads}]
~ \\ If one observer considers that three worldlines form a triad, all other observers agree.
\item[\CiteLemma{lem:plane-to-plane}]
~ \\  Suppose plane $P$ contains a $k$-triad whose slopes are either all finite or else all infinite. Then $\w{h}{k}[P]$ is contained in a plane.
\item[\CiteLemma{lem:straight-lines-are-observer-lines}]
~ \\ If infinite speeds occur, then all lines are observer lines.
\end{description}

\subsubsection{Proofs of the supporting lemmas}

\begin{defn}
Suppose \AxEField and \AxLine holds. If $\ell = \wl_k(h)$, we call the slope, $\slope(\ell)$, of line $\ell$ the \semph{speed} \lemph{of
  $h$ according to $k$}, \ie
\[
  \speed_k(h) \de \slope(\wl_k(h)) .
\]
\end{defn}

\begin{defn}
Recall that observer $k\in\IOb$ is moving according to observer $m\in\IOb$
if{}f $\w{m}{k}(\tunit)_s \neq \w{m}{k}(\origin)_s$ and at rest according to $m$ otherwise. We say that \lemph{observer
  $k\in\IOb$ }\semph{is moving instantaneously according to}\lemph{ observer
  $m$} if{}f $\w{m}{k}(\tunit)_t=\w{m}{k}(\origin)_t$.
\end{defn}

\begin{Lemma}{lem:speed}
Assume $\AxWvt$, $\AxEField$ and $\AxLine$. Then for every $m,k \in \IOb$, 
$\speed_m(k)$ is well-defined, and 
\begin{itemize}
\item $k$ is at rest according to $m$ if{}f $\speed_m(k)=0$,
\item $k$ is moving according to $m$ if{}f $\speed_m(k)\neq 0$, and
\item $k$ is moving instantaneously according to $m$ if{}f
  $\speed_m(k)=\infty$.
\end{itemize}
\end{Lemma}
\begin{proof}
By $\AxEField$ and $\AxLine$, it follows that $\speed_m(k)$ is
unambiguously defined for all $k$ and $m$.  The proof is
straightforward after noticing that $\w{m}{k}(\tunit) \neq
\w{m}{k}(\origin)$ which holds because $\w{m}{k}$ is a bijection by
\CiteLemma{lem:wvt}.
\end{proof}

\begin{Lemma}{lem:triads}
Suppose \AxEField, \AxWvt, \AxLine. Let $k, h \in \IOb$. If $T = \{\ell_1, \ell_2, \ell_3 \}$ is a $k$-triad, then 
$\w{h}{k}[T]:=\{ \w{h}{k}[\ell_1], \w{h}{k}[\ell_2], \w{h}{k}[\ell_3] \}$ is an $h$-triad.
\end{Lemma}
\begin{proof}
Each $\ell_i$ is a $k$-observer line, so by
\CiteLemma{lem:transformed-observer-lines}, each $\ell_i' =
\w{h}{k}[\ell_i]$ is an $h$-observer line (and hence a line). 
Because $\w{h}{k}$ is a bijection, we know that any two of the lines in
$\w{h}{k}[T]$ has non-empty intersection, and that they have three
distinct pairwise intersections in total.  It follows that the three lines
are coplanar and that their three pairwise intersection points are not
collinear.  That is, $\w{h}{k}[T]$ is an $h$-triad as claimed.
\end{proof}

\begin{Lemma}{lem:plane-to-plane}
Assume \AxEField, \AxWvt, \AxLine, \AxRelocate and \AxIso. Choose $k, h \in \IOb$, let $P$ be a  plane which contains a $k$-triad $\{\ell_1, \ell_2, \ell_3\}$, and suppose that the slopes of these lines are either all finite, or else all infinite. Then $\w{h}{k}[P]$ is contained in a  plane.
\end{Lemma}
\begin{figure}[h!btp] 
\small
\psfragfig[keepaspectratio,width=0.9\linewidth]{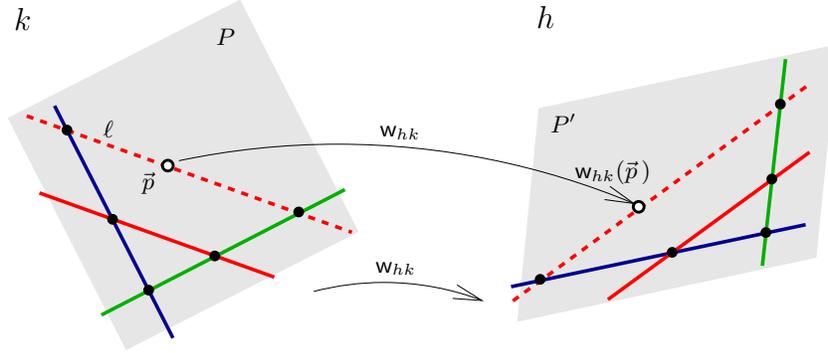}
{
\psfrag{P}[t][t]{$P$}
\psfrag{P'}[t][t]{$P'$}
\psfrag{w}[b][b]{$\w{h}{k}$}
\psfrag{p}[rt][rt]{$\vvp$}
\psfrag{wp}[rb][rb]{$\w{h}{k}(\vvp)$}
\psfrag{l}[b][b]{$\ell$}
\psfrag{k}[lt][lt]{\Large $k$}
\psfrag{h}[lt][lt]{\Large $h$}
\psfrag{wl}[rb][rb]{$\w{h}{k}[\ell]$}
}
\caption{Illustration for the proof of \CiteLemma{lem:plane-to-plane}}
\end{figure}
\begin{proof}
According to \CiteLemma{lem:triads}, the lines $\w{h}{k}[\ell_i]$ ($i = 1,2,3$) form an $h$-triad. We can therefore define $P'$, the plane spanned by this triad. We will prove that $\w{h}{k}[P] \subseteq P'$.

Choose any $\vvp \in P$.  If $\vvp$ lies on any of the lines $\ell_i$,
then the conclusion $\w{h}{k}(\vvp) \in P'$ is trivial.  Suppose,
then, that $\vvp$ does not lie on any of these lines.  Because the lines form a triad we can
draw a line $\ell$ through $\vvp$ which is parallel to one of the
lines (wlog, $\ell_1$) and which intersects the other two lines ($\ell_2$ and $\ell_3$) in distinct points.

We claim that $\ell \in \oblines(k)$. If all three lines have finite slope, this follows from \CiteTheorem{lem:observer-lines} because $\ell$ and $\ell_1$ have equal (hence finite) slopes and $\ell_1$ is a $k$-observer line. On the other hand, if all three lines (and hence also $\ell$) have infinite slope, this means there exist $t_1$, $t_2$ and $t_3$ such that all points on $\ell_i$ ($i = 1,2,3$) have time component $t_i$. But we know that the lines intersect one another, so we must have $t_1 = t_2 = t_3$. Since $\ell$ lies in the plane spanned by these lines it follows that points on $\ell$ share the same time component as points on $\ell_1$, and we can again apply \CiteTheorem{lem:observer-lines} to $\ell$ and $\ell_1$ to deduce that $\ell \in \oblines(k)$.

As claimed, therefore, $\ell$ is a $k$-observer line. Therefore,
$\ell$, $\ell_2$ and $\ell_3$ form a $k$-triad and
\CiteLemma{lem:triads} tells us that $\w{h}{k}[\ell]$,
$\w{h}{k}[\ell_2]$ and $\w{h}{k}[\ell_3]$ form an $h$-triad. It
follows that $\w{h}{k}[\ell]$ lies in the same plane as
$\w{h}{k}[\ell_2]$ and $\w{h}{k}[\ell_3]$, \ie $P'$, and hence
$\w{h}{k}(\vvp) \in \w{h}{k}[\ell] \subseteq P'$, as required.
\end{proof}

The following formula says that instantaneously moving observers exists.
\begin{newaxiom}
\item[\ul{\EInf}] There are observers $m,k\in\IOb$ such that
  $\w{m}{k}\take{\origin}_t=\w{m}{k}\take{\tunit}_t.$
\end{newaxiom}

\begin{Lemma}{lem:straight-lines-are-observer-lines}
Assume \AxEField, \AxWvt, \AxLine, \AxRelocate, \AxIso and \EInf.  Then for
any observer, every line is an observer line.
\end{Lemma}
\begin{figure}[h!btp] 
\small
\psfragfig[keepaspectratio,width=\linewidth]{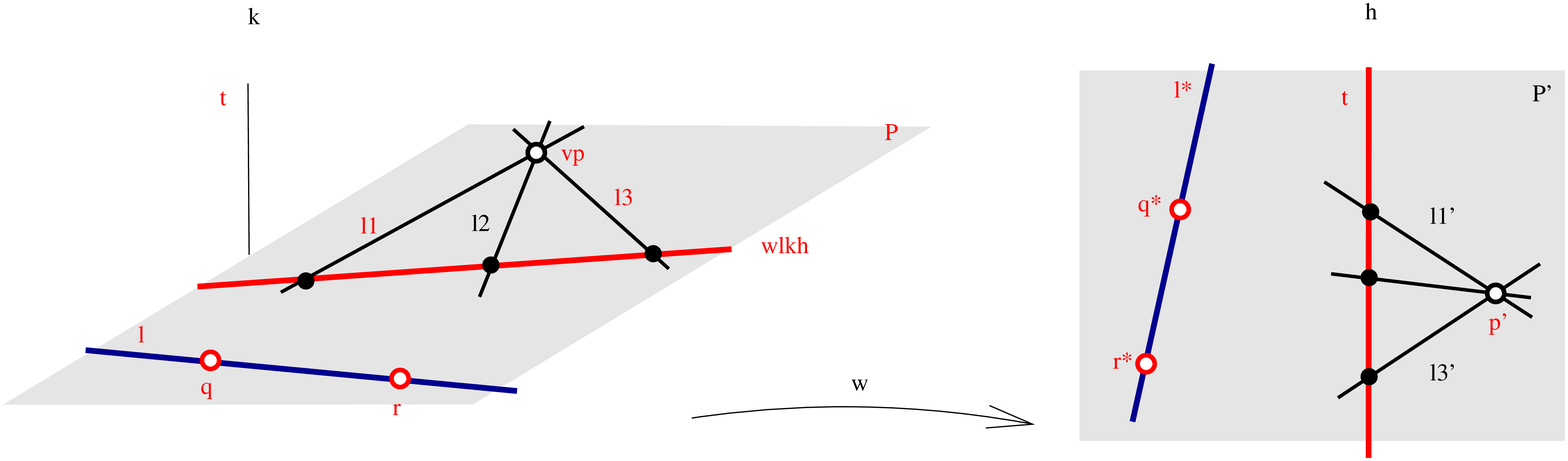}
{
\psfrag{P}[rt][rt]{$P$}
\psfrag{P'}[rt][rt]{$P'$}
\psfrag{l1}[rb][rb]{$\ell_1$}
\psfrag{l2}[rb][rb]{$\ell_2$}
\psfrag{l3}[lb][lb]{$\ell_3$}
\psfrag{l1'}[lb][lb]{$\ell_1'$}
\psfrag{l3'}[lt][lt]{$\ell_3'$}
\psfrag{l}[b][b]{$\ell$}
\psfrag{l*}[r][r]{$\ell^*$}
\psfrag{t}[r][r]{$\taxis$}
\psfrag{q}[t][t]{$\vvq$}
\psfrag{r}[t][t]{$\vvr$}
\psfrag{l}[b][b]{$\ell$}
\psfrag{q*}[r][r]{$\vvq^*$}
\psfrag{r*}[r][r]{$\vvr^*$}
\psfrag{t}[r][r]{$\taxis$}
\psfrag{vp}[l][l]{$\vvp$}
\psfrag{p'}[t][t]{$\vvp'$}
\psfrag{w}[b][b]{$\w{h}{k}$}
\psfrag{h}[t][t]{\Large $h$}
\psfrag{k}[t][t]{\Large $k$}
\psfrag{wlkh}[l][l]{$\wl_k(h)$}
}
\caption{\label{fig:linesareoblines}
Illustration for the proof of 
\CiteLemma{lem:straight-lines-are-observer-lines}}
\end{figure}
\begin{proof}  
Choose $k, h \in \IOb$ such that $\speed_k(h) = \infty$, and recall that this means that $\slope(\wl_k(h)) = \infty$. Thus, there exists some $t \in Q$ such that every point on $\wl_k(h)$ has time component $t$. Let $P$ be any `horizontal' plane containing $\wl_k(h)$, \ie all points in $P$ have this same time component $t$. 
Then every line in $P$ is in $\oblines(k)$ by \CiteTheorem{lem:observer-lines}
because every line in $P$ is of slope $\infty$.

Choose $\vvp \in P \setminus \wl_k(h)$, and notice that the plane $P$ is determined by $\vvp$ and $\wl_k(h)$. It follows from \CiteLemma{lem:plane-to-plane} that $\w{h}{k}[P]$ is contained in a plane containing both $\w{h}{k}(\vvp)$ and $\w{h}{k}[\wl_k(h)]$. In other words, if we define $\vvp' = \w{h}{k}(\vvp)$, observe that $\w{h}{k}[\wl_k(h)] = \taxis$, and define $P'$ to be the plane generated by $\vvp'$ and \taxis, then $\w{h}{k}[P] \subseteq P'$.  

\medskip
We will show first that the reverse inclusion also holds, so that $\w{h}{k}[P]$ is the whole of $P'$. To this end, choose three lines $\ell_i$ ($i=1,2,3$) in $P$ which pass through $\vvp$ and whose intersections with $\wl_k(h)$ are three distinct points; as observed above, these are all $k$-observer lines. Thus, if we define, for each $i = 1,2,3$, $\ell_i' := \w{h}{k}[\ell_i]$ then $\ell_1', \ell_2', \ell_3'$ and $\taxis$ (= $\w{h}{k}[\wl_k(h)]$) are all $h$-observer lines in $P'$. Since $\w{h}{k}$ is a bijection by \CiteLemma{lem:wvt}, all four of these lines are distinct and moreover, each $\ell_i'$ passes through
$\vvp'$, and they meet \taxis in three distinct points.

Since at most one of the lines $\ell_i'$ can have infinite slope (and $\slope(\taxis) = 0$), we
have therefore shown that there exists in $P'$ a $k$-triad of observer lines, all
with finite slope.  By \CiteLemma{lem:plane-to-plane}, it follows that
$\w{k}{h}[P'] \subseteq P$, and hence $P' \subseteq \w{h}{k}[P]$. Thus, $\w{h}{k}[P] = P'$, as claimed.

\medskip
Now we will prove that every line in $P'$ is in $\oblines(h)$.  Let
$\ell^* \subseteq P'$ be a line and let $\vvq^*, \vvr^*$ be two
distinct points on $\ell^*$.  Then $\vvq: = \w{k}{h}(\vvq^*)$,
$\vvr := \w{k}{h}(\vvr^*)$ are two distinct points in $P$ because
$\w{k}{h}[P'] \subseteq P$ and $\w{k}{h}$ is a bijection.  Let
$\ell$ be the line connecting $\vvq$ and $\vvr$.  Then $\ell$ lies in
$P$, and must therefore be in $\oblines(k)$.  Since $\w{h}{k}[\ell]
= \ell^*$, it follows by \CiteLemma{lem:transformed-observer-lines}
that $\ell^* \in \oblines(h)$ as claimed.

\medskip
Now we use the fact that $\taxis \subseteq P'$ to prove that
\emph{every} line is in $\oblines(h)$. Let $\ell$ be an arbitrary line. Then there is
some $\ell^* \subseteq P'$ which has the same slope as $\ell$ because $\taxis
\subseteq P'$ and therefore lines of every positive slope occur in $P'$ by
\CiteLemma{lem:triangulation}, while if $\slope(\ell) = 0$ we can take 
$\ell^* = \taxis$, and if $\slope(\ell) = \infty$ we can take 
$\ell^*$ to be the line joining $\vvp'$ to $((\vvp')_t, \vv 0)$. 
Moreover, by using translations `up or down' the time-axis as necessary, 
$\ell^*$ can be chosen such that
there are $\vvp \in \ell$, $\vvq \in \ell^*$ such that $\vvp_t = \vvq_t$.  We
know that $\ell^* \in \oblines(h)$ because every line in $P'$ is in
$\oblines(h)$.  But now $\ell \in \oblines(h)$ by
\CiteTheorem{lem:observer-lines}.  So $\oblines(h)$ is the set of all lines, as claimed.

\medskip
Finally, it is easy to see that because $\oblines(h)$ is the set of all lines for
one observer $h$, the same holds for every other observer $m$. 
For suppose $\ell'$ is a line, and choose distinct points $\vvp', \vvq' \in \ell'$. By \CiteLemma{lem:wvt}, the points $\vvp := \w{h}{m}(\vvp')$ and $\vvq := \w{h}{m}(\vvq')$ are again distinct, so they define a line $\ell$. As we've just seen, $\ell$ must be an $h$-observer line. It follows from \CiteLemma{lem:transformed-observer-lines} that $\w{m}{h}[\ell]$ is an $m$-observer line, and hence a line. This means that $\ell'$ and $\w{m}{h}[\ell]$ are both lines passing through the two points $\vvp' \neq \vvq'$, so they must be the same line. In other words, $\ell' = \w{m}{h}[\ell] \in \oblines(m)$, as claimed.
\end{proof}

\subsubsection{Main proof} We now complete the proof of \CiteTheorem{lem:line-to-line}.

\begin{defn}[Observer Planes]
Whenever a  plane $P$ contains at least one $k$-observer line, 
we shall say that $P$ is an \emph{observer plane for $k$}, 
or a $k$-observer plane. We write $\obplanes(k)$ for the set of all 
 $k$-observer planes.
\QED\end{defn}



\begin{proof}[Proof of \CiteTheorem{lem:line-to-line}]
  We have already noted that every worldview transformation $\w{k}{h}$ is a bijection; we will show first that they also take lines to lines.
  
  \medskip
  Suppose $m, m'$ are observers in motion relative to one another, \ie $\speed_m(m') > 0$ --- such observers
  exist by \EMovingIOb and \CiteLemma{lem:speed}. There are two cases to consider, depending
  on whether $\speed_m(m')$ can or cannot be infinite.

(Case 1: $\EInf$): If $m, m'$ can be chosen with $\speed_m(m') = \infty$, then \CiteLemma{lem:straight-lines-are-observer-lines} tells us that all lines belong to $\oblines(h)$ and we know that $\w{k}{h}$ takes observer lines to observer lines (which are again lines). So in this case, the result is immediate.

(Case 2: $\lnot\EInf$): Assume, therefore, that all observers move with finite speed relative to one
another (so that, given any observer $o$ and $\ell \in \oblines(o)$,
we have $\slope(\ell) \neq \infty$); in particular, $0 <
\speed_m(m') \neq \infty$.  Our proof will be given in four 
stages; we will show that 
\begin{itemize}
\item[(1)] if a plane $P$ contains a $k$-triad,
then $\w{h}{k}[P]$ is again a plane; 
\item[(2)] that for every observer
$o$ there is some $\ell \in \oblines(o)$ for which $\slope(\ell) \neq
0$; 
\item[(3)] if $P \in \obplanes(k)$ there exists a $k$-triad lying
entirely within $P$.  Items (1) and (3) imply that $\w{h}{k}$ maps
$k$-observer planes to $h$-observer planes.   
\item[(4)] Finally, we use this
information to show that every line can be obtained as
the intersection of two $k$-observer planes --- since the images of
these planes intersect in a line, the result then follows.
\end{itemize}

\textit{(1)} We prove that if a plane $P$ contains a $k$-triad, then
$\w{h}{k}[P]$ is a plane.  Let $\{\ell_1, \ell_2, \ell_3\}$ be a
$k$-triad contained in $P$, and for each $i=1,2,3$ define $\ell_i' :=
\w{h}{k}[\ell_i]$. Because all observer lines are assumed to have
finite slopes, \CiteLemma{lem:plane-to-plane} tells us that
$\w{h}{k}[P] \subseteq P'$, where $P'$ is the plane generated by
$\{\ell_1', \ell_2', \ell_3'\}$.  Since, by \CiteLemma{lem:triads},
$\{\ell_1', \ell_2', \ell_3'\}$ is likewise an $h$-triad contained in
$P'$ and comprising finite-slope lines, we can again apply
\CiteLemma{lem:plane-to-plane} to deduce that $\w{k}{h}[P']
\subseteq P$.  Consequently, $\w{h}{k}[P] = P'$, and
$\w{h}{k}[P]$ is a plane as claimed.

\textit{(2)} Next we show that for every observer $o$ there is some
$\ell \in \oblines(o)$ for which $\slope(\ell) \neq 0$. To this end,
let $\ell'$ be the line parallel to $\wl_m(m')$ which passes through
the origin $\origin$, and note that this line cannot be the time-axis
(which has slope 0). Since $\wl_m(m')$ is an $m$-observer line, so is
$\ell'$ (by \CiteTheorem{lem:observer-lines}). It follows that $\ell'$
and $\taxis = \wl_m(m)$ are non-identical intersecting $m$-observer
lines, whence $\w{o}{m}[\ell']$ and $\w{o}{m}[\taxis]$ are
non-identical intersecting $o$-observer lines. If these both had zero
slope, they would be the same line. So at least one of them has
non-zero slope and hence can be taken to be $\ell$.

\textit{(3)} Now we prove that for every $k$, if $P \in \obplanes(k)$
there exists a $k$-triad lying entirely in $P$.  Suppose $P \in
\obplanes(k)$, and choose some $k$-observer line $\ell = \wl_k(h)
\subseteq P$ and some $\vvp \in P\setminus\ell$, see
Figure~\ref{fig:LtoL3}.  Transforming to $h$'s worldview we have
$\w{h}{k}[\ell] = \w{h}{k}[\wl_k(h)] = \wl_h(h) = \taxis$ and
$\vvp' := \w{h}{k}(\vvp) \not\in \taxis$.  By (2), we know there is
some $\ell' \in \oblines(h)$ for which $\slope(\ell') \neq 0$, and by
assumption $\slope(\ell') \neq \infty$.  Thus, by
\CiteLemma{lem:triangulation} there exist lines $\ell_1'$, $\ell_2'$
passing through $\vvp'$ which have the same slope as $\ell'$, such
that $\{\taxis, \ell_1', \ell_2'\}$ is a $k$-triad (see
Figure~\ref{fig:LtoL3}), and we know that $\ell_1', \ell_2' \in
\oblines(h)$ by \CiteTheorem{lem:observer-lines}.  Taking $\ell_1 :=
\w{k}{h}[\ell_1']$ and $\ell_2 := \w{k}{h}[\ell_2']$, and
recalling that $\w{k}{h}[\taxis] = \ell$, it follows that all three
lines are $k$-observer lines, and together they form a $k$-triad lying
entirely within $P$ because their pairwise intersections comprise the
point $\vvp \not\in \ell$ together with two distinct points on $\ell$.

\begin{figure}[h!btp] 
\small
\psfragfig[keepaspectratio,width=\linewidth]{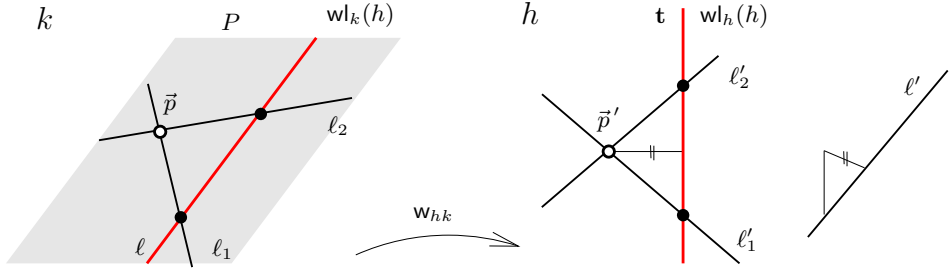}
{
\psfrag{p}[lb][lb]{$\vvp$}
\psfrag{p'}[b][b]{$\vvp'$}
\psfrag{l2'}[lt][lt]{$\ell_2'$}
\psfrag{l1'}[lb][lb]{$\ell_1'$}
\psfrag{l}[rb][rb]{$\ell$}
\psfrag{wl}[lb][lb]{$\wl_k(h)$}
\psfrag{P}[b][b]{$P$}
\psfrag{l'}[rb][rb]{$\ell'$}
\psfrag{k}[t][t]{\Large $k$}
\psfrag{h}[t][t]{\Large $h$}
\psfrag{l1}[lb][lb]{$\ell_1$}
\psfrag{l2}[lt][lt]{$\ell_2$}
\psfrag{wlh}[lt][lt]{$\wl_h(h)$}
\psfrag{t}[rt][rt]{$\taxis$}
\psfrag{w}[b][b]{$\w{h}{k}$}
}
\caption{\label{fig:LtoL3} Illustration for item (3) of the proof of
\CiteTheorem{lem:line-to-line}.}
\end{figure}

\medskip
Taken together, these results imply that whenever $P \in \obplanes(k)$, then $\w{h}{k}[P]$ is
a plane.

\medskip
\textit{(4)} Now let $k \in \IOb$.  We want to prove that any line can be
obtained as the intersection of two planes in $\obplanes(k)$.  To see this,
let $\ell$ be any line, and choose any $\vvp \in \ell$, see
Figure~\ref{fig:LtoL4}.  As we have just seen, we can also choose $\ell' \in
\oblines(k)$ such that $\slope(\ell') \neq 0$ and (by assumption)
$\slope(\ell') \neq \infty$.  Let $\ell_1$, $\ell_2$ be lines passing
through $\vvp$, having the same slope as $\ell'$, such that $\ell$, $\ell_1$
and $\ell_2$ are not co-planar (such lines can be obtained from $\ell'$ by a
combination of translation and spatial rotation). It follows from
\CiteTheorem{lem:observer-lines} that $\ell_1, \ell_2 \in \oblines(k)$.  For each $i=1,2$, let
$P_i$ be the plane containing $\ell_i$ and $\ell$.  Then $P_1$,
$P_2$ are $k$-observer planes and their intersection is $\ell$, as required.
\begin{figure}[h!btp] 
\small
\psfragfig[keepaspectratio,width=0.8\linewidth]{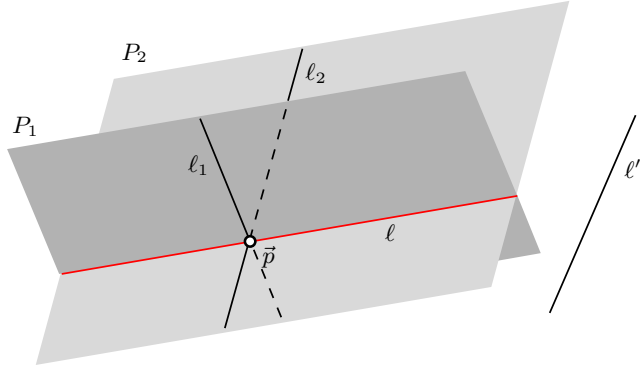}
{
\psfrag{l'}[lt][lt]{$\ell'$}
\psfrag{P1}[b][b]{$P_1$}
\psfrag{P2}[b][b]{$P_2$}
\psfrag{l}[t][t]{$\ell$}
\psfrag{l1}[rt][rt]{$\ell_1$}
\psfrag{l2}[lt][lt]{$\ell_2$}
\psfrag{p}[lt][lt]{$\vvp$}
}
\caption{\label{fig:LtoL4}
Illustration for item (4) of the proof of
\CiteTheorem{lem:line-to-line}.}
\end{figure}

It now follows, once again, that given any $k, h \in\IOb$, the
worldview transformation $\w{h}{k}$ is a bijection that takes lines to
lines. For if $\ell$ is any line, choose $k$-observer planes $P_1$,
$P_2$ such that $\ell = P_1 \cap P_2$. Since $\w{h}{k}$ is one-to-one,
$\w{h}{k}[\ell]=\w{h}{k}[P_1]\cap \w{h}{k}[P_2]$ and
$\w{h}{k}[P_1]\neq \w{h}{k}[P_2]$ (as $P_1\neq P_2$).  Since
$\w{h}{k}[P_1]$ and $\w{h}{k}[P_2]$ are distinct intersecting planes,
their intersection $\w{h}{k}[\ell]$ is a line.

\medskip
This completes the proof that lines are mapped to lines. The claim for planes
and hyperplanes now follows easily. Given a plane, choose three non-collinear
points. These determine three distinct intersecting lines and their images
determine the image plane. Likewise, we can choose four non-coplanar points in
a hyperplane whose images determine the image hyperplane.
\end{proof}



\subsection{The $tx$-Plane Lemma}

\InformalStatement{
If $\w{k}{m}$ maps the $tx$-plane to itself, then it also maps the $yz$-plane to itself; moreover, if $\w{k}{m}$ is linear, then there is some positive $\lambda$ such that
$|\w{k}{m}(\vvp)| = \lambda|\vvp|$ for all \vvp in the $yz$-plane.
}

\begin{defn}[Principal Observer]
We now fix one observer $o$ for the rest of the paper (the \emph{principal observer}) and define $$\IOb_o \de \{ k \in \IOb : \w{k}{o}(\origin) = \origin  \}$$ to be the set of observers who agree with $o$ (and hence each other) as to the location of the origin.
\QED\end{defn}

Analogously to the definition of the time-axis \taxis, the three spatial axes (\xaxis, \yaxis, and
\zaxis) are defined in the usual way as:
\begin{equation*}
 \xaxis \de \{(0,x,0,0) : x \in \Q\}, \quad
 \yaxis \de \{(0,0,y,0) : y \in \Q\}, \quad
 \zaxis \de \{(0,0,0,z) : z \in \Q\}.
\end{equation*}
   
We write $\plane(\taxis,\xaxis)$ for the $tx$-plane and
$\plane(\yaxis,\zaxis)$ for the $yz$-plane. More generally, if $\ell \neq
\ell'$ are intersecting lines, then $\plane(\ell,\ell')$ denotes the plane
containing $\ell$ and $\ell'$.

\begin{Theorem}{lem:txplane}
  Assume $\KIN + \AxIso$. Let $m,k\in\IOb_o$ such that
  $\w{k}{m}[\plane(\taxis,\xaxis)]= \plane(\taxis,\xaxis)$. Then
\begin{equation}
\label{eq1:txplane}
  \w{k}{m}[\plane(\yaxis,\zaxis)]=\plane(\yaxis,\zaxis)
\end{equation}
and
\begin{equation}
\label{eq2:txplane}
  \text{if } \vvq,\vvp\in\plane(\yaxis,\zaxis) \text{ and }
  |\vvp|=|\vvq|, \text{ then }
  \left|\w{k}{m}(\vvp)\right|=\left|\w{k}{m}(\vvq)\right|.
\end{equation}
Moreover, if $\w{k}{m}$ is also linear, then there is a positive
$\lambda\in\Q$ such that 
\begin{equation}\label{eq:txplane-lambda}
\left|\w{k}{m}(\vvp)\right|=\lambda|\vvp|
\end{equation}
for all $\vvp\in\plane(\yaxis,\zaxis)$.
\end{Theorem}

\subsubsection{Supporting lemmas}

The supporting lemmas can be informally described as:

\begin{description}
\item[\CiteLemma{lem:xy}]
~ \\ {A transformation is trivial if and only if it is a $\kappa$-isometry for at least two different choices of $\kappa$.}
\item[\CiteLemma{lem:iobo2}]
~ \\ Elementary results concerning worldview transformations involving members of $\IOb_o$.
\item[\CiteLemma{lem:affine}]
~ \\ Suppose $f$ is a bijection on $\Q^4$ taking lines to lines. Then there is an automorphism
$\varphi$ of \Q and an affine transformation $A$ such that $f = A \circ \widetilde{\varphi}$ (where $\widetilde{\varphi}$ is the coordinatewise extension of $\varphi$ to $\Q^4$).
\item[\CiteLemma{lem:equal-worldlines}] 
~ \\ If any one observer considers $m, m^* \in \IOb$ to have the same worldline, then all other observers do so as well.
\item[\CiteLemma{lem:colocate}]
~ \\ If two observers share the same worldline, the worldview transformation between them is trivial.
\end{description}

\subsubsection{Proofs of the supporting lemmas}

\begin{Lemma}{lem:xy}
  Assume that $(\Q,+,\cdot,0,1)$ is a field and choose $x,y\in\Q$ such
  that $x\neq y$. Then
  \begin{equation*}
    \triv=\kiso[x]\cap\kiso[y].
  \end{equation*}
In particular, every trivial transformation is a Euclidean isometry.
\end{Lemma}
\begin{proof}
($\subseteq$) Choose any $x,y \in \Q$, $T \in \triv$ and $\vvp = (t,\vv s) \in \Q^4$. We will show that $T \in \kiso[x]$. Without loss of generality we can assume that $T$ is linear (since it is the composition of a linear map with a translation, and all translations
are $x$-isometries). It follows that $T(\vvp) = T(t, \vv 0) + T(0,\vv s)$. However, because $T$ is trivial, we know that it fixes and preserves squared lengths in both \taxis and \saxis, so there exist $t'$, $\vv s'$ such that $T(t, \vv 0) = (t', \vv 0)$ and $T(0,\vv s) = (0, \vv s')$, where $|t|^2 = |t'|^2$ and $|\vv s|^2 = |\vv s'|^2$. It follows immediately that $\knorm[x]{\vv p} = |t|^2 - x|\vv s|^2 = |t'|^2 - x|\vv s'|^2 = \knorm[x]{T(\vv p)}$, \ie $T$ preserves squared
$\kappa$-lengths. 
It now follows that $T \in \kiso[x]$ when $x \neq 0$, and because $|\vv s|^2 = |\vv s'|^2$ no matter what the value of $t$, we also have $T \in \kiso[x]$ when $x = 0$. Finally, because $x$ can be any value in $\Q$ we also have $T \in \kiso[y]$, and hence $\triv \subseteq \kiso[x] \cap \kiso[y]$, as claimed.

($\supseteq$) To show the converse, choose any $x \neq y \in \Q$ and any
$T\in\kiso[x]\cap\kiso[y]$. We will show that $T \in \triv$.

Assume first that $T$ is linear. 
Choose any $\vvp = (t,\vv s) \in\Q^4$ and suppose $T(\vvp) = (t', \vv s')$.
Because $T$ is in both $\kiso[x]$ and $\kiso[y]$, we have both $\knorm[x]{T(\vvp)} = \knorm[x]{\vvp}$ and $\knorm[y]{T(\vvp)} = \knorm[y]{\vvp}$, \ie
\begin{align}
|t'|^2 - x|\vv s'|^2  &= |t|^2 - x|\vv s|^2 \text { and } \label{x} \\
|t'|^2 - y|\vv s'|^2  &= |t|^2 - y|\vv s|^2. \label{y}
\end{align}
Subtracting \eqref{y} from \eqref{x} gives
\[
(x-y)|\vv s'|^2  = (x-y)|\vv s|^2 
\]
whence division by $(x-y) \neq 0$ gives both
\begin{equation}\label{s}
|\vv s'|^2 =  |\vv s|^2
\end{equation}
and hence (by either \eqref{x} or \eqref{y})
\begin{equation}\label{t}
|t'|^2 = |t|^2 .
\end{equation}
Therefore,
\begin{align*}
  &\text{if } t=0, \text{ then } t'=0, \text{ and }\\
  &\text{if } \vv s = \vv 0, \text{ then } \vv s' = \vv 0,
\end{align*}
which together with \eqref{s} and \eqref{t} show that
$T\in\triv$.

If $T$ is not itself linear, notice that we can write $T = L \circ \tau$ where $\tau$ is a translation and $L$
is a linear $x$-isometry. Since $T \in \kiso[y]$ and $L = T \circ \tau^{-1}$ differs from $T$ 
only by a translation (and all translations are in $\kiso[y]$), we see that $L$ is in $\kiso[y]$ too. Thus, $L$ is a linear map in $\kiso[x]\cap\kiso[y]$ (in other words, the ``linear'' and ``translation'' parts of $T$ are the same in \kiso[x] as in \kiso[y]) whence it follows from what we have just shown that $L$ is trivial. Because $\tau$ is trivial, we now conclude that $T = L \circ \tau$ is itself trivial, as claimed.

In particular, we have $\triv = (\kiso[0] \cap \kiso[-1]) \subseteq \kiso[-1]$, \ie all trivial transformations are Euclidean isometries.
\end{proof}

\begin{Lemma}{lem:iobo2}
Assume \AxWvt. Let $k,h\in\IOb_o$ and $m\in\IOb$. Then (a)--(c) below hold.
\begin{itemize}
\item[(a)] $\w{k}{h}\take{\origin}=\origin$ and $\origin\in\wl_k(h)$.
\item[(b)] If $\w{k}{m}\take{\origin}=\origin$, then $m\in\IOb_o$.
\item[(c)] If $R:\Q^4\rightarrow\Q^4$, $R\take{\origin}=\origin$ and
$k\mappedto{R}{h} m$, then $m\in\IOb_o$.
\end{itemize}
\end{Lemma}
\begin{proof} The proof involves only straightforward applications of \CiteLemma{lem:wvt}, and we omit the details.
\end{proof}

\begin{Lemma}{lem:affine}
Assume $\Q = (\Q, +, \cdot, 0, 1, \leq)$ is a Euclidean field, 
and suppose $f \colon Q^4 \to \Q^4$ is a bijection taking 
lines to lines. Then there is an ordered-field automorphism $\varphi$ 
of \Q and an affine 
transformation $A$ on $\Q^4$ such that $f = A \circ \widetilde{\varphi}$, where 
$\widetilde{\varphi} \colon \Q^4 \to \Q^4$ is the map $\widetilde{\varphi} 
\colon (t,x,y,z) \mapsto  (\varphi(t), \varphi(x), \varphi(y), \varphi(z))$.
\end{Lemma}

\begin{proof}
By the Fundamental Theorem of Affine Geometry \cite[Thm.~2.6.3,
  p.~52]{Berger}, there is an automorphism $\varphi$ of
field $(\Q, +, \cdot, 0, 1)$ and an affine transformation $A$ such
that $f=A\circ \widetilde{\varphi}$. To complete the proof of the lemma,
we only have to show that $\varphi$ is order preserving, \ie
$\varphi(a)\le\varphi(b)$ if{}f $a\le b$. Since $x\le y$ if{}f $0\le
y-x$, it is enough to show that $0\le \varphi(z)$ if{}f $0\le z$ ---
and this follows directly from the Euclidean property, \ie $0\le d$ if{}f
$d=c^2$ for some $c\in\Q$.
\end{proof}

\begin{Lemma}{lem:equal-worldlines}
Assume \AxWvt.
Suppose $m,m^* \in \IOb$, and suppose $\wl_k(m) = \wl_k(m^*)$ for some 
$k \in \IOb$. Then $\wl_j(m) = \wl_j(m^*)$ for all $j \in \IOb$.
\end{Lemma}
\begin{proof}
By \CiteLemma{lem:wvt}, $\wl_{j}(m) = \w{j}{k}[\wl_{k}(m)]= 
\w{j}{k}[\wl_{k}(m^*)] = \wl_{j}(m^*)$ for all $j \in \IOb$.
\end{proof}

\begin{defn}
Let $m,m^*\in\IOb$. If $\wl_k(m) =\wl_k(m^*)$ for some $k\in\IOb$, we say that
$m$ and $m^*$ \semph{share the same worldline}.
\end{defn}

\begin{Lemma}{lem:colocate} 
Assume \AxWvt and let $m,m^*\in\IOb$. Suppose $m$ and $m^*$ share the same 
worldline. If 
$\AxColocate$ holds, then $\w{m}{m^*}\in\triv$. 
\end{Lemma}
\begin{proof}
Saying that $m$ and $m^*$ share the same worldline means that
$\wl_k(m) = \wl_k(m^*)$ for some $k \in \IOb$. By
\CiteLemma{lem:equal-worldlines}, this equation therefore holds for
\emph{all} choices of $k$, and in particular for $k = m$, \ie
$\wl_m(m) = \wl_m(m^*)$. The claim now follows immediately by
\AxColocate.
\end{proof}

\subsubsection{Main proof} We now complete the proof of \CiteTheorem{lem:txplane}.



\begin{proof}[Proof of \CiteTheorem{lem:txplane}]
Let $m,k\in\IOb_o$ such that $\w{k}{m}[\plane(\taxis,\xaxis)] = 
\plane(\taxis,\xaxis)$. 
By \CiteLemma{lem:iobo2}, 
$\w{m}{k}\take{\origin}=\w{k}{m}\take{\origin}=\origin$.

Let us first prove the following claim
\begin{equation}\label{csillag}
\parbox{.9\linewidth}{\centering
If $R\in\srot$ fixes $\plane(\taxis,\xaxis)$ pointwise, then there exists $k^*\in\IOb$ \\ 
such that (a) $\w{k}{k^*}=\w{k}{m}\circ R\circ \w{m}{k}$ and (b) $\w{k}{k^*}\in\triv$.
}
\end{equation}

\medskip
\emph{Proof of claim \eqref{csillag}}. (a) By
\CiteLemma{lem:relocation}, there exists some $k^*$ such that $k
\mappedto{R}{m} k^*$, \ie $\w{m}{k^*} = R \circ \w{m}{k}$. Hence,
$\w{k}{k^*}= \w{k}{m} \circ \w{m}{k^*} =\w{k}{m}\circ R \circ
\w{m}{k}$.  (b) By \CiteLemma{lem:worldline-relocation}, we have
$\wl_m(k^*) = R[\wl_m(k)]$, and because $\wl_m(k)=\w{m}{k}[\taxis]
\subseteq \plane(\taxis,\xaxis)$ and $R$ leaves
$\plane(\taxis,\xaxis)$ pointwise-fixed, we have that $R[\wl_m(k)] =
\wl_m(k)$.  Thus, $\wl_m(k^*) = R[\wl_m(k)] = \wl_m(k)$, \ie $k$ and
$k^*$ share the same worldline. So $\w{k}{k^*} \in \triv$ by
\CiteLemma{lem:colocate}. Thus, \eqref{csillag} holds.

\medskip
\emph{Proof of statement \eqref{eq1:txplane}}. Choose any
$\vvp\in\plane(\yaxis,\zaxis)$ and write $\vvp':=\w{k}{m}(\vvp)$. We
have to prove that $\vvp'\in\plane(\yaxis,\zaxis)$.

We will show that $\vvp'\cdot \vvq=0$ for every
$\vvq\in\plane(\taxis,\xaxis)$, whence it follows easily that
$\vvp'\in\plane(\yaxis,\zaxis)$. 

By \CiteLemma{lem:affine}, \CiteTheorem{lem:line-to-line} and the fact
that $\w{k}{m}(\origin)=\origin$, we know that $\w{k}{m}$ can be
written as a composition $ \w{k}{m} = L \circ \widetilde{\varphi} $ of
a linear transformation, $L$, and a map induced by a field
automorphism, $\varphi$.  Therefore, $\w{k}{m}(-\vvp) =
L(\widetilde{\varphi}(-\vvp)) = L(-\widetilde{\varphi}(\vvp)) =
-L(\widetilde{\varphi}(\vvp)) = -\w{k}{m}(\vvp)= -\vvp'$.

\begin{figure}[h!btp] 
\small
\psfragfig[keepaspectratio,width=\linewidth]{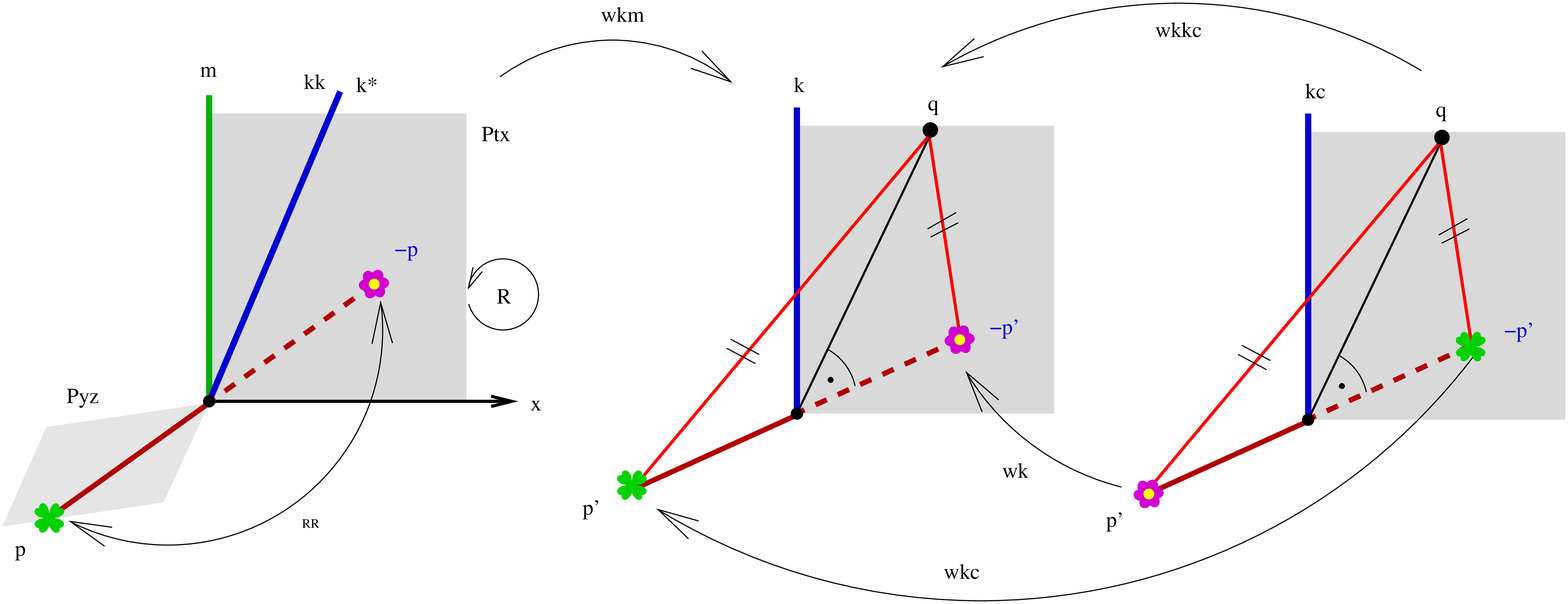}
{
\psfrag{q}[b][b]{$\vvq$}
\psfrag{m}[b][b]{\large $m$}
\psfrag{k}[b][b]{\large $k$}
\psfrag{kc}[b][b]{\large $k^*$}
\psfrag{kk}[rb][rb]{$k$}
\psfrag{k*}[lb][lb]{$k^*$}
\psfrag{x}[l][l]{$\xaxis$}
\psfrag{y}[t][t]{$\yaxis$}
\psfrag{z}[rt][rt]{$\zaxis$}
\psfrag{Ptx}[l][l]{$\plane(\taxis,\xaxis)$}
\psfrag{Pyz}[b][b]{$\plane(\yaxis,\zaxis)$}
\psfrag{R}[c][c]{$R$}
\psfrag{RR}[lt][lt]{$R$}
\psfrag{p}[rt][rt]{$\vvp$}
\psfrag{-p}[lb][lb]{$-\vvp$}
\psfrag{wkm}[b][b]{$\w{k}{m}$}
\psfrag{p'}[rt][rt]{$\vvp'$}
\psfrag{-p'}[lb][lb]{$-\vvp'$}
\psfrag{wkc}[b][b]{$\w{k}{k^*}$}
\psfrag{wk}[rt][rt]{$\w{k}{k^*}$}
\psfrag{wkkc}[b][b]{$\w{k}{k^*}$}
}
\caption{\label{fig:perpendicular} Illustration for the proof of
\eqref{eq1:txplane} of \CiteTheorem{lem:txplane}.}
\end{figure}

Let $R$ be the linear transformation that takes
$\tunit,\xunit,\yunit,\zunit$ to $\tunit,\xunit,-\yunit,-\zunit$,
respectively.  Then $R$ is a self-inverse spatial rotation that leaves
$\plane(\taxis,\xaxis)$ pointwise fixed and takes $\vvp$ to $-\vvp$,
see Figure~\ref{fig:perpendicular}.  So by \eqref{csillag}, there is
$k^*\in\IOb$ such that $\w{k}{k^*}\in\triv$ and
$\w{k}{k^*}=\w{k}{m}\circ R\circ \w{m}{k}$.

Let $\vvq\in\plane(\taxis,\xaxis)$ be arbitrary.  Now note that
$\w{m}{k}(\vvq)\in\plane(\taxis,\xaxis)$, hence
$R(\w{m}{k}(\vvq))=\w{m}{k}(\vvq)$. Note also that
$\w{k}{k^*}(\vvp')=-\vvp'$ and $\w{k}{k^*}(\vvq)=\vvq$ because
\[
  \w{k}{k^*}(\vvp') = \w{k}{m}(R(\w{m}{k}(\vvp'))) =
  \w{k}{m}(R(\vvp)) = \w{k}{m}(-\vvp) = - \vvp',
\]
\[
 \w{k}{k^*}(\vvq) =
 \w{k}{m}(R(\w{m}{k}(\vvq))) =  \w{k}{m}(\w{m}{k}(\vvq)) = \vvq.
\]

Now, because $\w{k}{k^*}$ is trivial, we know from \CiteLemma{lem:xy} that it is a Euclidean isometry. Moreover, because every trivial map is the composition of a linear map and a translation, and since it fixes \origin (because $\w{k}{m}$, $\w{m}{k}$ and $R$ all do so), $\w{k}{k^*}$ must be linear.

It follows that $|\vvq-\vvp'| = |\w{k}{k^*}(\vvq-\vvp')| = |\w{k}{k^*}(\vvq)- \w{k}{k^*}(\vvp')| =
|\vvq+\vvp'|$, whence
$(\vvq-\vvp')\cdot(\vvq-\vvp') = (\vvq+\vvp')\cdot
(\vvq+\vvp')$, and so $\vvp'\cdot\vvq=0$.  

Since this holds for
any $\vvq \in \plane(\taxis,\xaxis)$, in particular it holds for both
$\tunit$ and $\xunit$. Consequently, $\vvp'\in\plane(\yaxis,\zaxis)$
as claimed.

\begin{figure}[h!btp] 
\small
\psfragfig[keepaspectratio,width=\linewidth]{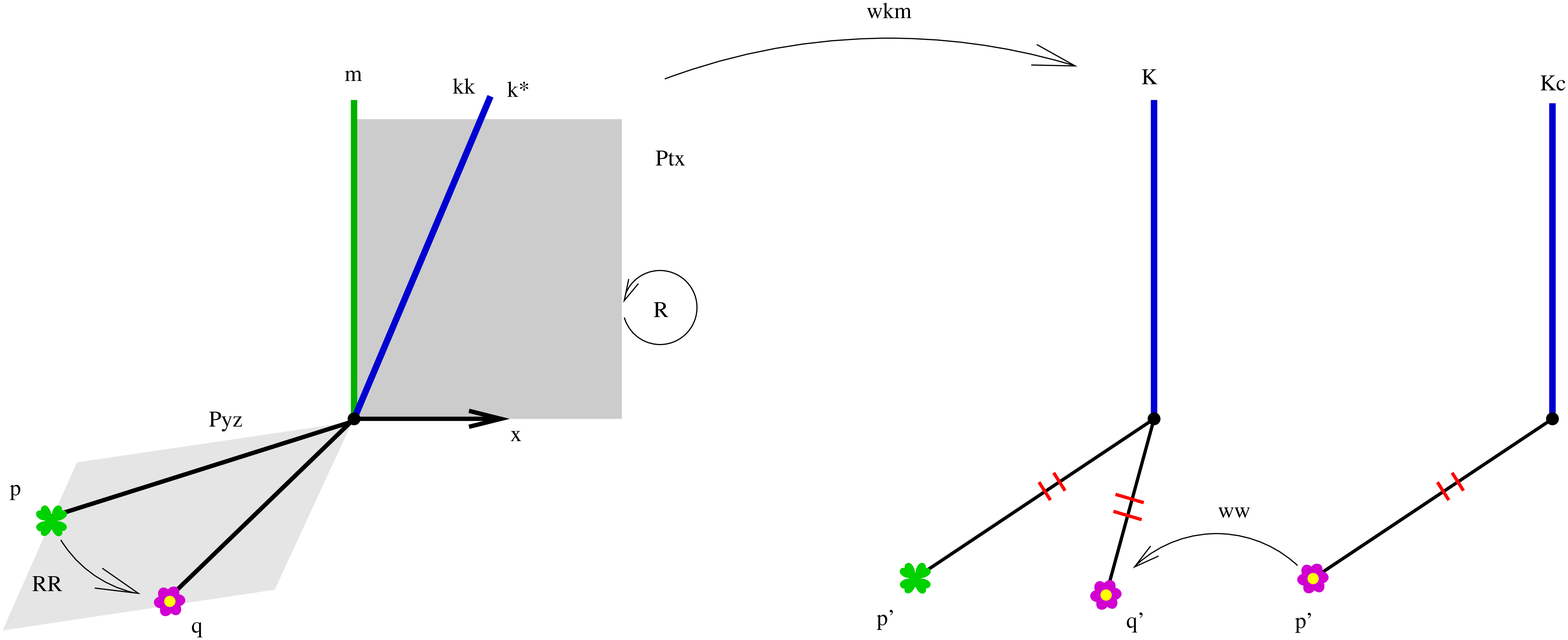}
{
\psfrag{x}[lt][lt]{$\xunit$}
\psfrag{wkm}[b][b]{$\w{k}{m}$}
\psfrag{m}[b][b]{\large $m$}
\psfrag{K}[b][b]{\large $k$}
\psfrag{Kc}[b][b]{\large $k^*$}
\psfrag{Ptx}[l][l]{$\plane(\taxis,\xaxis)$}
\psfrag{Pyz}[rb][rb]{$\plane(\yaxis,\zaxis)$}
\psfrag{ww}[b][b]{$\w{k}{k^*}$}
\psfrag{p'}[t][t]{$\vvp'$}
\psfrag{q'}[l][l]{$\vvq'$}
\psfrag{p}[rb][rb]{$\vvp$}
\psfrag{q}[lt][lt]{$\vvq$}
\psfrag{kk}[rb][rb]{$k$}
\psfrag{k*}[lb][lb]{$k^*$}
\psfrag{R}[c][c]{$R$}
\psfrag{RR}[rt][rt]{$R$}
}
\caption{\label{fig:equidistance} Illustration for the proof of
\eqref{eq2:txplane} of \CiteTheorem{lem:txplane}.}
\end{figure}

\medskip
\emph{Proof of statement \eqref{eq2:txplane}}. Let
$\vvp,\vvq\in\plane(\yaxis,\zaxis)$ and write $\vvp':=\w{k}{m}(\vvp)$
and $\vvq':= \w{k}{m}(\vvq)$.  Assume $\norm{\vvp}=\norm{\vvq}$. We
want to prove that $\norm{\vvp'}=\norm{\vvq'}$.  
By \CiteLemma{lem:vanrotacio}, there is a spatial rotation 
that takes $\xunit$ to $\xunit$ and $\vvp$ to $\vvq$.
Let $R'\in\srot$ be such a spatial rotation.
Then $R'$ leaves $\plane(\taxis,\xaxis)$ pointwise fixed
and takes $\vvp$ to $\vvq$.  By \eqref{csillag}, there is $k^*\in\IOb$
such that $\w{k}{k^*}\in\triv$ and $\w{k}{k^*}=\w{k}{m}\circ R'\circ
\w{m}{k}$, see Figure~\ref{fig:equidistance}. It follows that
\begin{equation*}
  \w{k}{k^*}(\vvp') = \w{k}{m}(R' (\w{m}{k}(\vvp'))) =
  \w{k}{m} (R'(\vvp)) = \w{k}{m}(\vvq) = \vvq'.
\end{equation*}
Finally, because $\w{k}{k^*}$ is trivial, \CiteLemma{lem:xy} tells us
that it is a Euclidean isometry. It now follows that $\norm{\vvp'} =
\norm{\w{k}{k^*}(\vvp')} = \norm{\vvq'}$, as claimed. Thus,
\eqref{eq2:txplane} holds.

\medskip
\emph{Proof of statement \eqref{eq:txplane-lambda}}. 
Now assume that $\w{k}{m}$ is linear. Let $\lambda:=\norm{\w{k}{m}
  (\yunit)}$. This $\lambda$ is positive since $\w{k}{m} (\yunit)\neq
\origin$ as $m,k\in\IOb_o$. We will prove that
$\left|\w{k}{m}(\vvp)\right|=\lambda\norm{\vvp}$ for every
$\vvp\in\plane(\yaxis,\zaxis)$. Clearly for $\vvp=\origin$ this holds, so assume that
$\vvp\in\plane(\yaxis,\zaxis) \setminus \{\origin\}$, and
note that
\[
\norm{\frac{\vvp}{\norm{\vvp}}}=1=\norm{\yunit}.
\]
Then, by \eqref{eq2:txplane}, 
\[
\norm{
\w{k}{m}
\left(\frac
{\vvp}
{\norm{\vvp}
}
\right)
}
=\norm{
\w{k}{m}(\yunit)
}
=
\lambda.
\]
Therefore,  by linearity of $\w{k}{m}$,
\[
\norm{
\w{k}{m}(\vvp)
}=
\norm{
\norm{
\vvp
}
\w{k}{m}
\left(
\frac{\vvp}{\norm{\vvp}}
\right)
}
=\lambda\norm{\vvp}.
\]
\end{proof}



\subsection{The Same-Speed Lemma}

\InformalStatement{
Suppose at least one observer considers $h$ and $k$ to be travelling with the
same speed. Then $\w{h}{k}$ is a $\kappa$-isometry for some $\kappa$.
}

\begin{Theorem}{lem:same-speed-hard}
Assume \KIN and \AxIso, and that $k,m,h \in \IOb_o$. If $\speed_m(k) = \speed_m(h)$, then 
\begin{itemize}
\item[(a)] there exists $\kappa$ such that $\w{h}{k}$ is a $\kappa$-isometry;
\item[(b)] $\speed_k(h) = \speed_h(k)$;
\item[(c)] $\speed_h(m) = \speed_k(m)$.
\end{itemize}
\end{Theorem}

\subsubsection{Supporting lemmas}

The supporting lemmas can be informally described as:

\begin{description}
\item[\CiteLemma{lem:iobo}]
~ \\ Every observer can be translated into $\IOb_o$.
\item[\CiteLemma{lem:letezikrotacio}]
~ \\ Every vertical plane can be rotated into the $tx$-plane.
\item[\CiteLemma{lem:lintriv-implies-same-speed}]
~ \\ If $\w{m}{m^*}$ is both linear and trivial, then every $j$ agrees that $m$ and $m^*$ are moving at the same speed, and likewise $m$ and $m^*$ agree on the speed of $j$.
\end{description}

\subsubsection{Proofs of the supporting lemmas}

\begin{Lemma}{lem:iobo}
Assume  \AxWvt and \AxRelocate. Given any $k \in \IOb$ there exists 
$k^o \in \IOb_o$ such that $\w{k^o}{k}$ is a translation.
\end{Lemma}
\begin{proof}
Let $T$ be the translation taking $\w{k}{o}(\origin)$ to the origin and let 
$k^o$ be an observer such that $\w{k^o}{k} = T$ 
(such an observer exists by \AxRelocate). 
Then $\w{k^o}{o}(\origin) = (\w{k^o}{k} \circ \w{k}{o}) \take{\origin}   
= T \take{\w{k}{o}(\origin)}   = \origin$, so $k^o \in \IOb_o$ as required.
\end{proof}

\begin{Lemma}{lem:letezikrotacio}
Assume $(\Q,+,\cdot,0,1,\leq)$ is a Euclidean field, that $P$ is a
 plane
in $\Q^4$ containing the time-axis $\taxis$, and that $\vvp \in P\setminus\taxis$.
Then there exists a spatial rotation $R$ that takes $P$
and $\vvp$ to $\plane(\taxis,\xaxis)$ and $(\vvp_t,\norm{\vvp_s},0,0)$, 
respectively.
\end{Lemma}
\begin{figure}[h!btp] 
\small
\psfragfig[keepaspectratio,width=0.9\linewidth]{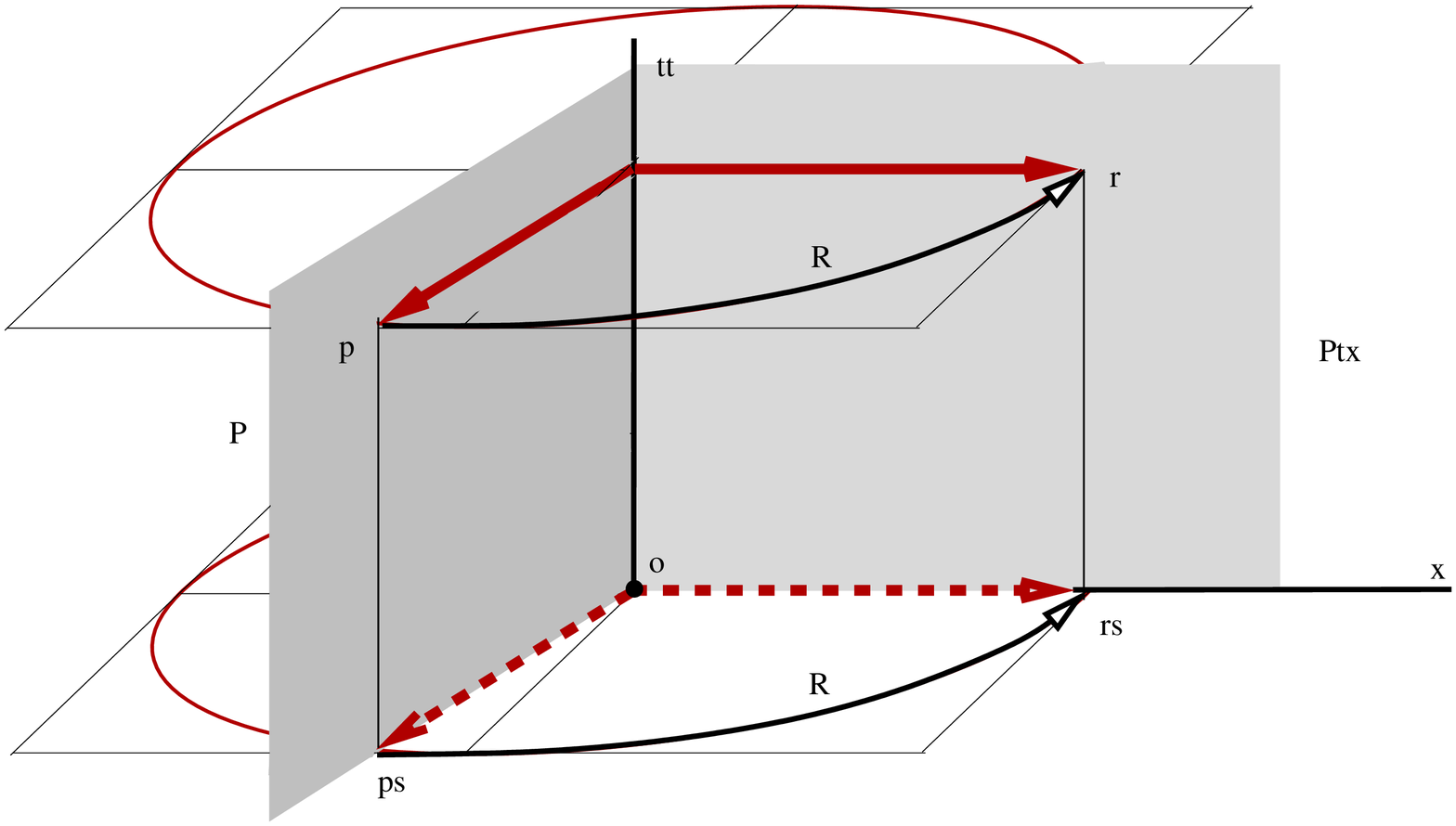}
{
\psfrag{p}[rt][rt]{$\vvp$}
\psfrag{R}[lb][lb]{$R$}
\psfrag{x}[lb][lb]{$\xaxis$}
\psfrag{tt}[l][l]{$\taxis$}
\psfrag{ps}[lt][lt]{$(0,\vvp_s)$}
\psfrag{rs}[lt][lt]{$(0,\norm{\vvp_s},0,0)$}
\psfrag{r}[l][l]{$(\vvp_t,\norm{\vvp_s},0,0)$}
\psfrag{Ptx}[l][l]{$\plane(\taxis,\xaxis)$}
\psfrag{P}[r][r]{$P$}
\psfrag{o}[lb][lb]{$\origin$}
}
\caption{\label{fig:rotation2} Illustration for
\CiteLemma{lem:letezikrotacio}.}
\end{figure}

\begin{proof}
By \CiteLemma{lem:vanrotacio},  there is
$R\in\srot$ which takes $(0,\vvp_s)$ to 
$(0,\norm{\vvp_s},0,0)$ and $\origin$ to \origin; 
see Figure~\ref{fig:rotation2}. It is easy to see
that this $R$ has the desired properties.
\end{proof}

\begin{Lemma}{lem:lintriv-implies-same-speed}
Assume \AxWvt and \AxEField and suppose $m, m^* \in \IOb$ and
$\w{m^*}{m}$ is a linear trivial transformation. Then
$\wl_j(m)=\wl_j(m^*)$ for every observer $j \in \IOb$. Furthermore, if
\AxLine is assumed, then $\speed_j(m)=\speed_j(m^*)$ and
$\speed_m(j)=\speed_{m^*}(j)$ for every
$j\in\IOb$.
\end{Lemma}
\begin{proof}
Recall that $\wl_{m^*}(m) = \w{m^*}{m}[\taxis]$.  Since $\w{m^*}{m}$
is a linear trivial transformation, we have $\w{m^*}{m}[\taxis] =
\taxis =\wl_{m^*}(m^*)$. Thus, $\wl_{m^*}(m) = \wl_{m^*}(m^*)$. Hence, for
every $j\in\IOb$,
$\wl_{j}(m)=\wl_{j}(m^*)$ by \CiteLemma{lem:equal-worldlines}. 

Now, assume
\AxLine and let $j\in\IOb$. Then $\speed_j(m)=\speed_j(m^*)$ since
$\wl_j(m)=\wl_j(m^*)$. It is easy to see that
$\slope(\ell)=\slope(f[\ell])$ holds for every trivial transformation $f$
and line $\ell$. Therefore,
$\speed_m(j)=\slope(\wl_m(j))=\slope(\w{m^*}{m}[\wl_m(j)])=\slope(\wl_{m^*}(j))
=\speed_{m^*}(j)$.
\end{proof}

\subsubsection{Main proof} We now complete the proof of \CiteTheorem{lem:same-speed-hard}.



\begin{proof}[Proof of \CiteTheorem{lem:same-speed-hard}]
Suppose $\speed_m(k) = \speed_m(h)$, where $m, k, h \in \IOb_o$.

\medskip
(a) If $\wl_m(k)=\wl_m(h)$, then $\w{h}{k}$ is a trivial
  transformation by \CiteLemma{lem:colocate}, hence it is a
  $\kappa$-isometry by \CiteLemma{lem:xy}.  
  
  Assume, therefore, that $\wl_m(k)\neq\wl_m(h)$.
  Because $k$ and $h$ have the same speed in $m$'s worldview, their
  worldlines have the same slope according to $m$. By
  \CiteLemma{lem:iobo2}, $\origin\in\wl_m(k)\cap\wl_m(h)$ because
  $m,k,h\in\IOb_o$. 
 
Let $\vvp_1\in\wl_m(k)$ and $\vvp_2\in\wl_m(h)$ be such that
$\vvp_1\neq\origin\neq\vvp_2$ and $(\vvp_1)_t=(\vvp_2)_t$, see
Figure~\ref{fig:same-speed-hard}.  Let $t^* := (\vvp_1)_t$ be the
common time component of $\vvp_1$ and $\vvp_2$.

Let $\vv s_1 := (\vvp_1)_s$ and $\vv s_2 := (\vvp_2)_s$.  Then $|\vv
s_1|=|\vv s_2|$ because lines $\wl_m(k)$ and $\wl_m(h)$ are of same slope. 
Thinking of $\vv s_1$ and $\vv s_2$ as points in $\Q^3$, let $\vv s^*$ be
the point mid-way between them, \ie $\vv s^* = (\vv s_1 + \vv s_2)/2$, and
let $\ell$ be a line in $Q^3$ passing through $\vv 0$ and $\vv s^*$.  If we
now define $\rho$ to be the map which rotates $Q^3$ through 180$^\circ$
about axis $\ell$, then the map $R$ given by $R(t,\vv s): = (t,\rho(\vv s))$
is a self-inverse spatial rotation.\footnote{\,We can define $\rho$ in the
usual way.  Given any $\vv s$ we decompose it into a sum $\vv s = \vv
s_\parallel + \vv s_\perp$ of components parallel and perpendicular to
$\ell$, respectively, and then $\rho(\vv s) = \vv s_\parallel - \vv
s_\perp$.}  

We claim that $R(\vvp_1) = \vvp_2$.  To see this, notice that the
points $\vv s_1$ and $\vv s_2$ form the base of an isosceles triangle
in $\Q^3$ whose vertex is $\vv 0$; it follows easily that the line
$\ell$ bisects and is orthogonal to the line joining $\vv s_1$ to $\vv
s_2$, whence the rotation $\rho$ about $\ell$ maps $\vv s_1$ to $\vv
s_2$ (and vice versa) in $Q^3$.  Thus, $R(\vvp_1) = R(t^*,\vv s_1) =
(t^*, \rho(\vv s_1)) = (t^*, \vv s_2) = \vvp_2$.  Since $R$ also fixes
\origin, it must take $\wl_m(k)$ to $\wl_m(h)$.  Point $\vv s^*$ is
fixed by $\rho$ because this point is on $\rho$'s axis of
rotation. Therefore, $(0,\vv s^*)$ is fixed by $R$.

So we have $R \in \srot$, $R[\wl_m(k)]=\wl_m(h)$, $R^{-1} = R$ and
$R(0,\vv s^*)=(0,\vv s^*)$.

\begin{figure}[h!btp] 
\small
\psfragfig[keepaspectratio,width=\linewidth]{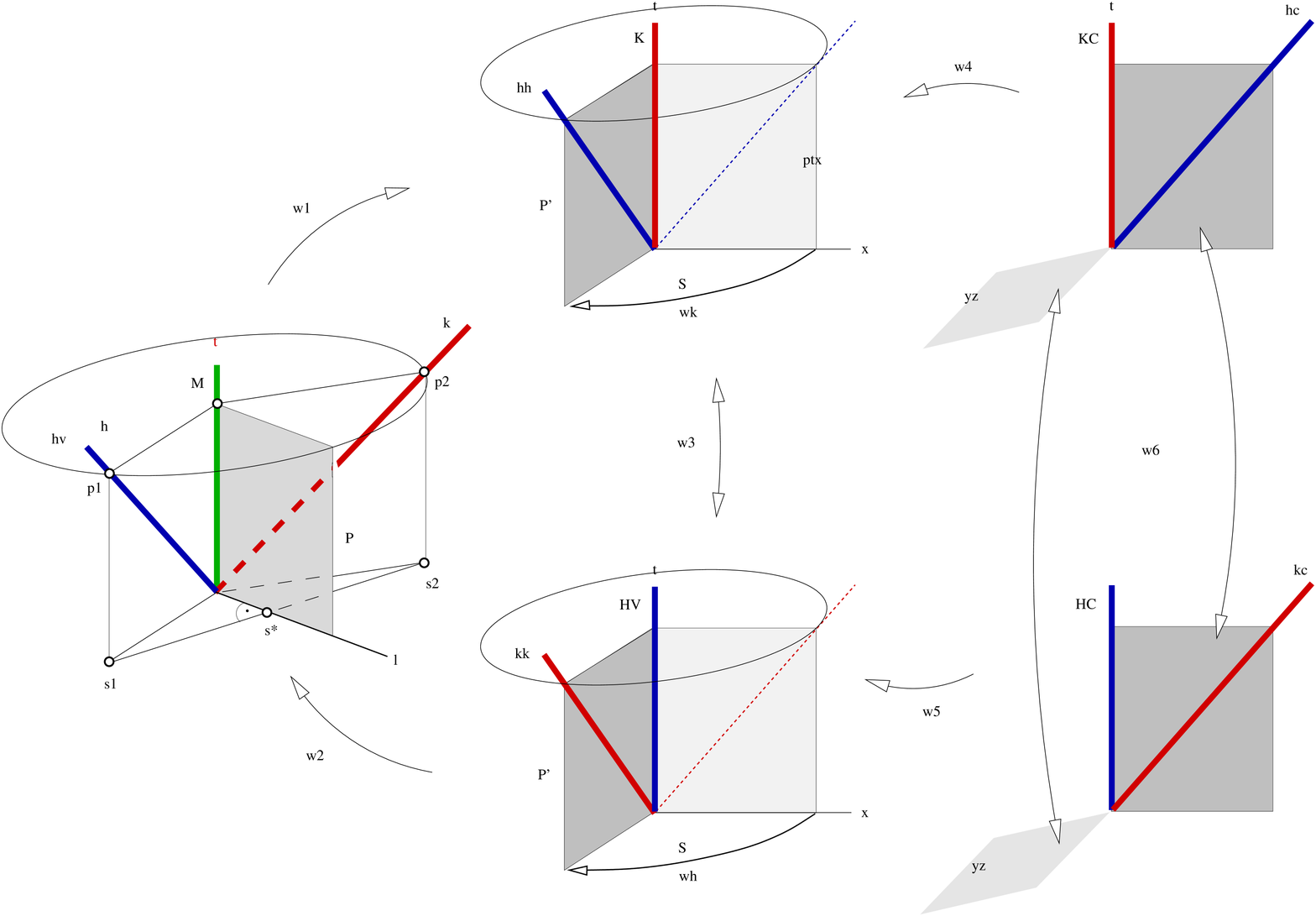}
{
\psfrag{p1}[rt][rt]{$\vv p_2$}
\psfrag{p2}[lt][lt]{$\vv p_1$}
\psfrag{s1}[t][t]{$\vv s_2$}
\psfrag{s2}[t][t]{$\vv s_1$}
\psfrag{s*}[t][t]{$\vv s_*$}
\psfrag{l}[l][l]{$\ell$}
\psfrag{x}[l][l]{$\xaxis$}
\psfrag{M}[r][r]{\large $m$}
\psfrag{K}[r][r]{\large $k$}
\psfrag{HV}[r][r]{\large $h'$}
\psfrag{KC}[r][r]{\large $k^*$}
\psfrag{HC}[r][r]{\large $h^*$}
\psfrag{h}[lb][lb]{$h$}
\psfrag{t}[b][b]{$\taxis$}
\psfrag{hv}[rt][rt]{$h'$}
\psfrag{k}[rb][rb]{$k$}
\psfrag{hh}[rb][rb]{$h'$}
\psfrag{kk}[rb][rb]{$k$}
\psfrag{hc}[rb][rb]{$h^*$}
\psfrag{kc}[rb][rb]{$k^*$}
\psfrag{P}[l][l]{$P$}
\psfrag{P'}[r][r]{$P'$}
\psfrag{ptx}[l][l]{$\plane(\taxis,\xaxis)$}
\psfrag{yz}[c][c]{$\plane(\yaxis,\zaxis)$}
\psfrag{w1}[rb][rb]{$\w{k}{m}$}
\psfrag{w2}[rt][rt]{$\w{m}{h'}$}
\psfrag{S}[b][b]{$S$}
\psfrag{wk}[t][t]{$\w{k}{k^*}$}
\psfrag{wh}[t][t]{$\w{h'}{h^*}$}
\psfrag{w4}[b][b]{$\w{k}{k^*}$}
\psfrag{w5}[t][t]{$\w{h'}{h^*}$}
\psfrag{w3}[r][r]{\shortstack[c]{$\w{h'}{k}$ \\ $ = $\\ $ \w{k}{h'}$}}
\psfrag{w6}[c][c]{\shortstack[c]{$\w{h^*}{k^*}$\\ $ = $ \\ $\w{k^*}{h^*}$}}
}
\caption{Illustration for the proof of \CiteTheorem{lem:same-speed-hard}
\label{fig:same-speed-hard}}
\end{figure}

Choose $h'\in\IOb_o$ such that $k \mappedto{R}{m} h'$.  Such $h'$
exists by \CiteLemma{lem:relocation} and \CiteLemma{lem:iobo2}.
By \CiteLemma{lem:worldline-relocation}, we have $\wl_m(h') = R[\wl_m(k)]$,
and since $R[\wl_m(k)] = \wl_m(h)$, we must have
\begin{equation} 
\label{eqn:ssh1}
   \wl_m(h) = \wl_m(h'),
\end{equation}
\ie $h$ and $h'$ share the same worldline. 
It follows, by \CiteLemma{lem:colocate} and $h,h'\in\IOb_o$, that 
\begin{equation} \label{eqn:ssh2}
   \w{h}{h'}\text{ is a linear trivial transformation}.
\end{equation}

Our goal is to prove that $\w{h}{k} \in \kiso$ for some $\kappa$. Since $\w{h}{k} =  \w{h}{h'} \circ \w{h'}{k} $ and (as we have just seen) $\w{h}{h'}$ is trivial, it is enough to prove that $\w{h'}{k} \in \kiso$ for some $\kappa$.

By $k \mappedto{R}{m} h'$, we have $\w{m}{h'} = R \circ
\w{m}{k}$. Thus,
\begin{equation} \label{eqn:xxx}
  \w{k}{h'} = \w{k}{m} \circ \w{m}{h'}  = \w{k}{m} \circ R \circ \w{m}{k}
\end{equation}
and
\[
  \w{h'}{k} = (\w{k}{m} \circ R \circ \w{m}{k})^{-1} = \w{k}{m} \circ R^{-1} \circ \w{m}{k} 
\]
whence (as $R^{-1}=R$)
\begin{equation} \label{eqn:ssh3}
   \w{h'}{k} = \w{k}{h'} \text{, and thus } \wl_k(h') = \wl_{h'}(k).
\end{equation}

Let $P$ be the plane containing $(0,\vv s^*)$ and \taxis.  Since
$(0,\vv s^*)$ and \taxis are pointwise fixed by $R$, it follows that
the whole of $P$ is likewise fixed pointwise by $R$; see
Figure~\ref{fig:same-speed-hard}.

We claim that $\w{h'}{k}$ ($= \w{k}{h'}$) leaves the plane $\w{k}{m}[P]$ pointwise fixed. 
To see this, choose any $\vvp \in \w{k}{m}[P]$. By \eqref{eqn:xxx}, 
$\w{k}{h'}(\vvp) = (\w{k}{m} \circ R \circ \w{m}{k})(\vvp)$. But $\w{m}{k}(\vvp) \in \w{m}{k}[\w{k}{m}[P]] = P$, so $R( \w{m}{k}(\vvp) ) = \w{m}{k}(\vvp)$. It follows that
\[
  \w{k}{h'}(\vvp) = (\w{k}{m} \circ \w{m}{k}  )(\vvp) = \vvp
\]
as stated.

We know that $\w{h'}{k}$ is a bijective collineation by
\CiteTheorem{lem:line-to-line} and that it leaves $\origin$ fixed by
\CiteLemma{lem:iobo} because $h', k \in \IOb_o$.  So, by
\CiteLemma{lem:affine}, $\w{h'}{k}$ is a linear transformation
composed with a map induced by a field automorphism.  But since
$\w{h'}{k}$ leaves the plane $\w{k}{m}[P]$ pointwise fixed, the
automorphism component must be the identity, and we deduce that
$\w{h'}{k}$ is a linear transformation.

By $\wl_m(k)\neq\wl_m(h)=\wl_m(h')$ and
\CiteLemma{lem:equal-worldlines}, we have that
$\wl_{h'}(k)\neq\wl_{h'}(h')=\taxis$. By \CiteLemma{lem:iobo2}, we
have that $\origin\in\wl_k(h')$.  Let $P'$ be the plane determined by
the time-axis and $\wl_k(h')$ ($= \wl_{h'}(k)$) and let $S$ be a
spatial rotation that takes the $tx$-plane to $P'$, see
Figure~\ref{fig:same-speed-hard}.  Such a rotation exists by
\CiteLemma{lem:letezikrotacio}.  Choose $k^*$, $h^*$ such that
$\w{k}{k^*} = \w{h'}{h^*} = S$ (these exist by \AxRelocate).  Then
\begin{equation}
\label{eqn:wk^*j*}
  \w{h^*}{k^*} = \w{h^*}{h'} \circ \w{h'}{k}  \circ  \w{k}{k^*}   
= S^{-1} \circ \w{h'}{k} \circ S
\end{equation}
and hence
\begin{equation*} 
  \w{k^*}{h^*} = (S^{-1} \circ \w{h'}{k} \circ S)^{-1} = S^{-1} \circ \w{h'}{k} \circ S 
\end{equation*}
because $\w{h'}{k} = \w{k}{h'}$. Therefore, $\w{h^*}{k^*} =
\w{k^*}{h^*}$ and $\w{h^*}{k^*}$ is a linear transformation
since $S^{-1}$, $\w{h'}{k}$, and $S$ are linear.

To prove that there is $\kappa$ such that $\w{h'}{k} \in \kiso$, it
is therefore enough to show that there is $\kappa$ such that
$\w{h^*}{k^*} \in \kiso$, because spatial rotations $S, S^{-1} \in
\kiso$ for every $\kappa$.

The worldview transformation $\w{h'}{k}$ leaves plane $P'$ fixed
because it takes $\taxis$ and $\wl_{k}(h')$ to $\wl_{h'}(k)$ and
$\taxis$, respectively, and $P'$ is the unique plane that contains
$\taxis$ and $\wl_{k}(h')=\wl_{h'}(k)$.  By this and
(\ref{eqn:wk^*j*}), we have that $\w{h^*}{k^*}$ maps the $tx$-plane to
itself.  Hence, by \CiteTheorem{lem:txplane} $\w{h^*}{k^*}$ also takes
the $yz$-plane to itself and there is $\lambda > 0$ such that for
every $\vvp \in \plane(\yaxis,\zaxis)$, $\norm{\w{h^*}{k^*}(\vvp)} =
\lambda \norm{\vvp}$.  But now, for every $\vvp \in
\plane(\yaxis,\zaxis)$, we have
\[
  \norm{\vvp} = \norm{(\w{k^*}{h^*} \circ \w{h^*}{k^*})(\vvp)}
      = \norm{ (\w{h^*}{k^*} \circ \w{h^*}{k^*} )(\vvp) }
      = \lambda^2 \norm{\vvp}.
\]
Thus, $\lambda^2 = 1$, whence $\lambda = 1$ (as $\lambda > 0$).

This means that $\w{h^*}{k^*}$ preserves Euclidean length in 
$\plane(\yaxis,\zaxis)$.

We have proven so far that $\w{h^*}{k^*} = \w{k^*}{h^*}$, 
that $\w{h^*}{k^*}$ is a linear transformation taking 
$\plane(\taxis,\xaxis)$ to $\plane(\taxis,\xaxis)$ and 
$\plane(\yaxis,\zaxis)$ to $\plane(\yaxis,\zaxis)$, and that it 
preserves Euclidean length in $\plane(\yaxis,\zaxis)$. 
It remains to show that $\w{h^*}{k^*} \in \kiso$.

We have already seen that $\origin\in \wl_{h'}(k)\neq \taxis$.  Thus,
$\speed_{h'}(k)\neq 0$.  By \CiteLemma{lem:lintriv-implies-same-speed}
and the fact that $\w{h'}{h^*}$ and $\w{k}{k^*}$ are spatial rotations
(hence linear trivial transformations), we have that
$\speed_{h^*}(k^*)=\speed_{h'}(k^*)=\speed_{h'}(k)$.  Thus,
$\speed_{h^*}(k^*)\neq 0$.

We will choose $\kappa$ so that
\[
  \knorm{ \w{h^*}{k^*}(\tunit) }^2 = 1.
\]
We can do this because we know that $ \w{h^*}{k^*}(\tunit) \in \plane(\taxis,\xaxis)$, 
so we can write $\w{h^*}{k^*}(\tunit) = (t_e,x_e,0,0)$ for some $t_e$ and 
$x_e$, and we know that $x_e \neq 0$ 
because $\speed_{h^*}(k^*) \neq 0$ and
$\origin,\w{h^*}{k^*}(\taxis)\in\wl_{h^*}(k^*)$. 
So we can take $\kappa := (t_e^2-1)/x_e^2$, because then
\[
  \knorm{ \w{h^*}{k^*}(\tunit) }^2 = t_e^2 - \kappa x_e^2 = t_e^2 - \frac{(t_e^2-1)}{x_e^2} x_e^2 = 1 ,
\]
as required. 

It follows that $\knorm{\w{h^*}{k^*}(\vvp)}^2 = \knorm{\vvp}^2$ 
for every $\vvp \in \plane(\taxis,\xaxis)$, \ie $\w{h^*}{k^*}$ preserves
$\kappa$-length in the $tx$-plane.
To see why, let $\vvp\in\plane(\taxis,\xaxis)$.
Notice that $\vvp$ can be written as some linear combination 
$\vvp = \lambda \tunit + \mu \w{h^*}{k^*}(\tunit)$. 
From this and the fact that 
$\w{h^*}{k^*} = \w{k^*}{h^*}$ is a linear transformation, we have
\[
  \w{h^*}{k^*}(\vvp) = \w{h^*}{k^*}(\lambda \tunit + \mu \w{h^*}{k^*}(\tunit))
                  = \lambda \w{h^*}{k^*}(\tunit) + \mu \tunit.
\]
Writing $\vvp_\dagger = \w{h^*}{k^*}(\vvp)$ and recalling that $\w{h^*}{k^*}(\tunit) = (t_e,x_e,0,0)$, we have
\[
     \vvp  = \lambda (1,0,0,0) + \mu (t_e,x_e,0,0)  \qquad \text{ and } \qquad
     \vvp_\dagger = \lambda (t_e,x_e,0,0) + \mu (1,0,0,0)
\]
and now direct calculation (using $\kappa = (t_e^2-1)/x_e^2)$ shows that
\[
  \knorm{ \vvp }^2
     = (\lambda + \mu t_e)^2 - \frac{(t_e^2-1)}{x_e^2} \mu^2 x_e^2 
    = \lambda^2 +2t_e\lambda\mu + \mu^2
\]
and likewise
\[
 \knorm{ \vvp_\dagger }^2
     = (\lambda t_e + \mu)^2                       - \frac{(t_e^2-1)}{x_e^2} \lambda^2 x_e^2
      = \lambda^2 + 2t_e \lambda \mu + \mu^2 ,
\]
whence $\knorm{\vvp^2} = \knorm{\vvp_\dagger}^2=\knorm{\w{h^*}{k^*}(\vvp)}$ 
as claimed.

Next, we are going to prove that $\w{k^*}{h^*}$ preserves the
$\kappa$-length. To prove this, let $\vvp = (t,x,y,z)$ be an arbitrary
point in $\Q^4$ and let $(\hat t, \hat x, \hat y, \hat z) =
\w{h^*}{k^*}(\vvp)$.  By linearity, we have
\[
  (\hat t, \hat x, \hat y, \hat z) = \w{h^*}{k^*}(t,x,y,z) 
                = \w{h^*}{k^*}(t,x,0,0)  + \w{h^*}{k^*}(0,0,y,z) ,
\]
whence $(\hat t, \hat x, 0,0) = \w{h^*}{k^*}(t,x,0,0)$ and $(0,0,
\hat y, \hat z) = \w{h^*}{k^*}(0,0,y,z)$, because $\w{h^*}{k^*}$
preserves both the $tx$- and $yz$-planes. We also have that
\[
   \hat t^2 - \kappa \hat x^2 = t^2 - \kappa x^2  
\qquad \text{ and } \qquad 
   \hat y^2 + \hat z^2      = y^2 + z^2 
\]
because $\w{h^*}{k^*}$ preserves the $\kappa$-length in the $tx$-plane
and preserves the Euclidean length in the $yz$-plane.  It follows
immediately that
\[
   (\hat t^2 - \kappa \hat x^2) - \kappa (\hat y^2 + \hat z^2) = (t^2 - \kappa x^2) - \kappa (y^2 + z^2) ,
\]
or in other words, $\knorm{\vvp}^2 = \knorm{\w{h^*}{k}(\vvp)}$, and so 
$\w{h^*}{k^*}$ preserves the $\kappa$-length.

Therefore, if $\kappa\neq 0$, then $\w{h^*}{k^*}$ is a linear
$\kappa$-isometry, so $\w{h^*}{k^*} \in \kiso$, and we are done.

Suppose, finally, that $\kappa = 0$. We will prove that $\w{h^*}{k^*}$
is a linear $0$-isometry.  Recall that $\w{h^*}{k^*}(\tunit) =
(t_e,x_e,0,0)$ and $\kappa=(t_e^2-1)/x_e^2$. Since $\kappa = 0$, we
have $t_e = \pm 1$, and hence $\w{h^*}{k^*}(\tunit) = (\pm 1, x_e, 0,
0)$. Thus, $(0,x_e,0,0) = \w{h^*}{k^*}(\tunit) \mp \tunit $. This and
the fact that $\w{h^*}{k^*}$ is both linear and self-inverse now
yields
\begin{eqnarray}
\label{x-coord}  \w{h^*}{k^*}(0,x_e,0,0) &=&  \w{h^*}{k^*}( \w{h^*}{k^*}(\tunit) ~\mp~ \tunit ) \\
\nonumber	                        &=&  \w{h^*}{k^*}(\w{h^*}{k^*}(\tunit)) ~\mp~ \w{h^*}{k^*}(\tunit) \\
\nonumber	                        &=&  \tunit ~\mp~ \w{h^*}{k^*}(\tunit)  \\
         \nonumber                    &=&  ~\mp~(0,x_e,0,0).
\end{eqnarray}

Writing $f := \w{h^*}{k^*}$ we have already shown that $f$ preserves
$\kappa$-length, so for $\kappa =0$ we have $f(\vvp)_t^2 =
\knorm[0]{f(\vvp)}^2 = \knorm[0]{\vvp}^2 = \vvp_t ^2$ for every
$\vvp\in\Q^4$.  By \eqref{eqn:kiso-defn}, it only remains to show that
$\norm{ f(\vvp)_s }^2 = \norm{ \vvp_s }^2$ when $\vvp_t = 0$.  However, we
know that $f$ maps the $yz$-plane to itself and preserves Euclidean
length in that plane, and that it simply reverses or preserves
$x$-coordinates by \eqref{x-coord}.  Hence, $f$ also preserves 
Euclidean length in the $xyz$-hyperplane. Thus, $\w{h^*}{k^*}$ is a
linear $0$-isometry.

This completes the proof of (a).

\medskip
\textit{Proof of (b).}  By \eqref{eqn:ssh2} (which says that
$\w{h}{h'}$ is a linear trivial transformation) and by
\CiteLemma{lem:lintriv-implies-same-speed}, for every $j\in\IOb$, we have that
\begin{eqnarray}
\label{eqn:ssh4}
\speed_j(h) & = & \speed_j(h')\text{ and }\\
\label{eqn:ssh5}
\speed_h(j) & = & \speed_{h'}(j),
\end{eqnarray}  
and so
\begin{equation*}
   \speed_k(h) \stackrel{\text{\tiny \eqref{eqn:ssh4}}}{\longeq}
   \speed_k(h') \stackrel{\text{\tiny \eqref{eqn:ssh3}}}{\longeq}
   \speed_{h'}(k) \stackrel{\text{\tiny \eqref{eqn:ssh5}}}{\longeq}
   \speed_h(k)
\end{equation*}
as required.

\medskip
\textit{Proof of (c).}  First we show that
\begin{equation} \label{eqn:ssh6}
   \wl_k(m) = \wl_{h'}(m).
\end{equation}
To do so, recall that $\w{m}{h'} = R \circ \w{m}{k}$ (by $k \mappedto{R}{m} h'$). It follows that $\w{h'}{m} = \w{k}{m} \circ R^{-1}$, and hence (because the time-axis $\taxis$ is fixed under spatial rotations),
\[
   \wl_{h'}(m) = \w{h'}{m}[\taxis] =   (\w{k}{m} \circ R^{-1})[\taxis]  =  \w{k}{m}[\taxis] = \wl_k(m)
\]
as claimed.  Consequently,
\begin{equation*}
  \speed_k(m)  \stackrel{\text{\tiny \eqref{eqn:ssh6}}}{\longeq}  \speed_{h'}(m) 
              \stackrel{\text{\tiny \eqref{eqn:ssh5}}}{\longeq}  \speed_h(m)   .
\end{equation*}

This completes the proof.
\end{proof}



\subsection{Fundamental Lemma}

\begin{Theorem}{lem:fundamental}
Assume $\KIN + \AxIso + \lnot\EInf$. Then for every $k,m\in\IOb_o$ with $\speed_k(m)>0$, there is a positive $\varepsilon \in \Q$ such that for every non-negative $v \leq \speed_k(m)+\varepsilon$, there is some $h\in\IOb_o$ with $\speed_k(h) = v$ and $\speed_m(k) = \speed_m(h)$.
\end{Theorem}

\begin{figure}[!htb] 
\psfragfig[keepaspectratio, width=\textwidth]{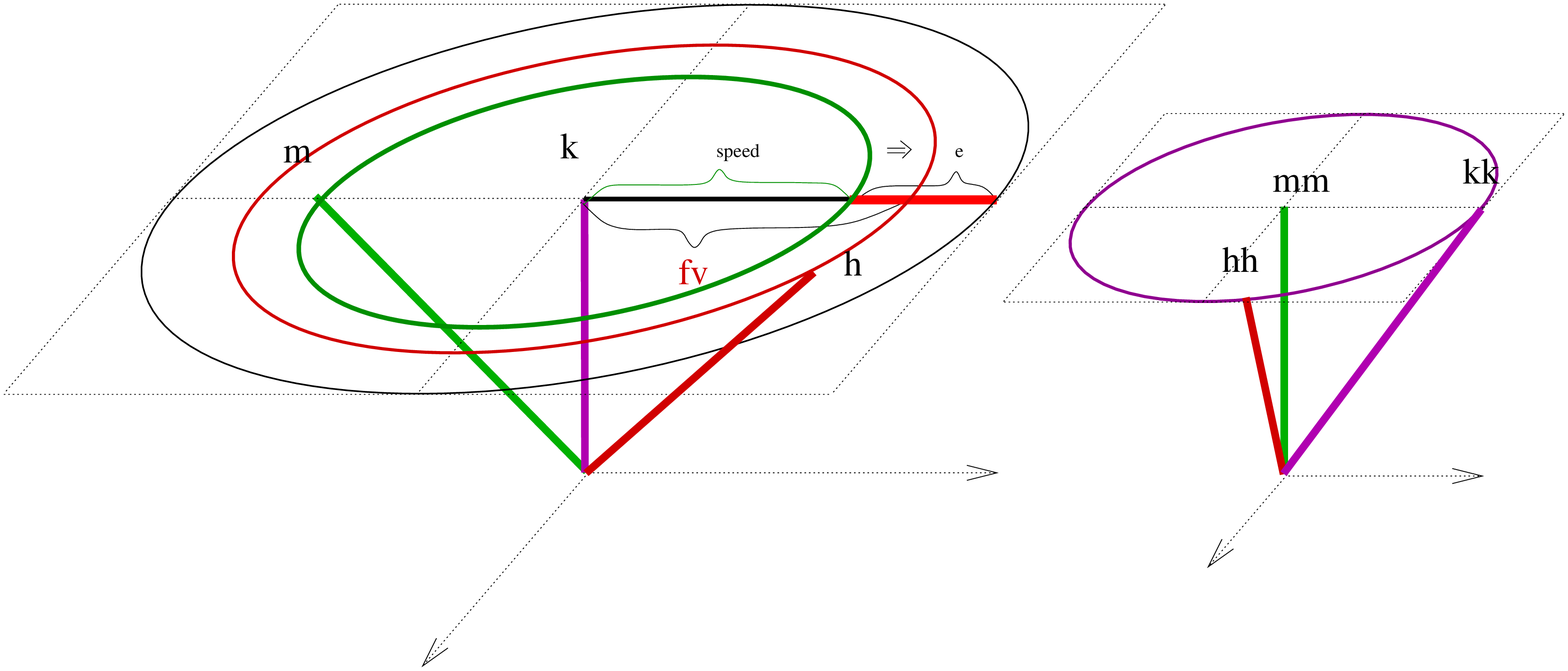}
{
\psfrag{speed}[b][b]{\scriptsize $\speed_k(m)>0$}
\psfrag{mm}[b][b]{$m$}
\psfrag{kk}[b][b]{$k$}
\psfrag{hh}[b][b]{$h$}
\psfrag{k}[b][b]{$\forall k$}
\psfrag{m}[b][b]{$\forall m$}
\psfrag{h}[t][t]{$\exists h$}
\psfrag{fv}[t][t]{$\forall v$}
\psfrag{e}[b][b]{$\exists\varepsilon$}
}
\caption{Figure illustrating \CiteTheorem{lem:fundamental}.}
\label{fig:fundamental}
\end{figure}

\begin{figure}[!htb] 
\small
\psfragfig[keepaspectratio, width=\textwidth]{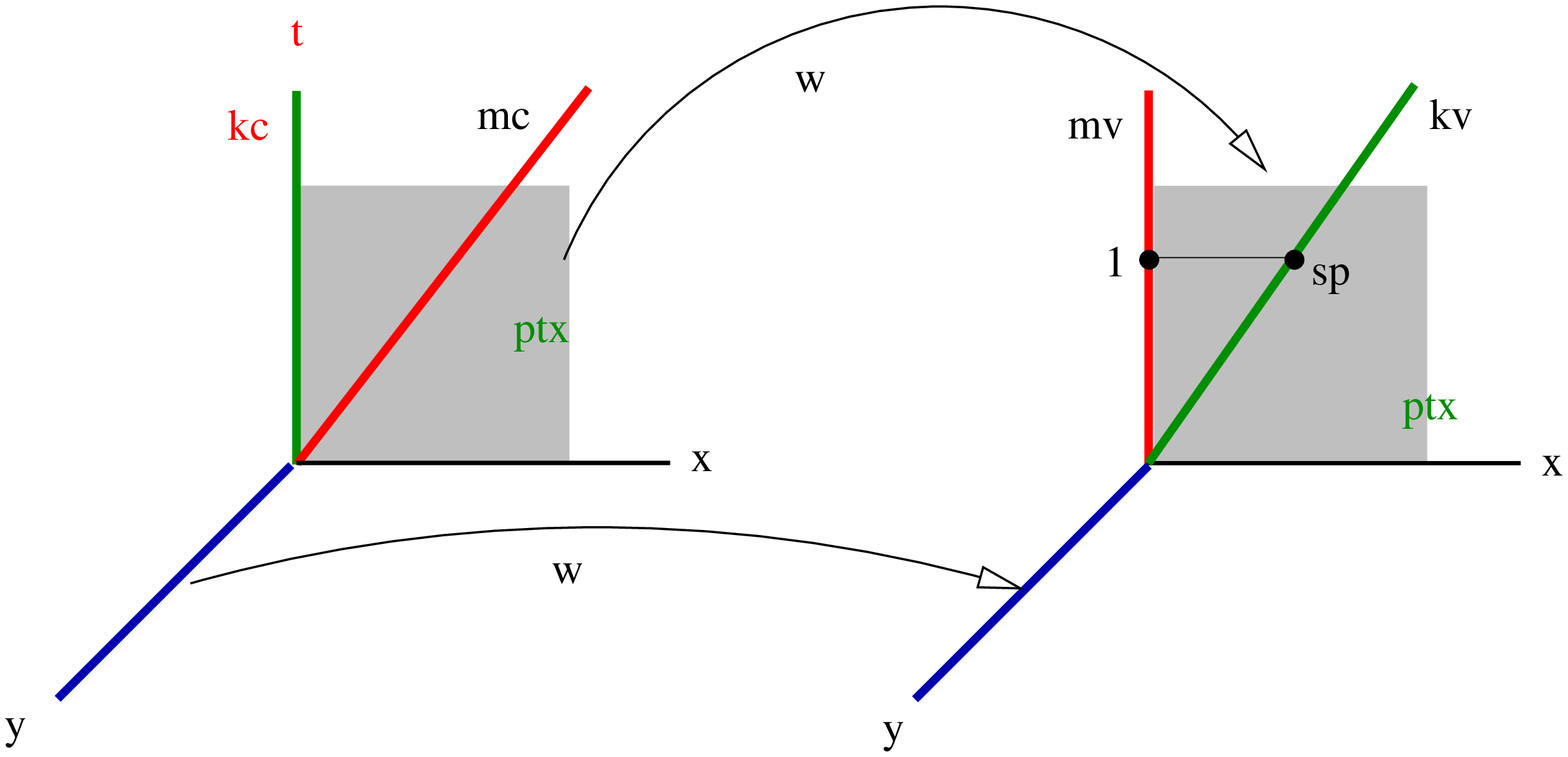}
{
\psfrag{kc}[r][r]{$k^*$}
\psfrag{mc}[rb][rb]{$m^*$}
\psfrag{ptx}[l][l]{$\plane(\taxis,\xaxis)$}
\psfrag{y}[rt][rt]{$\yaxis$}
\psfrag{x}[l][l]{$\xaxis$}
\psfrag{t}[b][b]{$\taxis$}
\psfrag{1}[r][r]{$1$}
\psfrag{sp}[lt][lt]{ $(1,\speed_m(k),0,0)$}
\psfrag{w}[t][t]{$\w{m^*}{k^*}$}
\psfrag{mv}[r][r]{$m^*$}
\psfrag{kv}[lt][lt]{$k^*$}
}
\caption{Illustration for \CiteLemma{lem:configuration}.}
\label{fig:lem-configuration} 
\end{figure}

We first show that observers can be found which satisfy certain standard
configurations; see Figure~\ref{fig:lem-configuration}.

\subsubsection{Supporting lemmas}

The supporting lemmas can be informally described as:

\begin{description}
\item[\CiteLemma{lem:configuration}]
~ \\ If two observers $k$ and $m$ are moving at any speed $u >0$ relative to
one another, there are `rotated versions' $k^*$ and $m^*$ of those observers
which agree with each other as to where the $tx$-plane and the $y$-axis are
located.  Moreover, if $u$ is finite, then $m^*$ considers $k^*$
to be moving in the  positive direction of the $x$-axis.

\item[\CiteLemma{lem:quadratic}]
~ \\ This is a purely technical lemma stating that the Intermediate Value Theorem holds for functions of the form $f(x) = \sqrt{F(x)/G(x)}$ where $F$ and $G$ are quadratic polynomials over \Q.
\end{description}

\subsubsection{Proofs of the supporting lemmas}

\begin{Lemma}{lem:configuration}
 Assume $\KIN + \AxIso$.  Given any $k, m
\in \IOb_o$ satisfying $\speed_m(k)\neq 0$, there exist $k^*, m^* \in \IOb_o$
such that 
 \begin{itemize}
   \item[(a)] $\w{k^*}{k}$ and $\w{m^*}{m}$ are spatial rotations,
hence\footnote{\,by  \CiteLemma{lem:lintriv-implies-same-speed}}\\
$\speed_{m^*}(k^*)=\speed_{m}(k)$,\\
$\speed_{k^*}(h)=\speed_{k}(h)$ and
$\speed_{m^*}(h)=\speed_{m}(h)$ for every $h\in\IOb$;
   \item[(b)] $\w{k^*}{m^*}[\plane(\taxis,\xaxis)] = \plane(\taxis,\xaxis)$; 
    \item[(c)] $\w{k^*}{m^*}[\yaxis] = \yaxis$;
    \item[(d)] $k^*$ moves in the positive direction of the
$x$-axis according to $m^*$, \ie \\
$\big(1,\speed_{m^*}(k^*),0,0\big)\in  \wl_{m^*}(k^*)$ and
$\origin\in\wl_{m^*}(k^*)$ if
 $\speed_{m^*}(k^*)\neq\infty$.
\end{itemize}
\end{Lemma}

\begin{proof}
Let us recall that, by \CiteTheorem{lem:line-to-line}, worldview
transformations are bijections taking lines to lines and planes to planes. 

We know that $\wl_k(m)$ and $\taxis$ are distinct lines, because
$\speed_k(m)\neq 0$.  Since, by \CiteLemma{lem:iobo2}, they meet at
the origin, we know that $\plane(\taxis,\wl_k(m))$ is a well-defined
plane, and because this plane contains the time-axis, by
\CiteLemma{lem:letezikrotacio} there must exist a spatial rotation
about \taxis which takes $\plane(\taxis,\xaxis)$ to
$\plane(\taxis,\wl_k(m))$. By \AxRelocate and (b) of
\CiteLemma{lem:iobo2}, there is some $k^* \in \IOb_o$ for which this
rotation equals $\w{k}{k^*}$, so that
\begin{equation}
\label{eq1:configuration}
   \w{k}{k^*}[ \plane(\taxis, \xaxis)] = \plane(\taxis,\wl_k(m)),
\end{equation}
see the left-top of Figure~\ref{fig:configuration}.

According to \CiteLemma{lem:letezikrotacio} there is also a spatial rotation $R$ that takes
$\plane(\taxis,\wl_m(k))$ to $\plane(\taxis,\xaxis)$; moreover, if $\speed_m(k)\neq \infty$, 
we can choose $\vvp\in\wl_m(k)$ such that $\vvp_t=1$ and require of $R$ that $R(\vvp) 
= (1,\norm{\vvp_s},0,0)$. In this case, because $\origin,\vvp\in\wl_m(k)$ and 
$\vvp_t=1$, we have $\speed_m(k)=\slope(\wl_m(k))=\norm{\vvp_s}$, and so
\begin{equation}\label{eq:r-of-p}
R(\vvp)=(1,\speed_m(k),0,0).
\end{equation}

\medskip
Now let $m'\in\IOb_o$ be such that $\w{m'}{m}=R$ (such an $m'$ exists by
\AxRelocate and (b) of \CiteLemma{lem:iobo2}). We will show that $\w{m'}{k^*}$ fixes both the $tx$-plane and the $yz$-plane. By definition, 
\begin{equation}
\label{eq2:configuration}
   \w{m'}{m}[ \plane(\taxis,\wl_m(k)) ] = \plane(\taxis,\xaxis) 
\end{equation}
see the left-bottom of Figure~\ref{fig:configuration}. If
$\speed_m(k)\neq\infty$, by $\vvp\in\wl_m(k)$, we have that
$\w{m'}{m}(\vvp)\in\wl_{m'}(k)$. Combining this with \eqref{eq:r-of-p}
tells us that
\begin{equation}
\label{equj:configuration}
(1,\speed_m(k),0,0)\in\wl_{m'}(k) \text{ if $\speed_m(k)\neq\infty$}.
\end{equation}
\begin{figure}[h!btp] 
\small
\psfragfig[keepaspectratio,width=\linewidth]{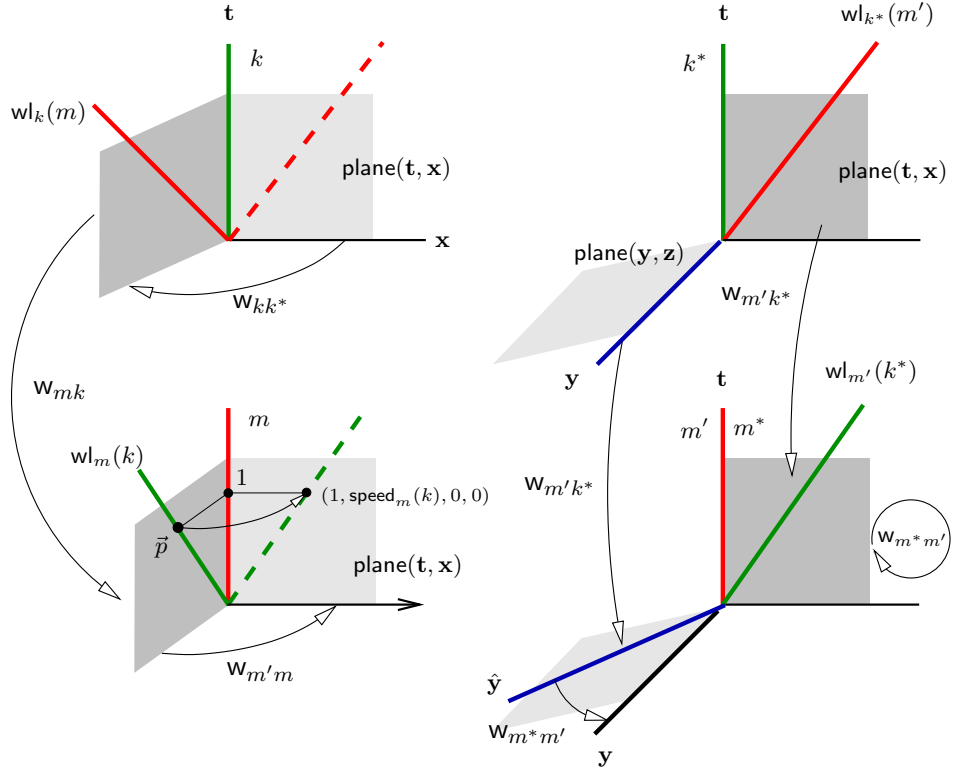}
{
\psfrag{mc}[l][l]{$m^*$}
\psfrag{1}[lb][lb]{$1$}
\psfrag{p}[rt][rt]{$\vvp$}
\psfrag{sp}[lt][lt]{\scriptsize $(1,\speed_m(k),0,0)$}
\psfrag{wm1}[b][b]{$\wl_{k^*}(m')$}
\psfrag{wk1}[b][b]{$\wl_{m'}(k^*)$}
\psfrag{k}[l][l]{$k$}
\psfrag{m}[l][l]{$m$}
\psfrag{kc}[r][r]{$k^*$}
\psfrag{mv}[r][r]{$m'$}
\psfrag{w4}[r][r]{\large $\w{m'}{k^*}$}
\psfrag{wm}[rt][rt]{$\wl_k(m)$}
\psfrag{t}[b][b]{$\taxis$}
\psfrag{x}[l][l]{$\xaxis$}
\psfrag{w2}[t][t]{\large $\w{k}{k^*}$}
\psfrag{w1}[l][l]{\large $\w{m}{k}$}
\psfrag{wk}[rt][rt]{$\wl_m(k)$}
\psfrag{w3}[t][t]{\large $\w{m'}{m}$}
\psfrag{text}[l][l]{$\w{m'}{k^*}=\w{m'}{m}\circ\w{m}{k}\circ\w{k}{k^*}$}
\psfrag{ptx}[l][l]{$\plane(\taxis,\xaxis)$}
\psfrag{pyz}[lb][lb]{$\plane(\yaxis,\zaxis)$}
\psfrag{yk}[rb][rb]{$\hat{\yaxis}$}
\psfrag{yy}[lt][lt]{$\yaxis$}
\psfrag{y}[rt][rt]{$\yaxis$}
\psfrag{w6}[c][c]{$\w{m^*}{m'}$}
\psfrag{w5}[rt][rt]{\large $\w{m^*}{m'}$}
}
\caption{\label{fig:configuration} Illustration for the proof of 
\CiteLemma{lem:configuration}}
\end{figure}

Notice next that the world-view transformation $\w{m}{k}$ takes $\taxis$ to
$\wl_m(k)$ and $\wl_k(m)$ to $\taxis$, respectively. Therefore,
\begin{equation}
\label{eq3:configuration}
\w{m}{k}[\plane(\taxis,\wl_k(m))] = \plane(\taxis,\wl_m(k)),
\end{equation}
see the left-hand side of Figure~\ref{fig:configuration}.
By \eqref{eq1:configuration}, 
\eqref{eq3:configuration}, \eqref{eq2:configuration}, 
and the fact that $\w{m'}{k^*}=\w{m'}{m}\circ\w{m}{k}\circ\w{k}{k^*}$, we have
that
\[
\w{m'}{k^*}[\plane(\taxis,\xaxis)] = \plane(\taxis,\xaxis).
\]
By \CiteTheorem{lem:txplane}, it follows that
$\w{m'}{k^*}[\plane(\yaxis,\zaxis)]$ $=$ $\plane(\yaxis,\zaxis)$. Thus, 
$\w{m'}{k^*}$ fixes both the $tx$-plane and the $yz$-plane, as claimed.

\medskip
Now write $\hat\yaxis := \w{m'}{k^*}[\yaxis]$, and note that
$\hat\yaxis \subseteq \plane(\yaxis,\zaxis)$ because $\w{m'}{k^*}$
preserves this plane.  We can find a spatial rotation which fixes the
$tx$-plane pointwise and takes $\hat\yaxis$ to $\yaxis$ because of the
following.  Let $\vvq\in\hat\yaxis$ and $\vvq'\in\yaxis$ be such that
$\norm{\vvq}=\norm{\vvq'}\neq 0$.  Then $\vvq\cdot\tunit=\xunit\cdot
\vvq=\vvq'\cdot\tunit =\xunit\cdot\vvq'=0$ because
$\vvq,\vvq'\in\plane(\yaxis,\zaxis)$.  Therefore, by
\CiteLemma{lem:vanrotacio} there is a spatial rotation that takes
$\vvq$ to $\vvq'$ and $\xunit$ to itself.  By \AxRelocate and
\CiteLemma{lem:iobo2}, there is some $m^* \in \IOb_o$ such that
$\w{m^*}{m'}$ is this spatial rotation, see the right-bottom of
Figure~\ref{fig:configuration}.

Notice that $\w{m^*}{m'}$ maps $\hat\yaxis$ to $\yaxis$
(because it fixes \origin and maps $\vvq \in \hat\yaxis$ to 
$\vvq' \in \yaxis$) and fixes $\plane(\taxis,\xaxis)$ pointwise
because it fixes $\tunit$ and $\xunit$.

In summary, we have so far shown that $\w{m^*}{m'}$ and $\w{m'}{m}$ are spatial rotations; and that $\w{m^*}{m'}$ and $\w{m'}{k^*}$ both fix the $tx$-plane and the $yz$-plane.

\medskip
\textit{Proof of (a).} The transformation $\w{k^*}{k}$ is a spatial
rotation by definition.  Since $\w{m^*}{m} =
\w{m^*}{m'} \circ \w{m'}{m}$ is a composition of two spatial
rotations, it is also a spatial rotation. 
By \CiteLemma{lem:lintriv-implies-same-speed}, 
$\speed_{m^*}(k^*)=\speed_{m}(k^*)=\speed_{m}(k)$, 
$\speed_{k^*}(h)=\speed_{k}(h)$ and
$\speed_{m^*}(h)=\speed_{m}(h)$ for every $h\in\IOb$.

\medskip
\textit{Proof of (b).}  Since $\w{m^*}{k^*} = \w{m^*}{m'} \circ
\w{m'}{k^*}$ and both $\w{m^*}{m'}$ and $\w{m'}{k^*}$ fix the
$tx$-plane, $\w{m^*}{k^*}$ and its inverse $\w{k^*}{m^*}$ also fix the
$tx$-plane.

\medskip
\textit{Proof of (c).} We have $\yaxis = \w{m^*}{m'}[\hat\yaxis] =
\w{m^*}{m'}[\w{m'}{k^*}[\yaxis]] = \w{m^*}{k^*}[\yaxis]$, so
$\w{m^*}{k^*}$ and its inverse $\w{k^*}{m^*}$ fix the $y$-axis.

\medskip
\textit{Proof of (d).}  It is already clear that
$\origin\in\wl_{m^*}(k^*)$, by \CiteLemma{lem:iobo2}.  We need to show
that $(1,\speed_{m^*}(k^*),0,0)\in\wl_{m^*}{(k^*)}$ as well.
  
  By \eqref{eq2:configuration} and
$\wl_{m'}(k)=\w{m'}{m}[\wl_m(k)]$, we have that
$\wl_{m'}(k)\subseteq\plane(\taxis,\xaxis)$. Because
$\w{m^*}{m'}$ fixes $\plane(\taxis,\xaxis)$ pointwise
and takes $\wl_{m'}(k)$ to
$\wl_{m^*}(k)$, we therefore have 
     $\wl_{m^*}(k)=\wl_{m'}(k)$. 
By
\CiteLemma{lem:lintriv-implies-same-speed}, 
     $\wl_{m^*}(k^*)=\wl_{m^*}(k)$ 
because
$\w{k^*}{k} \in \srot$ is a linear trivial transformation.  Therefore,
$\wl_{m^*}(k^*) = \wl_{m^*}(k) = \wl_{m'}(k)$.  Now assume that
$\speed_m(k)\neq\infty$. Then  \eqref{equj:configuration} tells us that
$(1,\speed_m(k),0,0) \in \wl_{m'}(k) = \wl_{m^*}{(k^*)}$. By (a),
$\speed_m(k)=\speed_{m^*}(k^*)$. Therefore, 
$(1,\speed_{m^*}(k^*),0,0) \in \wl_{m^*}{(k^*)}$, 
as required.

\medskip
This completes the proof.
\end{proof}

\begin{rem}
Using the fact that any real-closed field is elementarily
equivalent to the field of real numbers (\ie they satisfy the same
first-order logic formulas), it is easy to show that an ordered field
is real-closed if{}f it satisfies the Intermediate Value Theorem for
every polynomial function. However, for arbitrary ordered fields (\eg the field $\mathbb{Q}$ of rationals) the
Intermediate Value Theorem can fail even for quadratic functions: if
$F(x)=x^2-2$, then despite the fact that $F(0) < 0 < F(2)$ there is 
no $c \in \mathbb{Q}$ for which $F(c)=0$. 

\end{rem}

In the proof of \CiteTheorem{lem:fundamental} below, 
we will need the following lemma stating
that the Intermediate Value Theorem holds for
a specific class of algebraic functions defined over Euclidean fields.

\begin{Lemma}{lem:quadratic}
Assume \ax{AxEField}, and let $F$ and $G$ be quadratic
functions on \Q.\footnote{$F:\Q\rightarrow\Q$ is called a \semph{quadratic
    function} if there are $p,q,r\in\Q$ such that $F(x)=px^2+qx+r$ for
  every $x\in\Q$.} Let $a < b$ be values in \Q and suppose $F(x)
\geq 0$ and $G(x) > 0$ for all $x \in [a,b]$. Let $g :
     [a,b] \to \Q$ be the function $g(x) := \sqrt{F(x)/G(x)}$. Then given any $y$ between $g(a)$ and $g(b)$,
     there exists $c \in [a, b]$ such that $g(c) = y$.
\end{Lemma}
\begin{proof}
If $g(a) = y$ or $g(b) = y$ the proof is trivial, so suppose that $y$ lies strictly between $g(a)$ and $g(b)$ and consider the quadratic function $p(x) = F(x) - y^2G(x) \equiv [g(x)^2 - y^2] G(x)$. 
Because $y^2$ lies strictly between $g(a)^2$ and $g(b)^2$, the values $p(a) = [g(a)^2 - y^2]G(a)$ and $p(b) = [g(b)^2 - y^2]G(b)$ are both non-zero and have opposite signs. 

We will show that there exists some $c \in (a,b)$ for which $p(c) = 0$. 
Because $p$ is quadratic, it can be written in the form 
$p(x) = \alpha x^2 + \beta x + \gamma$. 
We know that $p$ is not constant because $p(a) \neq p(b)$, so $\alpha$ 
and $\beta$ cannot both be zero. If $\alpha = 0$, then 
$\beta \neq 0$ and $p(x) = \beta x + \gamma$ is a linear function for 
which a suitable $c$ can trivially be found. Suppose, then, that $\alpha \neq 0$. 
Then we can rewrite $p$ as $p(x) = \alpha\left[ ( x + \beta/2\alpha)^2 - ( \beta^2 - 4\alpha\gamma)/4\alpha^2 )\right]$,
and now the fact that $p(x)$ can be both positive and negative implies immediately that the discriminant $\Delta := (\beta^2 - 4\alpha\gamma)$ is positive, whence $p$ can be factorised over $\Q$ with the usual quadratic roots $x_1 := (-\beta + \sqrt\Delta)/2\alpha$ and $x_2 := (-\beta - \sqrt\Delta)/2\alpha$. Writing $p(x) = \alpha(x - x_1)(x - x_2)$ it is now easy to see from $p(a)p(b) < 0$ that at least one of these roots must lie strictly between $a$ and $b$, and we set $c$ equal to this root.

Given the definition of $p$ it now follows from $p(c) = 0$ that $0 = [g(c)^2 - y^2] G(c)$. Because $G$ is positive on $[a,b]$ we can divide through by $G(c)$, whence $g(c)^2 = y^2$. By construction, however, we know that $g(x) \ge 0$ for all $x \in [a,b]$, so both $g(c)$ and $y$ (which lies between $g(a)$ and $g(b)$) are non-negative. We have therefore found a value $c \in (a,b)$ satisfying $g(c) = y$, as required.

\end{proof}

\subsubsection{Main proof} We now complete the proof of \CiteTheorem{lem:fundamental}.



\begin{proof}[Proof of \CiteTheorem{lem:fundamental}]

Choose any $k, m \in \IOb_o$ satisfying $\speed_k(m) > 0$.  Then
$\taxis$ and $\wl_k(m)$ are distinct lines intersecting in $\origin$.
Therefore, their $\w{m}{k}$-images, $\wl_m(k)$ and $\taxis$, are
distinct intersecting lines. Hence, $\speed_m(k)>0$.  By
\CiteLemma{lem:configuration} and $\lnot\EInf$, we can assume that
\begin{itemize}
\item $\w{k}{m}[\plane(\taxis,\xaxis)] = \plane(\taxis,\xaxis)$; 
\item $\w{k}{m}[\yaxis] = \yaxis$; and
\item $k$ moves in the positive direction of the
$x$-axis according to $m$, \ie \\
$(1,\speed_{m}(k),0,0)\in  \wl_{m}(k)$ and
$\origin\in\wl_{m}(k)$.
\end{itemize}

\begin{figure}[h!btp] 
\small
\psfragfig[keepaspectratio,width=\linewidth]{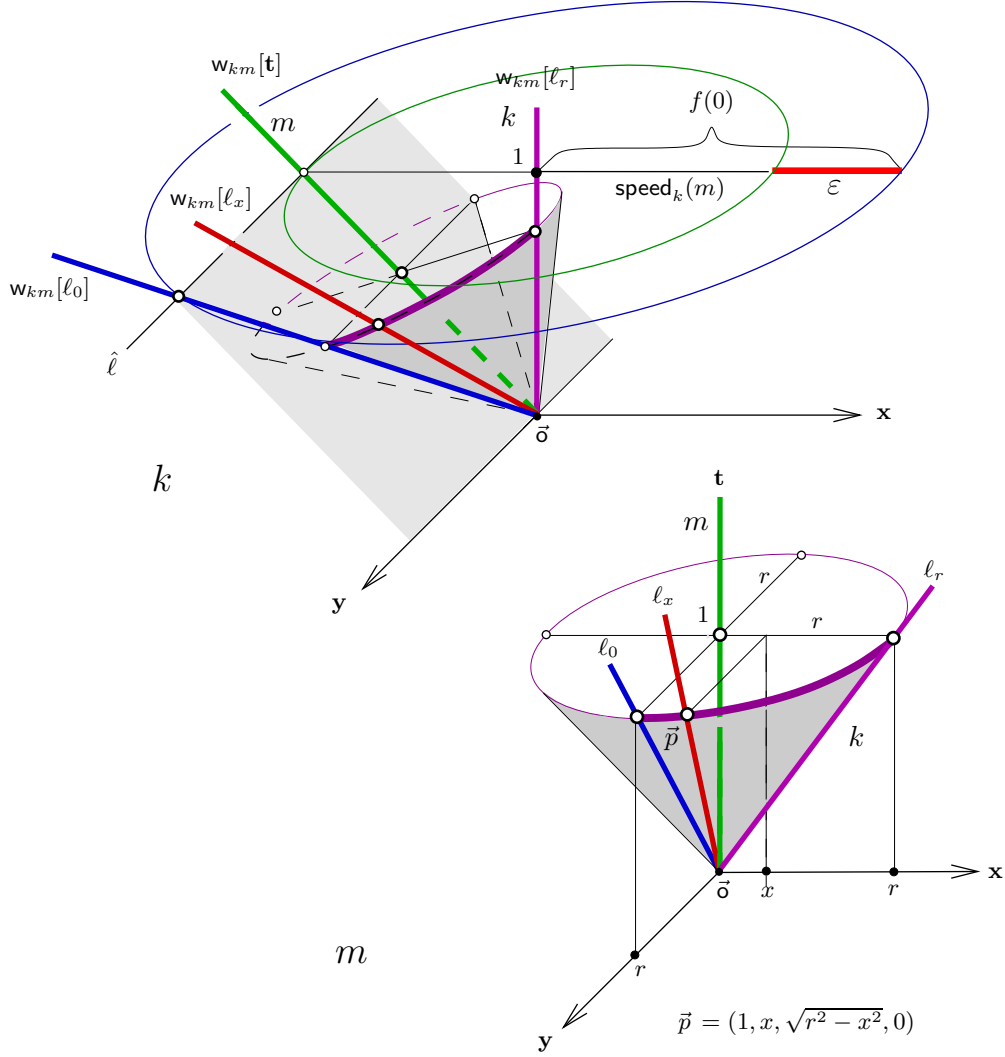}
{
\psfrag{o}[lt][lt]{$\origin$}
\psfrag{1}[rb][rb]{$1$}
\psfrag{p}[rt][rt]{\normalsize $\vvp$}
\psfrag{text}[lt][lt]{$\vvp=(1,x,\sqrt{r^2-x^2},0)$}
\psfrag{e}[t][t]{\Large $\varepsilon$}
\psfrag{wl0}[rt][rt]{$\w{k}{m}[\ell_0]$}
\psfrag{lk}[lt][lt]{$\hat{\ell}$}
\psfrag{wlx}[b][b]{$\w{k}{m}[\ell_x]$}
\psfrag{wlr}[b][b]{$\w{k}{m}[\ell_r]$}
\psfrag{mcc}[lb][lb]{\large $m$}
\psfrag{kc}[r][r]{\large $k$}
\psfrag{kcc}[lt][lt]{\large $k$}
\psfrag{wt}[lb][lb]{$\w{k}{m}[\taxis]$}
\psfrag{t}[b][b]{$\taxis$}
\psfrag{11}[rb][rb]{$1$}
\psfrag{speed}[t][t]{$\speed_k(m)$}
\psfrag{f0}[b][b]{$f(0)$}
\psfrag{x}[l][l]{$\xaxis$}
\psfrag{y}[rt][rt]{$\yaxis$}
\psfrag{K}[c][c]{\LARGE $k$}
\psfrag{M}[c][c]{\LARGE $m$}
\psfrag{-r}[t][t]{$-r$}
\psfrag{r}[t][t]{$r$}
\psfrag{x1}[t][t]{$x$}
\psfrag{l+r}[b][b]{$\ell_r$}
\psfrag{mc}[r][r]{\large $m$}
\psfrag{w}[b][b]{$\w{m}{k}[\taxis]$}
\psfrag{lx}[b][b]{$\ell_x$}
\psfrag{l0}[b][b]{$\ell_0$}
\psfrag{xx}[r][r]{$\xaxis$}
}
\caption{Illustration for the proof of \CiteTheorem{lem:fundamental}
\label{fig:fundamental-newproof}}
\end{figure}

Let $r := \speed_{m}(k)$, and note that $r \neq \infty$ 
by $\lnot\EInf$. Then $(1,r,0,0) \in \wl_{m}(k)$.

For each $x\in [0,r]$, let $\ell_x$ be the line containing $\origin$
and the point $( 1, x,\sqrt{r^2-x^2},0 )$. Observe that
$\slope(\ell_x) = r$ for all such $x$, and that $\ell_r=\wl_{m}(k)$;
see Figure~\ref{fig:fundamental-newproof}.  Since $\wl_{m}(k)$ is an
$m$-observer line, by \CiteTheorem{lem:observer-lines} every $\ell_x$
is an $m$-observer line, hence by
\CiteLemma{lem:transformed-observer-lines} every $\w{k}{m}[\ell_x]$ is
a $k$-observer line. It follows from $\lnot\EInf$ that the function
$f:[0,r] \to \Q$ given by
\[
f(x) := \slope(\w{k}{m}[\ell_x]).
\]
is well-defined, and it is easy to see that 
\[
f(r)=\slope(\w{k}{m}[\ell_r])=
\slope(\w{k}{m}[\wl_{m}(k)])=\slope[\taxis]=0.
\]
We will prove that $f(0) > \speed_k(m)$.

Recall that $\w{k}{m}$ is a bijection taking planes to 
planes by \CiteTheorem{lem:line-to-line}.
Since $\ell_0 \subseteq \plane(\taxis,\yaxis)$ and $\w{k}{m}$ fixes 
the $y$-axis, we have 
\[
  \w{k}{m}[\ell_0] \subseteq 
\plane(\w{k}{m}[\taxis],\w{k}{m}[\yaxis]) = 
\plane(\w{k}{m}[\taxis],\yaxis) .
\]
Let us write $\hat P :=\plane(\w{k}{m}[\taxis],\yaxis)$.

Because $\slope(\w{k}{m}[\taxis])=\speed_{k}(m)$ cannot be infinite (by
$\lnot\EInf$), there exists some $\vv s \in \Q^3$ and such that $(1,\vv s)
\in \w{k}{m}[\taxis]$.  And because \taxis is a subset of the $tx$-plane
(which is fixed by $\w{k}{m}$), we know that $\w{k}{m}[\taxis]
\subset \plane(\taxis,\xaxis)$.  Thus, the $y$- and $z$-components of $\vv s$
must both be zero, and there exists some $\hat x \in \Q$ with $(1,\hat
x,0,0) \in \w{k}{m}[\taxis]$.

\medskip
By \CiteLemma{lem:iobo2}, we have $\w{k}{m}\take{\origin}=\origin$, so we know 
that $\origin \in \w{k}{m}[\taxis] \subseteq \hat P$. 
It follows that $\hat P = \plane(\w{k}{m}[\taxis],\yaxis)$ is the unique plane
containing both the origin and the line ${\hat\ell} := \{ (1,\hat
x,y,0) : y \in \Q \}$, and every line in this plane which has finite
slope and passes through the origin must intersect $\hat\ell$ at some
point $(1,\hat x, y, 0)$ where $y \in \Q$. The line of this form with
the smallest slope is the one which minimises the value of ${\hat x}^2
+ y^2$, and since this is minimal precisely when $y=0$ the line in
this plane through the origin which has the least slope is
$\w{k}{m}[\taxis]$.  At the same time, we know that
$\w{k}{m}[\ell_0]$ is a line in this plane, and that
$\w{k}{m}[\ell_0] \neq \w{k}{m}[\taxis]$ because
$\w{k}{m}$ is a bijection and $\ell_0 \neq \taxis$.
Hence, $ \slope(\w{k}{m}[\ell_0]) > \slope
  (\w{k}{m}[\taxis])$.
Therefore, we have
\[
  f(0) = \slope(\w{k}{m}[\ell_0]) > \slope (\w{k}{m}[\taxis]) = 
\speed_{k}(m).
\] 
Thus, $f(0)>\speed_k(m)$ as claimed.

Let $\varepsilon = f(0)-\speed_k(m)$. We will prove that for this
choice of $\varepsilon$ the conclusion of the lemma holds, \ie that
for every non-negative $v \le \speed_k(m)+\varepsilon$ there is
$h\in\IOb_o$ such that $\speed_k(h)=v$ and $\speed_m(k)=\speed_m(h)$.

\medskip
To prove this, choose any $v\in\Q$ satisfying
$0\leq v \le \speed_k(m)+\varepsilon=f(0)$, and  
recall that $f(r)=0$.  Thus,
\begin{equation}
\label{equnop1-alt}
f(0) \geq v \geq f(r).
\end{equation}
We will use \CiteLemma{lem:quadratic} to prove that 
\begin{equation}
\label{allitas}
\text{there is $x\in [0,r]$ such that $f(x)=v$.}
\end{equation}  

\medskip
We know from \CiteTheorem{lem:line-to-line} that $\w{k}{m}$ is a 
bijection taking lines to lines.  It also preserves the
origin since $m,k\in\IOb_o$. Hence, by 
\CiteLemma{lem:affine}, there exists some linear transformation $L$ and 
automorphism $\varphi$ of $(\Q,+,\cdot,0,1,\leq)$ for which 
$\w{k}{m} = L \circ \widetilde\varphi$.

By construction, $\widetilde\varphi$ maps each coordinate axis to itself, so
it takes $\plane(\taxis,\xaxis)$ to $\plane(\taxis,\xaxis)$ and $\yaxis$ to
$\yaxis$.  We have already seen that $\w{k}{m}$ does likewise, and so
the same must be true of $L$.

We can therefore find  $a,b,c,d,\lambda\in\Q$ with $\lambda\neq 0$  such that, 
for every $t,x,y\in\Q$,
\begin{equation*}
  \w{k}{m}(t, x,y,0) = ( a\varphi(t) + b\varphi(x),  c\varphi(t) + d\varphi(x), 
\lambda \varphi(y), 0 ).
\end{equation*}
As $\varphi$ is an automorphism of $(\Q,+,\cdot,0,1,\leq)$,  
it follows that $\varphi(1)=1$; that for every  $x\in [0,r]$ we have $\varphi(x)\leq \varphi(r)$;  and that
 \[
   \w{k}{m} \left( 1,x,\sqrt{r^2-x^2},0 \right) 
	= \left( a+b\varphi(x), c+d\varphi(x),\lambda{\sqrt{\varphi{(r)}^2-
\varphi(x)^2}}, 0 \right).
\]
By definition, for every $x\in[0,r]$, $\ell_x$ is the line containing
$\origin$ and $\left( 1,x,\sqrt{r^2-x^2},0 \right)$; therefore,
$\w{k}{m}[\ell_x]$ is the line containing $\origin$ and $\w{k}{m} \left(
1,x,\sqrt{r^2-x^2},0 \right)$, and $f(x)\in\Q$ is the slope of this line. 
Since this slope cannot be infinite we have, for all $x \in [0,r]$, that
\begin{equation}\label{eq:abphi}
a+b\varphi(x)\neq 0
\end{equation}
and hence
\[
   f(x) = \sqrt{ \frac{\left( c+d\varphi(x) \right)^2 + \lambda^2 
\left( \varphi{(r)}^2-\varphi(x)^2 \right)}
                      { \left( a+ b\varphi (x) \right)^2 }
				       } .
\]

\noindent
Let $F:[0,\varphi(r)]\rightarrow \Q$ and
$G:[0,\varphi(r)]\rightarrow\Q$ be the quadratic functions defined by
\begin{eqnarray*}
F(y) &:=& \left( c+dy \right)^2 + \lambda^2 \left( \varphi{(r)}^2-y^2 \right) \\
G(y) &:=& (a + by)^2,
\end{eqnarray*}
and consider any $y \in [0,\varphi(r)]$. Because $(\varphi{(r)}^2-y^2) \geq 0$, it follows immediately that $F(y) \geq 0$. Moreover, $G(y) > 0$, because $\varphi$ is an ordered-field automorphism, whence $\varphi^{-1}(y) \in [0,r]$, and so by \eqref{eq:abphi} we have $a+by = a+b\varphi(\varphi^{-1}(y)) \neq 0$. So, if we now define $g(y) = \sqrt{F(y)/G(y)}$, then $g$ is of the correct form for \CiteLemma{lem:quadratic} to be applied over the interval $[0,\varphi(r)]$.

Because $f = g \circ \varphi$, it follows from \eqref{equnop1-alt} and $\varphi(0) = 0$ that
\[
g(0) \geq v \geq g(\varphi{(r)}). 
\]
By \CiteLemma{lem:quadratic}, there therefore exists some $y \in [0,\varphi(r)]$ with $g(y) = v$. 
Taking $x = \varphi^{-1}(y)$ now shows that there exists $x\in[0,r]$ 
satisfying $f(x)=v$, and \eqref{allitas} holds as claimed.

\medskip
Accordingly, let $\tilde x \in [0,r]$ be such that 
$f(\tilde x)= \slope(\w{k}{m}[\ell_{\tilde x}])=v$. 
Then $\ell_{\tilde x}$ is a line satisfying 
$\slope(\ell_{\tilde x}) = r = \speed_{m}(k)$ and 
$\slope(\w{k}{m}[\ell_{\tilde x}])$ 
= $f({\tilde x})$ = $v$. Since $\ell_{\tilde x}$ is an $m$-observer line, 
there exists  $h\in\IOb_o$ with $\wl_{m}(h)=\ell_{\tilde x}$, 
and hence
\begin{itemize}
\item $\wl_{k}(h)=\w{k}{m}[\ell_{\tilde x}]$,
\item $\speed_{m}(h) = \slope(\wl_{m}(h)) = \slope(\ell_{\tilde x}) = r = 
\speed_{m}(k)$, and 
\item $\speed_{k}(h) = \slope(\wl_{k}(h)) = \slope(\w{k}{m}[\ell_{\tilde x}]) 
= v$.
\end{itemize}
This is exactly what we had to 
prove, viz.\ there exists some $h$ with $\speed_k(h) = v$ and 
$\speed_m(k) = \speed_m(h)$.
\end{proof}



\subsection{Main Lemma}

\begin{Theorem}{lem:main}
Assume $\KIN+\AxIso$. Then there is $k\in\IOb_o$ and $\kappa\in\Q$ such that
\begin{equation}
\label{eq-main}
\{\w{m}{k}\: :\: m\in\IOb_o\}\subseteq\kiso.
\end{equation}
\end{Theorem}

\subsubsection{Supporting lemmas}

The supporting lemmas can be informally described as:

\begin{description}
\item[\CiteLemma{lem:same-speed-easy}]
~ \\ If $m$ considers $k$ and $h$ to be moving at the same speed and $\w{m}{k}$ is a $\kappa$-isometry, then so is $\w{m}{h}$.
\item[\CiteLemma{lem:rest}]
~ \\ Two observers are at rest with respect to one another if and only if the transformation between them is trivial.
\item[\CiteLemma{lem:observer-origin}]
~ \\ Given any point on an observer's worldline, we can find an observer with the same worldline which regards that point as its origin.
\item[\CiteLemma{lem:median-observer}]
~ \\ Given any two observers, 
there is a third observer which sees them both moving with the same
speed.
\item[\CiteLemma{lem:k-is-unique}]
~ \\ If two observers are moving relative to one another, there exists a unique value $\kappa$ for which the transformation between them is a $\kappa$-isometry.
\end{description}

\subsubsection{Proofs of the supporting lemmas}

\begin{Lemma}{lem:same-speed-easy}
Assume $\KIN + \AxIso$, and let $k, h, m \in \IOb_o$. If $\speed_m(k) = \speed_m(h)$ and $\w{m}{k} \in \kiso$, then $\w{m}{h} \in \kiso$.
\end{Lemma}
\begin{proof}
By \CiteLemma{lem:2vanrotacio}, there exists a spatial rotation 
$R$ taking $\wl_m(k)$ to $\wl_m(h)$, 
and by \CiteLemma{lem:relocation} there is some observer $k^*$ satisfying $k \mappedto{R}{m} k^*$. 
Since $\w{m}{k^*} = R \circ \w{m}{k} $ and $R[\wl_m(k)] = \wl_m(h)$, it follows that 
$\wl_m(k^*) = R[\wl_m(k)] = \wl_m(h)$, so that $k^*$ and $h$ share the same worldline.
By \CiteLemma{lem:colocate}, $\w{k^*}{h}$ is therefore trivial, and hence a $\kappa$-isometry.
It now follows that $\w{m}{h} = \w{m}{k^*} \circ \w{k^*}{h}   = R \circ \w{m}{k} \circ \w{k^*}{h} $ is a composition of $\kappa$-isometries, so $\w{m}{h} \in \kiso$ as claimed.
\end{proof}

\begin{Lemma}{lem:rest}
  Assume \KIN. For all observers $k,m\in \IOb$, we have
  \begin{equation*}
    k \text{ is at rest according to } m \quad \text{ if{}f } \quad \w{m}{k}\in\triv. 
  \end{equation*}
\end{Lemma}

\begin{proof}
$(\Rightarrow)$ Suppose first that $k$ is at rest according to $m$, \ie
$\w{m}{k}(\origin)_s=\w{m}{k}(\tunit)_s$. We will show that $\w{m}{k}\in\triv$.

Recall that $\wl_m(k)$ is a  line
(by \ax{AxLine}) and notice that
$\w{m}{k}(\origin),\w{m}{k}(\tunit)\in\w{m}{k}[\taxis]=\wl_m(k)$. Hence,
$\wl_m(k)$ is parallel to
$\taxis$ (because it is a  line containing two distinct points, 
$\w{m}{k}(\origin)$ and $\w{m}{k}(\tunit)$, whose spatial components are 
identical), and
it passes through $\w{m}{k}(\origin)$.

Next, according to \ax{AxRelocate} we can find an observer $m' \in\IOb$ for which $\w{m}{m'}$ is the translation 
taking \origin to $\w{m}{k}(\origin)$. Because it is a translation, $\w{m}{m'}$ necessarily takes \taxis to a 
line parallel to \taxis; and because this line is $\w{m}{m'}[\taxis] = \wl_m(m')$, we see that $\wl_m(m')$ is 
parallel to \taxis. Moreover, because $\taxis$ contains $\origin$, we know that $\w{m}{k}(\origin) = \w{m}{m'}(\origin) 
\in \wl_m(m')$, whence $\wl_m(m')$ is also a line parallel to \taxis that passes through $\w{m}{k}(\origin)$.

Since $\wl_m(k)$ and $\wl_m(m')$ are parallel lines which share a common point, 
they must be the same (world)line, so $\w{m'}{k} \in \triv$
by \CiteLemma{lem:colocate}. At the same time we know that $\w{m}{m'} \in \triv$, because it is a translation.
It therefore follows by composition that $\w{m}{k}=\w{m}{m'}\circ\w{m'}{k}\in\triv$, as claimed.

$(\Leftarrow)$ To prove the converse, suppose that $\w{m}{k} \in \triv$. We need to show that $k$ is at rest according to
$m$, \ie $\w{m}{k}(\tunit)_s=\w{m}{k}(\origin)_s$. But this is obvious because every trivial transformation maps \taxis to a line parallel to \taxis.
\end{proof}

\begin{rem}
It follows easily from \CiteLemma{lem:rest} and the fact that \triv is a group under composition that ``being at rest according to'' is an equivalence relation 
on observers, and ``moving according to'' is a symmetric relation.
\end{rem}

\begin{Lemma}{lem:observer-origin} Assume \AxEField, \AxWvt and
\AxRelocate.
If $\ell \in \oblines(k)$ and $\vvp \in \ell$, then there exists some 
$h \in \IOb$ for which $\w{k}{h}(\origin) = \vvp$ and $\wl_k(h) = \ell$.
\end{Lemma}
\begin{proof}
Choose $h'\in\IOb$ such that $\wl_k(h')=\ell$. By $\vvp\in\wl_k(h')$, 
we have $\w{h'}{k}(\vvp)\in\w{h'}{k}[\wl_k(h')]=\wl_{h'}(h')=\taxis$. Let
$h\in\IOb$ be such that $\w{h'}{h}$ is the translation by vector 
$\w{h'}{k}(\vvp)$. Such $h$ exists by \AxRelocate. 
Translation $\w{h'}{h}$ fixes $\taxis$
because $\w{h'}{k}(\vvp)\in\taxis$. Then
$\wl_k(h)=\w{k}{h}[\taxis]=\w{k}{h'}[\w{h'}{h}[\taxis]]=\w{k}{h'}[\taxis]=\wl_k(h')=\ell$
and
$\w{k}{h}(\origin)=\w{k}{h'}(\w{h'}{h}(\origin))=\w{k}{h'}(\w{h'}{k}(\vvp))=\vvp$
as claimed.  
\end{proof}

\begin{Lemma}{lem:median-observer}
Assume \KIN, \AxIso, and $\lnot\EInf$. Then given any $k, m \in \IOb_o$, 
there exists some $h \in \IOb_o$ for which $\speed_h(k) = \speed_h(m)$.
\end{Lemma}
\begin{proof}
If $\speed_k(m) = 0$, the result follows trivially by choosing $h=k$,
so suppose $\speed_k(m) > 0$.  By applying \CiteTheorem{lem:fundamental}
choosing $v=\speed_k(m)$, there exists $h\in\IOb_o$ such that
\begin{eqnarray}
  \speed_k(h)   & = & \speed_k(m) \label{median2} \\
\speed_m(k)  & = &  \speed_m(h). \label{median1}
\end{eqnarray}
Applying \CiteTheorem{lem:same-speed-hard} to \eqref{median1} tells us that
\begin{eqnarray}
  \speed_k(h) &=& \speed_h(k) \label{median3} \\
 \speed_h(m)    &=& \speed_k(m)  \label{median4}
\end{eqnarray}
and so
\begin{equation*}
  \speed_h(k) \stackrel{\text{\tiny
      \eqref{median3}}}{\longeq} \speed_k(h)  \stackrel{\text{\tiny
      \eqref{median2}}}{\longeq} \speed_k(m) \stackrel{\text{\tiny
      \eqref{median4}}}{\longeq} \speed_h(m)
\end{equation*}
as claimed.
\end{proof}

\begin{Lemma}{lem:k-is-unique}
Assume \ax{AxEField} and let $m,k\in\IOb$ be observers such that $k$ is moving 
according to
$m$ and $\w{m}{k}\in\kiso$. Then $\kappa$ is uniquely determined by:
\begin{equation}
\label{eq:kappa}
\kappa=\frac{\left|\w{m}{k}\take{\tunit}_t-
      \w{m}{k}\take{\origin}_t\right|^2-1}{\left|\w{m}{k}\take{\tunit}_s-
      \w{m}{k}\take{\origin}_s\right|^2}.
\end{equation}
\end{Lemma}
\begin{proof} Let $f:\Q^4\rightarrow\Q^4$ be the linear
part of $\w{m}{k}$, \ie
$f(\vvp):=\w{m}{k}(\vvp)-\w{m}{k}(\origin)$. Then $f$ is a linear
$\kappa$-isometry, so it preserves $\kappa$-length. Hence,
$1=\knorm{\tunit}=\knorm{f(\tunit)}=f(\tunit)^2_t-\kappa
|f(\tunit)_s|^2=|\w{m}{k}(\tunit)_t-\w{m}{k}(\origin)_t|^2-\kappa
|\w{m}{k}(\tunit)_s-\w{m}{k}(\origin)_s|^2$. We have that
$\w{m}{k}(\tunit)_s\neq\w{m}{k}(\origin)_s$ because $k$ is moving
according to $m$. Thus, \eqref{eq:kappa} follows by reorganizing the
equality above.
\end{proof}

\subsubsection{Main proof} We now complete the proof of \CiteTheorem{lem:main}.



\begin{proof}[Proof of \CiteTheorem{lem:main}]
There are two cases to consider:
Case 1: $\lnot\EInf$ holds.
Case 2: $\EInf$ holds.

\medskip\noindent
\textit{Proof of Case 1: Assume $\lnot\EInf$.}

Suppose $\hat k,\hat m$ are any observers in $\IOb_o$. According to
\CiteLemma{lem:median-observer}, there is some $\hat h \in \IOb_o$
such that $\speed_{\hat h}(\hat k) = \speed_{\hat h}(\hat m)$. By
\CiteTheorem{lem:same-speed-hard}, $\w{\hat m}{\hat k}$ is a
$\kappa$-isometry for some $\kappa \in \Q$.  This shows that every
worldview transformation between two observers in $\IOb_o$ is a
$\kappa$-isometry for some $\kappa$, and by \CiteLemma{lem:k-is-unique},
this $\kappa$ is unique if the two observers are moving relative to
each other (however, even this unique $\kappa$ may vary with the
choice of the two observers.)

\medskip
Suppose, then, that $k\in\IOb_o$. We will show that $\kappa$ can be found such that \eqref{eq-main} holds.

Notice first that if any observer $m \in \IOb_o$ is at rest relative to $k$, then \CiteLemma{lem:rest} tells us that $\w{m}{k}$ is trivial, thus it is a $\kappa$-isometry for every $\kappa \in \Q$ by \CiteLemma{lem:xy}. So we only need to consider observers which are moving relative to $k$.

Suppose, therefore, that $m_1,m_2\in\IOb_o$ are two observers, and that at least one is moving according to
$k$. Without loss of generality we can assume that $0$ $<$ $\speed_k(m_1)$ and $\speed_k(m_2)$ $\leq$ $\speed_k(m_1)$. We have
already seen that 
\begin{equation}
\label{main1}
\w{m_1}{k}\in\kiso[\tilde\kappa]
\end{equation} 
for some unique $\tilde\kappa$. We will show that
$\w{m_2}{k}\in\kiso[\tilde\kappa]$ as well. We have already seen that this is the case if $\speed_k(m_2) = 0$, so we can assume that $0 < \speed_k(m_2)$.

By \CiteTheorem{lem:fundamental}, choosing $v=\speed_k(m_2)$ and
$m=m_1$, there exists $h \in \IOb_o$ such that
\begin{eqnarray} 
  \speed_k(h) &=& \speed_k(m_2)  \label{main2b} \\
  \speed_{m_1}(k) &=& \speed_{m_1}(h) \label{main2a} 
  \end{eqnarray}
It follows from \CiteLemma{lem:same-speed-easy} with \eqref{main1} and 
\eqref{main2a} that 
\begin{equation} \label{main3}
  \w{m_1}{h} \in \kiso[\tilde\kappa]
\end{equation}
and hence (by \eqref{main1}) that
\begin{equation} \label{main4}
  \w{k}{h} = \w{k}{m_1} \circ \w{m_1}{h}=\w{m_1}{k}^{-1}\circ\w{m_1}{h} \in \kiso[\tilde\kappa] .
\end{equation}
Applying \CiteLemma{lem:same-speed-easy} with \eqref{main2b} and 
\eqref{main4} now tells us 
that $\w{k}{m_2} \in \kiso[\tilde\kappa]$. But then 
\[
\w{m_2}{k}\in\kiso[\tilde\kappa]
\]
as claimed.

Finally, let $m\in\IOb_o$ be arbitrary. As we have shown, no matter whether $m$ is at rest or in motion relative to $k$, there is some $\kappa_m$ such that
$\w{m_1}{k}$ and $\w{m}{k}$ are both in $\kiso[\kappa_m]$. But because $m_1$ is moving relative to $k$ this $\kappa_m$ is unique for $m_1$, so we must have $\kappa_m = \tilde\kappa$. Thus, taking $\kappa := \tilde\kappa$ ensures that \eqref{eq-main} holds as claimed.

\medskip\noindent
\textit{Proof of Case 2: Assume \EInf.}

By \CiteLemma{lem:straight-lines-are-observer-lines}, every observer 
considers every line to be the worldline of an observer, so in particular
any `horizontal' line through \origin is an observer line. 
By \CiteLemma{lem:observer-origin}, therefore, there exists $h \in \IOb_o$ satisfying 
$\speed_o(h) = \infty$.

Recall that \saxis is the spatial hyperplane $\{ (0,x,y,z) : x,y,z \in \Q \}$; 
let us consider $\w{o}{h}[\saxis]$. By \CiteTheorem{lem:line-to-line}, this is a 3-dimensional subspace of $\Q^4$ which contains $\w{o}{h}(\origin) = \origin$ (because $h \in \IOb_o$). 
It follows that the subspace formed by the intersection of $\saxis$ with 
$\w{o}{h}[\saxis]$ must be at least 1-dimensional and so there is some 
line $\ell$ such that $\origin \in \ell \subseteq \w{o}{h}[\saxis] \cap \saxis$.
See Figure~\ref{mainthminfty-fig}.

\begin{figure}[h!btp] 
\small
\psfragfig[keepaspectratio,width=\linewidth]{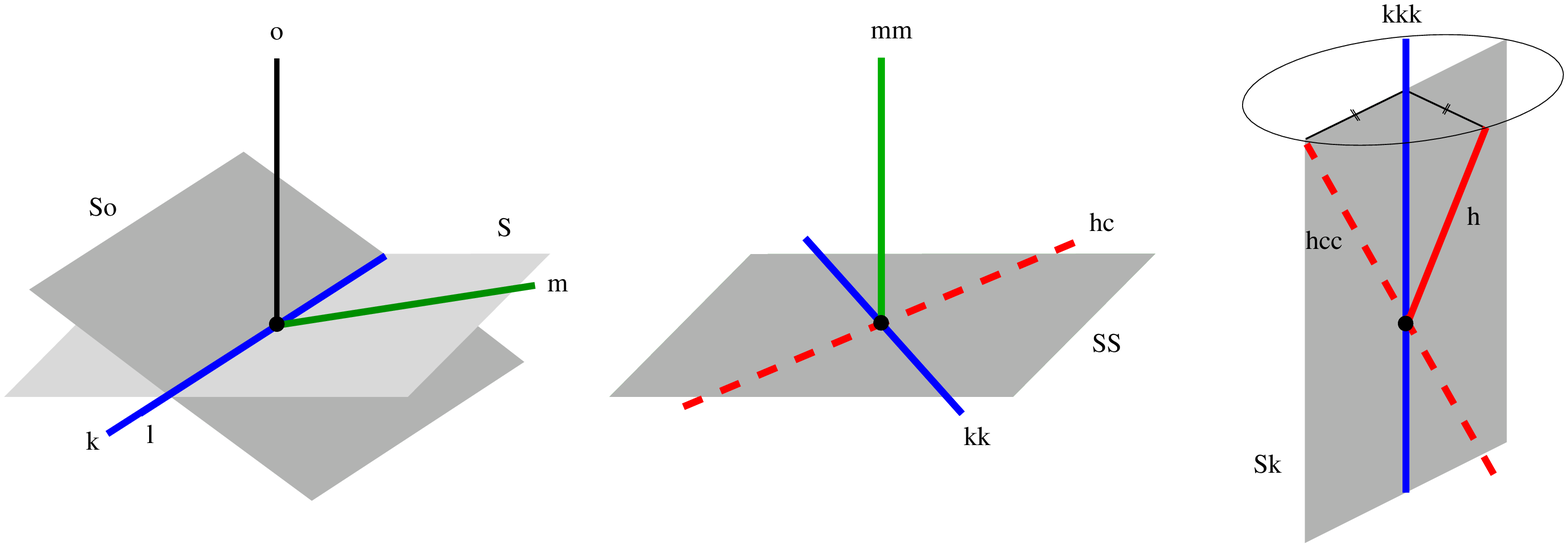}
{
\psfrag{o}[b][b]{$o$}
\psfrag{S}[b][b]{$\saxis$}
\psfrag{k}[rt][rt]{$k$}
\psfrag{m}[l][l]{$h$}
\psfrag{mm}[b][b]{$h$}
\psfrag{kk}[t][t]{$k$}
\psfrag{So}[rb][rb]{$\w{o}{h}[\saxis]$}
\psfrag{S}[b][b]{$\saxis$}
\psfrag{SS}[lt][lt]{$\saxis$}
\psfrag{hc}[lb][lb]{$m^*$}
\psfrag{hcc}[rt][rt]{$m^*$}
\psfrag{kkk}[b][b]{$k$}
\psfrag{Sk}[r][r]{$\w{k}{h}[\saxis]$}
\psfrag{h}[lt][lt]{$m$}
\psfrag{l}[lt][lt]{$\ell$}
}
\caption{\label{mainthminfty-fig} Illustration for the proof of
\CiteTheorem{lem:main} if $\EInf$ is assumed.}  
\end{figure}

Because every observer considers every line to be an observer line, $o$
considers $\ell$ to be an observer line, so there exists some $k$ such that
$\ell = \wl_o(k)$. By \CiteLemma{lem:observer-origin}, we can choose this $k$
to be in $\IOb_o$. Since $\ell\subseteq \saxis$, we have $\speed_o(k) = \infty$.
It follows that $\speed_o(k) = \speed_o(h)$ (both are infinite), whence
\CiteTheorem{lem:same-speed-hard} tells us that $\w{h}{k} \in \kiso$ for some
$\kappa$. Let us fix such a $\kappa$.  We will prove that \eqref{eq-main}
holds for this $\kappa$.

\medskip
To do this, we first switch from $o$'s worldview to $h$'s. By
construction, 
we know that $\wl_o(k) = \ell \subseteq
\w{o}{h}[\saxis]$, so by applying $\w{h}{o}$, 
we have  
\begin{equation} \label{case3.1}
\wl_h(k) \subseteq \saxis,
\end{equation}
and hence $\speed_h(k)=\infty$.

\medskip
Now let $m$ be any observer $m \in \IOb_o$.

In the particular case when $\wl_h(m) \subseteq \saxis$, we must
have $\speed_h(m) = \infty$ because all points in $\saxis$ have the same
time coordinate. In this case, we have $\speed_h(k) =
\speed_h(m)$, and since we know that $\w{h}{k} \in \kiso$,
\CiteLemma{lem:same-speed-easy} tells us that $\w{h}{m}$ (hence also $\w{m}{h}$) is a
$\kappa$-isometry as well.  It now follows by composition, in this special case, that $\w{m}{k} = \w{m}{h}
\circ \w{h}{k}$ is a $\kappa$-isometry, as required.

\medskip
Now consider things more generally from $k$'s point of view.  As
before, $\w{k}{h}[\saxis]$ is a hyperplane, and we know from
\eqref{case3.1} that $\wl_h(k)\subseteq \saxis$. It follows that
\[
  \taxis = \wl_k(k) = \w{k}{h}[\wl_k(h)] \subseteq \w{k}{h}[\saxis]
\]
so $\w{k}{h}[\saxis]$ contains the time-axis $\taxis$.

We can therefore find a line $\ell$ such that $\origin \in \ell
\subseteq \w{k}{h}[\saxis]$ and $\slope(\ell) = \speed_k(m)$. For if
$\speed_k(m) = \infty$ we can choose the line through \origin in
$\w{k}{h}[\saxis]$ that is perpendicular to \taxis, and if
$\speed_k(m) = 0$ we can take $\ell = \taxis$. For the remaining case,
where $0 < \speed_k(m) < \infty$, choose any point $\vvp \in
\w{k}{h}[\saxis]\setminus\taxis$. By \CiteLemma{lem:triangulation}, we
can find a line of slope $\speed_k(m)$ in $\w{k}{h}[\saxis]$ which
meets \taxis, and a translation along \taxis can then be applied to
find a parallel line (also in $\w{k}{h}[\saxis]$) that passes through
\origin.

Because all lines are observer lines, $\ell$ is an observer line; and by \CiteLemma{lem:observer-origin} there is some $m^* \in \IOb_o$
for which $\wl_k(m^*) = \ell \subseteq \w{k}{h}[\saxis]$. But this means that $\wl_h(m^*) = \w{h}{k}[\wl_k(m^*)] \subseteq \w{h}{k}[\w{k}{h}[\saxis]] = \saxis$ and hence, as we saw in the special case above, $\w{m^*}{k} \in \kiso$. But now \CiteLemma{lem:same-speed-easy} tells us that from
$\speed_k(m) = \slope(\ell) = \speed_k(m^*)$ and $\w{k}{m^*} \in \kiso$ we can deduce
$\w{k}{m} \in\kiso$. Therefore, for arbitrary $m\in\IOb_o$,
$\w{m}{k}\in\kiso$, \ie \eqref{eq-main} holds.
\end{proof}


\section{Proofs of the main theorems}
\label{sec:main-proofs}

\begin{proof}[Proof of \CiteTheorem{thm:characterisation}]
If $\lnot\EMovingIOb$ is assumed, then  $\W
\subseteq \triv$ by \CiteLemma{lem:rest}, hence $\mathbb{W} \subseteq \kiso$ for every
$\kappa$ by \CiteLemma{lem:xy}.

Assume $\EMovingIOb$.  Let $k\in\IOb_o$ and $\kappa$ be such
that \eqref{eq-main} in \CiteTheorem{lem:main} holds, \ie $\{\w{m}{k}
: m\in\IOb_o\}\subseteq\kiso$.  Then by \CiteLemma{lem:k-is-unique} it
is enough to prove that the worldview transformations are
$\kappa$-isometries.

To prove that worldview transformations are $\kappa$-isometries,
choose any observers $m_1,m_2 \in \IOb$.  By \CiteLemma{lem:iobo}, we
can find $m^o_1,m^o_2 \in \IOb_o$ for which $\w{m_1}{m^o_1}$ and
$\wrev{m_2}{m^o_2}$ are translations and hence $\kappa$-isometries. As
$\w{m_1^o}{k}$ and $\w{m_2^o}{k}$ are also $\kappa$-isometries, so it
follows that
\begin{equation}
\label{wm1m2}
  \w{m_1}{m_2} =
\w{m_1}{m_1^o}\circ\w{m_1^o}{k}\circ\w{k}{m_2^o}\circ\w{m_2^o}{m_2}=
\w{m_1}{m_1^o}\circ\w{m_1^o}{k}\circ\w{m_2^o}{k}^{-1}\circ\w{m_2^o}{m_2}
\end{equation}
is a $\kappa$-isometry.
\end{proof} 


\begin{proof}[Proof of \CiteTheorem{lem:group}]
  Let us first prove that
  \begin{equation}\label{eq:Wk=G}
    \W_k=\G, \text{ for every } k\in\IOb.
  \end{equation}
  To do so, let $k\in\IOb$. Then, by 
  the definition of $\W_k$ and the construction of $\M_{G}$,
  \begin{equation*}
    \W_k=\{\w{k}{h}:h\in\IOb\}=\{k\circ h^{-1}: h\in \G\}=k\circ
    \G^{-1}=\G
  \end{equation*}
  because $\G$ is a group. Thus, \eqref{eq:Wk=G} holds.

\textit{(a)} By construction of $\M_{G}$,
we have $\w{k}{k}=k\circ k^{-1}=Id$ and $\w{m}{h}\circ\w{h}{k}=m\circ
h^{-1}\circ h\circ k^{-1}=m\circ k^{-1}=\w{m}{k}$ for every
$m,k,h\in\IOb=\G$. Thus, \AxWvt holds.  By \eqref{eq:Wk=G}, we
have that $\W_k=\W_h$ for every $k,h\in \IOb$, which is a trivial
reformulation of \AxSpr.  Finally, also by \eqref{eq:Wk=G}, we have
$\W=\bigcup_{k\in\IOb}\W_k=\G$.

\textit{(b)} A trivial reformulation of \AxRelocate is that
$\srot\cup\tran\subseteq \W_k$ for all $k\in\IOb$, which, by
\eqref{eq:Wk=G}, is equivalent to $\srot\cup\tran\subseteq\G$ in
$\M_{\G}$.

\textit{(c)} By definition of worldline, a trivial reformulation of
\AxLine is that $g[\taxis]$ is a line for every $g\in\W$. We know from
(a) that $\W=\G$, hence the statement holds.

\textit{(d)} We know from (a) that \AxWvt holds, hence by
\CiteLemma{lem:wvt}, $\wl_k(k)=\taxis$ for every $k\in\IOb$. Recall
that by definition of worldline $\wl_k(k'):=\w{k}{k'}[\taxis]$ for
every $k,k'\in\IOb$.  Hence, for every $k,k'\in\IOb$,
$\wl_k(k)=\wl_k(k')$ is equivalent to $\w{k}{k'}[\taxis]=\taxis$ in
$\M_{\G}$.  Therefore, \AxColocate holds in $\M_{\G}$ if{}f $g\in\triv$
whenever $g\in\W$ and $g[\taxis]=\taxis$.  We know from (a) that
$\W=\G$, hence the statement holds.
\end{proof}

\begin{proof}[Proof of \CiteTheorem{thm:group-instantiation}]
From \CiteLemma{lem:group}(a-c), it is clear that \AxWvt, \AxSpr,
\AxLine and \AxRelocate all hold, and that $\W = \G$. To see that
\AxColocate also holds, suppose $g\in\cpoi\cup\ceucl\cup\gal$
satisfies $g[\taxis]=\taxis$. We will show that $g \in \triv$, whence
the result follows by \CiteLemma{lem:group}(d).

To this end, write $g = T \circ L$ as a composition of a translation
$T$ and linear $\kappa$-isometry $L$, and recall that a linear map is
trivial if and only if it fixes (setwise) both the time-axis and the
present simultaneity, and preserves squared lengths in both.  We will
show that $L$ has these properties.

To see that $L[\taxis] = \taxis$, note that $T(\origin) = T(L(\origin)) =
g(\origin) \in \taxis$, whence $T$ must be a translation along the $t$-axis. 
Thus, $g$ and $T$ both fix \taxis setwise, whence so does $L = T^{-1} \circ
g$.  

To see that $L$ preserves squared length in $\taxis$, choose arbitrary
$t \in \Q$.  Since $L[\taxis] = \taxis$ there is some $t' \in \Q$ such
that $L(t,\vv 0) = (t',\vv 0)$, and now $\knorm{L(t,\vv 0)}^2 =
\knorm{(t,\vv 0)}^2$ forces $t' = \pm t$.  Thus, $L$ preserves squared
lengths in \taxis.

If $\kappa=0$, then $L$ fixes the present simultaneity $\saxis$ and
preserves the square lengths in it by definition.  To see that the
same statement holds if $\kappa\neq 0$, choose arbitrary $\vv s \in
\Q^3$ and define $t^* \in \Q$ and $\vv s^* \in \Q^3$ by $(t^*, \vv
s^*) := L(0,\vv s)$.  Then by linearity
\[
 L( 1, \vv s) = (\pm 1 + t^*, \vv s^*) \qquad \text{ and } \qquad
 L( 1,-\vv s) = (\pm 1 - t^*,-\vv s^*).
\]
Because $\knorm{(1, \vv s)}^2 = \knorm{(1,-\vv s)}^2$ and $L$ is a
linear $\kappa$-isometry, we have that $\knorm{L( 1, \vv
  s)}^2=\knorm{L(1,-\vv s)}^2$, which implies that $(1 + t^*)^2 = (1 -
t^*)^2$ and hence $t^* = 0$.  Thus, $L(0,\vv s)=(0,\vv s^*)$, \ie $L$
maps \saxis to itself.  If $\kappa\neq 0$, $\knorm{(0,\vv
  s)}^2=\knorm{L(0,\vv s)}^2=\knorm{(0,\vv s^*)}$ implies that
$\norm{\vv s}^2=\norm{\vv s^*}^2$. Hence, $L$ preserves the square
lengths in $\saxis$.

As claimed, therefore, $L$ is a linear map which fixes both the
time-axis and the present simultaneity, and preserves squared lengths
in both, whence it is linear trivial and $g = T \circ L$ is trivial.
As outlined above, it now follows that \AxColocate also holds, and
that hence $\M_{\G}$ is a model in which $\KIN+\AxSpr$ holds and
$\W=\G$.
\end{proof}


\begin{proof}[Proof of \CiteTheorem{thm:classification1}]
Assume that $\G$ is a group satisfying the conditions. We will prove
that statements (i) and (ii) are equivalent.

Assume that (i) holds.  By \CiteTheorem{lem:group}, $\M_\G$ is a
model of $\KIN+\AxSpr$ (and hence also $\KIN+\AxIso$) for which
$\W=\G$. Then (ii) follows by \CiteTheorem{thm:characterisation}.

Assume that (ii) holds. Then by \CiteTheorem{thm:group-instantiation}
  $\M_{\G}$ is a model of $\KIN+\AxSpr$ for which $\W=\G$. Then (i)
  follows by \CiteTheorem{lem:group}.
\end{proof}

\begin{proof}[Proof of \CiteTheorem{thm:classification2}]
  Assume $\KIN+\AxIso$. It is clear that at least one of cases (1)-(4)
  holds. First we show the consequences of the cases and then from
  those we show that they are mutually exclusive.

(Cases 1-3) If $k, m \in \IOb$ are at rest relative to each other, then because
$\w{m}{k}$ is trivial by \CiteLemma{lem:rest}, it is also a Euclidean isometry
by \CiteLemma{lem:xy}. Thus, for all observers $k$ and $m$ we have
\begin{equation}\label{gcsillag}
   \text{ if }\w{m}{k}(\tunit)_s=\w{m}{k}(\origin)_s,\text{ then
}\norm{\w{m}{k}(\tunit)_t-\w{m}{k}(\origin)_t}=1.
\end{equation}

We claim we can choose
$k^*$ and $m^*$ such that $\w{m^*}{k^*}(\tunit)_s \neq \w{m^*}{k^*}(\origin)_s$. This is true by definition
if \EMovingAccurateClock holds, and follows from \eqref{gcsillag} if either \ESlowClock or \EFastClock holds
because in each of these cases we can choose $m^*,k^*$ such that $\norm{\w{m^*}{k^*}(\tunit)_t-\w{m^*}{k^*}(\origin)_t} \neq 1$. 

It follows that $\EMovingIOb$ holds in all three cases, and so by \CiteTheorem{thm:characterisation}, there is a unique $\kappa$ such that $\W\subseteq \kiso$. Recall from \eqref{eq:kappa} that $\kappa$ can be determined from the motion of any two observers moving relative to one another by
\[
    \kappa=\frac{\left|\w{m}{k}\take{\tunit}_t-
      \w{m}{k}\take{\origin}_t\right|^2-1}{\left|\w{m}{k}\take{\tunit}_s-
      \w{m}{k}\take{\origin}_s\right|^2} .
\]
So, given our choice of $m^*,k^*$ (and the definitions of \EFastClock, \ESlowClock and \EMovingAccurateClock) we have
\begin{align*}
\ESlowClock 
  &\Rightarrow \norm{\w{m^*}{k^*}(\tunit)_t- \w{m^*}{k^*}(\origin)_t}^2 > 1 
  &\Rightarrow \kappa > 0 \\
\EFastClock 
  &\Rightarrow \norm{\w{m^*}{k^*}(\tunit)_t- \w{m^*}{k^*}(\origin)_t}^2 < 1 
  &\Rightarrow \kappa < 0 \\
\EMovingAccurateClock 
  &\Rightarrow \norm{\w{m^*}{k^*}(\tunit)_t- \w{m^*}{k^*}(\origin)_t}^2 = 1 
  &\Rightarrow \kappa = 0
\end{align*}

Because \eqref{eq:kappa} holds for any two relatively moving
observers it now follows from the uniqueness of $\kappa$ that $\ESlowClock \Rightarrow \AMovingClockSlow$,
$\EFastClock \Rightarrow \AMovingClockFast$ and $\EMovingAccurateClock \Rightarrow \AClockAccurate$.

Finally, to complete the proof of cases (1-3) it is enough to note that
\begin{align*}
  \kappa > 0 &\qquad\Rightarrow\qquad \kiso = \cpoi \text{ where $c = \sqrt{\nicefrac{1}{\kappa}}$; } \\
  \kappa < 0 &\qquad\Rightarrow\qquad \kiso = \ceucl \text{ where $c = \sqrt{\nicefrac{-1}{\kappa}}$; } \\
  \kappa = 0 &\qquad\Rightarrow\qquad \kiso = \gal .
\end{align*}

(Case 4). If $\lnot\EMovingIOb$ holds, then all worldview transformations are trivial by \CiteLemma{lem:rest},
so $\W \subseteq \triv$ as claimed.

The four cases are clearly mutually exclusive, because the situations
$$(\AMovingClockSlow+\EMovingIOb),\quad
(\AMovingClockFast+\EMovingIOb),$$
$$\EMovingAccurateClock\quad\text{ and}\quad \lnot\EMovingIOb$$ 
are mutually exclusive.
\end{proof}


\begin{proof}[Proof of \CiteTheorem{thm:consistency}]
(Cases 1-3) 
By \CiteTheorem{thm:group-instantiation} and \eqref{eq:tartalmazasok}, there are models $\M_P$, $\M_E$ and
$\M_G$ of $\KIN+\AxSpr$ such that the set of worldview transformations
are respectively $\poi$, $\eucl$ and $\gal$. In all three models, there
are $m,k\in\IOb$ such that
$\w{m}{k}\take{\tunit}_s\neq\w{m}{k}\take{\origin}_s$ because if
$\W=\poi$ or $\W=\eucl$ or $\W=\gal$, then it can be easily seen that
there is $f\in\W$ such that $f\take{\tunit}_s \neq f\take{\origin}_s$.
Let such $m$ and $k$ be fixed. Then \EMovingIOb holds. Thus, by
\CiteTheorem{thm:characterisation}, there is a unique $\kappa$ such that the set
of worldview transformations is a subset of $\kiso$. This $\kappa$ is
positive ($\kappa=1$) in $\M_P$, negative ($\kappa=-1$) in $\M_E$ and
$0$ in $\M_G$. Then by equation \eqref{eq:kappa} in
\CiteLemma{lem:k-is-unique} it can be seen that $\ESlowClock$ holds in
$\M_P$, $\EFastClock$ holds in $\M_E$ and \EMovingAccurateClock holds
in $\M_G$.

(Case 4) It remains to prove that $\KIN+\AxSpr+\lnot\EMovingIOb$ has a model.
Let $\M_T$ be a model of $\KIN+\AxSpr$ such that $\W=\triv$. Such
$\M_T$ exists by \CiteTheorem{thm:group-instantiation} and \eqref{eq:tartalmazasok}.  Let us notice that for
any $f\in\triv$, $f\take{\tunit}_s = f\take{\origin}_s$. Therefore,
for every $m,k\in\IOb$, $\w{m}{k}\take{\tunit}_s =
\w{m}{k}\take{\origin}_s$, and this means that $\lnot\EMovingIOb$
holds in $\M_T$.
\end{proof}

\section{Discussion}
\label{sec:discussion}

In this paper, we have presented an essentially elementary description
of what can be deduced about the geometry of $(1+3)$-dimensional
spacetime from isotropy if we restrict ourselves to first-order
logic and make as few background assumptions as reasonably
possible. Nonetheless, there is potential to go further, as even our
own very simple assumptions can potentially be weakened while still
providing a physically relevant description. The history of the field
has shown repeatedly that authors have inadvertently made unconscious,
and sometimes unnecessary, assumptions, and it would be foolish to
assume that we are necessarily immune to this problem. We have
accordingly started a programme of painstakingly machine-verifying our
results using interactive theorem provers \cite{StannettNemeti}, but
this programme remains very much in its infancy. In the meantime,
therefore, we have been as explicit as possible at all stages of our
proofs.

We began by noting that, in the elementary framework advocated in this paper 
there are reasons why it is no longer appropriate to assume that the ordered field \Q of 
numbers used when recording physical measurements is the field $\mathbb{R}$ of 
real numbers. Partly this is because practical measurements can never achieve 
more than a few decimal points of accuracy, and partly because the field
$\mathbb{R}$ cannot be uniquely characterised in terms of the first-order
sentences it satisfies. But as we have also shown, it is simply not necessary to make
the assumption. As long as \Q allows the taking of square roots of non-negative
values, all of our results hold.

Our results tell us, subject to a small number of very basic axioms,
that the worldview transformations that characterise kinematics in
isotropic spacetime form a group \W of $\kappa$-isometries for some
$\kappa$. In contrast to earlier studies, we have not needed to assume
the full special principle of relativity, but have shown instead that
the strictly weaker assumption that space is isotropic is already
enough to entail these results. We accordingly obtain four basic
possibilities: the universe is not static (there are moving
  observers) and $\W$ is a subgroup of either \poi, \eucl or \gal, or
 the universe is static (all observers are at rest
with respect to one another) and $\W \subseteq \triv$.

As usual (if moving observers exist) we can identify which kind of spacetime we are in by considering whether moving clocks run slow or fast or remain accurate. But because we have not restricted ourselves to $\Q = \mathbb{R}$, we have allowed for the possibility that
the structure of \Q may be somewhat more complicated than usually assumed (for example,
there is no reason why \Q should not contain infinite or infinitesimal values). This in turn means that 
the topological structure of $\Q^4$ need not satisfy the usual theorems of $\mathbb{R}^4$, nor the
symmetry group $\sym(\Q^4)$ those of $\sym(\mathbb{R}^4)$. Even so, we have shown that all `reasonable'
subgroups \G of $\sym(\Q^4)$ can occur as the transformation group $\W$ in some associated model $\M_{\G}$. In other words, assuming that $\Q = \mathbb{R}$ has inadvertently imposed severe and unnecessary limitations on the set of models investigated in earlier papers.

Nonetheless, many questions remain to be answered. Which of our results still hold, for example, if we remove the requirement for \Q to be Euclidean? Are square roots essential, and if not, how can this be interpreted physically? For example, when $\kappa > 0$ the value $\kappa$ corresponds to a model in which the speed of light is given by $c = \sqrt{1/\kappa}$, but what happens if $\kappa$ has no square root? Presumably this would be a model in which light signals cannot exist, since they would need to travel with non-existent speed. Some familiar expressions might still be meaningful, for example $\sqrt{1 - v^2/c^2}$ can be rewritten as $\sqrt{1 - \kappa v^2}$, but even so, how does time dilation `work' if $v$ is a value for which $\sqrt{1 - \kappa v^2}$ is undefined?

There is also the issue of dimensionality. Our initial investigations suggest that all of the
proofs presented here go through for dimensions $d \geq (1+3)$, but can fail for $d = (1+1)$. But do they hold
for $d = (1+2)$?  The answer appears to be yes if we allow trivial transformations to reverse the direction of time --- but is this inclusion of reflections essential? We simply do not know.

\bibliographystyle{amsalpha}
\bibliography{LogRel2019}

\end{document}

%% file: header.tex
\newcommand{\semph}[1]{\emph{\textbf{#1}}}
\newcommand{\lemph}[1]{\emph{#1}}

\newcommand{\InformalStatement}[1]{}

\newcommand{\InsertBeforeTheoremProof}[1]{}

\theoremstyle{definition}
\newtheorem{thm}{Theorem}[section]

\newtheorem{lem}{Lemma}[subsection]
\newtheorem{cor}[lem]{Corollary}
\newtheorem{prop}[lem]{Proposition}
\newtheorem{claim}[lem]{Claim}
\newtheorem{defn}[lem]{Definition}

\newtheorem{rem}{Remark}[section]

\numberwithin{equation}{section}

\newenvironment{newaxiom}
 {\smallskip\begin{description}}
 {\end{description}\smallskip}

\newcommand{\MATH}[1]{\ensuremath{#1}\xspace}
\newcommand{\MATHBF}[1]{\MATH{\mathbf{#1}}}
\newcommand{\MATHIT}[1]{\MATH{\mathit{#1}}}
\newcommand{\MATHBB}[1]{\MATH{\mathbb{#1}}}
\newcommand{\MATHSF}[1]{\MATH{\mathsf{#1}}}
\newcommand{\MATHCAL}[1]{\MATH{\mathcal{#1}}}

\newcommand{\ax}[1]{\MATHSF{#1}}
\newcommand{\EInf}{\ax{\exists\infty Speed}}
\newcommand{\KIN}{\ax{KIN}}

\newcommand{\Ax}[1]{\ax{Ax{#1}}}

\newcommand{\AxIso}{\Ax{Isotropy}}
\newcommand{\AxSpr}{\Ax{SPR}}
\newcommand{\AxColocate}{\Ax{Colocate}}
\newcommand{\AxEField}{\Ax{EField}}
\newcommand{\AxRelocate}{\Ax{Relocate}}

\newcommand{\AxWvt}{\Ax{Wvt}}
\newcommand{\AxLine}{\Ax{Line}}

\newcommand{\M}{\MATHCAL{M}}

\newcommand{\dU}[1]{\MATH{#1^{\uparrow}}}
\newcommand{\dD}[1]{\MATH{#1^{\downarrow}}}
\newcommand{\dP}[1]{\MATH{#1_{+}}}
\newcommand{\dN}[1]{\MATH{#1_{-}}}
\newcommand{\dST}[1]{\MATH{#1_{\oplus}}}
\newcommand{\dTS}[1]{\MATH{#1_{\ominus}}}
\newcommand{\dUP}[1]{\MATH{#1^{\uparrow}_{+}}}
\newcommand{\dDP}[1]{\MATH{#1^{\downarrow}_{+}}}
\newcommand{\dUN}[1]{\MATH{#1^{\uparrow}_{-}}}
\newcommand{\dDN}[1]{\MATH{#1^{\downarrow}_{-}}}
\newcommand{\dPU}[1]{\MATH{#1^{\uparrow}_{+}}}
\newcommand{\dPD}[1]{\MATH{#1^{\downarrow}_{+}}}
\newcommand{\dNU}[1]{\MATH{#1^{\uparrow}_{-}}}
\newcommand{\dND}[1]{\MATH{#1^{\downarrow}_{-}}}

\newcommand{\CreateDecorations}[2]{
  \expandafter\newcommand\csname #1U\endcsname{\dU{#2}}
  \expandafter\newcommand\csname #1D\endcsname{\dD{#2}}
  \expandafter\newcommand\csname #1P\endcsname{\dP{#2}}
  \expandafter\newcommand\csname #1N\endcsname{\dN{#2}}
  \expandafter\newcommand\csname #1UP\endcsname{\dUP{#2}}
  \expandafter\newcommand\csname #1UN\endcsname{\dUN{#2}}
  \expandafter\newcommand\csname #1DP\endcsname{\dDP{#2}}
  \expandafter\newcommand\csname #1DN\endcsname{\dDN{#2}}
  \expandafter\newcommand\csname #1PU\endcsname{\dPU{#2}}
  \expandafter\newcommand\csname #1PD\endcsname{\dPD{#2}}
  \expandafter\newcommand\csname #1NU\endcsname{\dNU{#2}}
  \expandafter\newcommand\csname #1ND\endcsname{\dND{#2}}
  \expandafter\newcommand\csname #1ST\endcsname{\dST{#2}}
  \expandafter\newcommand\csname #1TS\endcsname{\dTS{#2}}
}

\CreateDecorations{k}{\kiso}
\CreateDecorations{p}{\poi}
\CreateDecorations{e}{\eucl}
\CreateDecorations{g}{\gal}

\newcommand{\poi}{\MATHSF{Poi}}
\newcommand{\gal}{\MATHSF{Gal}}
\newcommand{\eucl}{\MATHSF{Eucl}}

\newcommand{\cpoi}{\MATH{c\mathsf{Poi}}}
\newcommand{\ceucl}{\MATH{c\mathsf{Eucl}}}

\newcommand{\srot}{\MATHSF{SRot}}
\newcommand{\tran}{\MATHSF{Trans}}
\newcommand{\iso}{\MATHSF{Iso}}
\newcommand{\sym}{\MATHSF{Sym}}
\newcommand{\kiso}[1][\kappa]{\MATH{{}_{#1}\iso}}

\newcommand{\triv}{\MATHSF{Triv}}
\newcommand{\speed}{\MATHSF{speed}} 

\newcommand{\mappedto}[2]{\MATH{\stackrel{#1}{\leadsto}_{ #2}}}

\newcommand{\G}{\MATHSF{G}} 
\newcommand{\GH}{\MATHSF{H}} 

\newcommand{\take}[1]{\MATH{\left(#1\right)}}
\newcommand{\takeset}[1]{\MATH{\left[{#1}\right]}}

\newcommand{\entrypar}[1]{\parbox{.7\columnwidth}{\smallskip{}{#1}\smallskip{}}}

\newcommand{\tick}{\MATH{\checkmark}}
\newcommand{\ptick}{\phantom{\tick}}

\newcommand{\oblines}{\MATHSF{ObLines}}
\newcommand{\obplanes}{\MATHSF{ObPlanes}}

\newcommand{\plane}{\MATHSF{plane}}
\newcommand{\abs}[1]{\MATH{\left|{#1}\right|}}
\newcommand{\norm}[1]{\abs{#1}}
\newcommand{\knorm}[2][\kappa]{\MATH{\left\|{#2}\right\|_{#1}}}

\newcommand{\saxis}{\MATHBF{S}}
\newcommand{\taxis}{\MATHBF{t}}
\newcommand{\xaxis}{\MATHBF{x}}
\newcommand{\yaxis}{\MATHBF{y}}
\newcommand{\zaxis}{\MATHBF{z}}
\newcommand{\Q}{\MATHIT{Q}}
\newcommand{\IOb}{\MATHIT{IOb}}

\newcommand*{\vv}[1]{\MATH{\vec{\mkern0mu#1}\,}}
\newcommand{\origin}{\vv{\MATHSF{o}}}
\newcommand{\tunit}{\vv{\MATHSF{t}}}
\newcommand{\xunit}{\vv{\MATHSF{x}}}
\newcommand{\yunit}{\vv{\MATHSF{y}}}
\newcommand{\zunit}{\vv{\mathsf{z}}}

\newcommand{\vvp}{\vv{p}}
\newcommand{\vvq}{\vv{q}}
\newcommand{\vvr}{\vv{r}}
\newcommand{\vvv}{\vv{v}}
\newcommand{\vvw}{\vv{w}}

\newcommand{\de}{\MATH{\stackrel{\text{\tiny def}}{=}}}

\newcommand{\wl}{\MATHSF{wl}}
\newcommand{\longeq}{\,\scalebox{2}[1]{=}\,}

\newcommand{\wrev}[2]{\MATHSF{w}_{{#2}{#1}}}
\newcommand{\w}[2]{\MATHSF{w}_{{#1}{#2}}} 

\newcommand{\slope}{\MATHSF{slope}}
\newcommand{\id}{\MATHSF{Id}}

\newcommand{\after}{\MATH{\circ}}

\newcommand{\eg}{e.g.,\ }
\newcommand{\ie}{i.e.\ }

\newcommand{\W}{\MATHBB{W}}
\newcommand{\QED}{\hfill\ensuremath{\Box}}

\newcommand{\EMovingIOb}{\ax{\exists MovingIOb}}

\newcommand{\ESlowClock}{\ax{\exists SlowClock}}
\newcommand{\EFastClock}{\ax{\exists FastClock}}
\newcommand{\EMovingAccurateClock}{\ax{\exists MovingAccurateClock}}
\newcommand{\AClockAccurate}{\ax{\forall ClockAccurate}}
\newcommand{\AMovingClockFast}{\ax{\forall MovingClockFast}}
\newcommand{\AMovingClockSlow}{\ax{\forall MovingClockSlow}}

\newcommand{\MakeLemmaReference}[2]
{ 
  \expandafter\gdef\csname lemName#2\endcsname{#1}
}

\newcommand{\LemmaName}[1]{\csname lemName#1\endcsname}
\newcommand{\CiteLemma}[2][Lemma]{{#1}~\ref{#2} (\LemmaName{#2})}

\newcommand{\CiteTheorem}[1]{\CiteLemma[Theorem]{#1}}

\newenvironment{Lemma}[1]
  {\begin{lem}[\LemmaName{#1}] \label{#1}}
  {\end{lem}}
\newenvironment{Theorem}[1]
  {\begin{thm}[\LemmaName{#1}] \label{#1}}
  {\end{thm}}

%% file: axspr.tikz
\newcommand{\coordsys}[1]{
\draw[->,>=latex] (0,0) to (0,3) node[left,black] {$#1$};
\draw[->,>=latex] (0,0) to (3,0); 
\draw[->,>=latex] (0,0) to (-2,-1);
}

\tikzstyle{eltolas}=[cm={0.3,-.1,0.15,.3,(1.1,1.2)}]

\pgfmathsetmacro{\x}{8}
\pgfmathsetmacro{\y}{5.6}

\begin{tikzpicture}[scale=0.5]
\small
\begin{scope}[shift={(0,\y)},very thick]
\coordsys{\forall k}
\begin{scope}[eltolas,thick,blue]
\coordsys{h}
\end{scope}
\node (k) at  (2,1.5){}; 
\end{scope}

\begin{scope}[shift={(\x,\y)},very thick,blue]
\coordsys{\forall h}
\node (k') at  (-1.2,1.5){}; 
\end{scope}

\begin{scope}[shift={(0,0)},very thick]
\coordsys{\forall k^*}
\begin{scope}[eltolas,dashed,thick,blue],
\coordsys{h^*}
\end{scope}
\node (h) at  (2,1.5){}; 
\end{scope}

\begin{scope}[shift={(\x,0)},dashed, very thick,blue]
\coordsys{\exists h^*}
\node (h') at  (-1.2,1.5){}; 
\end{scope}

\draw[->,ultra thick] (k') to[out=150,in=30]node[above]{$\w{k}{h}$} (k);
\draw[->,ultra thick] (h') to[out=150,in=30]node[above]{$\w{k^*}{h^*}$} (h);

\node at (\x/2,3*\y/4){$\w{k}{h}=\w{k^*}{h^*}$};

\end{tikzpicture}

%% file: Wk2.tikz
\newcommand{\coordsys}[1]{
\draw[->,>=latex] (0,0) to (0,3) node[left,black] {$#1$};
\draw[->,>=latex] (0,0) to (3,0); 
\draw[->,>=latex] (0,0) to (-2,-1);
}

\tikzstyle{eltolas}=[cm={-0.3,-.1,-0.15,.3,(1.6,0.7)}]
\tikzstyle{eltolas2}=[cm={0.3,-.2,0.25,.3,(1.3,2.3)}]
\tikzstyle{eltolas3}=[cm={0.2,.1,-0.1,.2,(1.4,-1.2)}]

\pgfmathsetmacro{\x}{8}
\pgfmathsetmacro{\y}{5}
\pgfmathsetmacro{\s}{.6}

\begin{tikzpicture}[scale=0.35]

\draw[gray,fill=gray!10] (\x/1.5,0.8) ellipse [x radius=1.5*\x,y radius=1.4*\y];
\node at (1.5*\x,1.45*\y) {\Large $\mathbb{W}_k$};
\begin{scope}[shift={(-.3,0)},ultra thick,scale=1.2]
\coordsys{k}
\begin{scope}[eltolas,thick,red!95!black]
\coordsys{b}
\end{scope}
\begin{scope}[eltolas2,thick,green!80!black]
\coordsys{a}
\end{scope}
\begin{scope}[eltolas3,thick,blue]
\coordsys{c}
\end{scope}
\node (bk) at  (2,-1.5){}; 
\node (tk) at  (2.3,3.3){}; 
\node (rk) at (2.7,1){};
\end{scope}

\begin{scope}[shift={(\x,\y)},very thick,scale=\s,green!80!black]
\coordsys{a}
\node (a) at  (-2,.5){}; 
\end{scope}

\begin{scope}[shift={(1.3*\x,0)},very thick,scale=\s,red!95!black]
\coordsys{b}
\node (b) at  (-1.5,1.2){}; 
\end{scope}

\begin{scope}[shift={(0.8*\x,-\y)}, very thick,scale=\s,blue]
\coordsys{c}
\node (c) at  (-1.5,1){}; 
\end{scope}

\draw[->,ultra thick] (a) to[out=180,in=60]node[above left]{$\w{k}{a}$} (tk);
\draw[->,ultra thick] (b) to[out=150,in=30]node[above left]{$\w{k}{b}$} (rk);
\draw[->,ultra thick] (c) to[out=170,in=-60]node[below left]{$\w{k}{c}$} (bk);

\node at (1.4*\x,-0.7*\y) {\Huge \ldots};


\end{tikzpicture}